\documentclass[twocolumn]{aastex62}
\usepackage{graphicx}
\usepackage{natbib}
\usepackage{amsmath,amsthm,amssymb}
\usepackage{color}
\usepackage{ulem}
\usepackage{epstopdf}
\newcommand{\gaiat}{Gaia~DR2}
\newcommand{\PS}{\protect \hbox {Pan-STARRS1}}
\newcommand{\spitzer}{Spitzer}                                    

\newcommand{\yps}{\ensuremath{y_{\rm P1}}}

\newcommand{\jtwo}{\ensuremath{J_{\rm 2MASS}}}
\newcommand{\kabstwo}{\ensuremath{M_{Ks,{\rm 2MASS}}}}

\newcommand{\ymko}{\ensuremath{Y_{\rm MKO}}}
\newcommand{\jmko}{\ensuremath{J_{\rm MKO}}}
\newcommand{\mjmko}{\ensuremath{M_{\jmko}}}
\newcommand{\hmko}{\ensuremath{H_{\rm MKO}}}
\newcommand{\mhmko}{\ensuremath{M_{\hmko}}}
\newcommand{\kmko}{\ensuremath{K_{\rm MKO}}}
\newcommand{\jhmko}{\ensuremath{(J-H)_{\rm MKO}}}
\newcommand{\jkmko}{\ensuremath{(J-K)_{\rm MKO}}}

\newcommand{\yjm}{$y_{\rm P1}-J_{\rm MKO}$}
\newcommand{\ykm}{$y_{\rm P1}-K_{\rm MKO}$}
\newcommand{\jmkowa}{$J_{\rm MKO}-W1$}
\newcommand{\jmkowb}{$J_{\rm MKO}-W2$}
\newcommand{\ywa}{$y_{\rm P1}-W1$}
\newcommand{\wawb}{$W1-W2$}

\newcommand{\fldg}{\mbox{\textsc{fld-g}}}
\newcommand{\intg}{\mbox{\textsc{int-g}}}
\newcommand{\vlg}{\mbox{\textsc{vl-g}}}

\newcommand{\rchi}{\ensuremath{\chi^2_\nu}}

\newcommand{\teff}{\ensuremath{T_{\rm eff}}}
\newcommand{\dphot}{\ensuremath{d_{\rm phot}}}

\newcommand{\plx}{\ensuremath{\varpi}}
\newcommand{\mjup}{\ensuremath{M_{\rm Jup}}}           
\newcommand{\um}{$\mu$m}
\newcommand{\vmax}{\ensuremath{V/V_{\rm max}}}
\newcommand{\exvmax}{\ensuremath{\langle V/V_{\rm max}\rangle}}
\newcommand{\dponesig}{\ensuremath{\dphot\le 25+1\sigma_{\rm dist}}}

\newcommand{\varngoodplx}{348}
\newcommand{\varnvollim}{369}
\newcommand{\varnfromukirt}{128}
\newcommand{\varnfromgaia}{121}
\newcommand{\varnfromlit}{120}
\newcommand{\varnsingle}{301}
\newcommand{\varnbinary}{44}
\newcommand{\varbinfraccorr}{$9.9\%\pm1.6\%$}

\newcommand{\varnbinldwarfs}{27}
\newcommand{\varbinlfraccorr}{$15.1\%\pm3.2\%$}

\newcommand{\varnbintdwarfs}{17}
\newcommand{\varbintfraccorr}{$5.9\%\pm1.6\%$}
\newcommand{\varnyoung}{22}
\newcommand{\varyoungfraccorr}{$5.5\%\pm1.2\%$}

\newcommand{\varsubdfraccorr}{$2.6\%\pm1.6\%$}
\newcommand{\varnwidecomp}{27}
\newcommand{\varwidecompfraccorr}{$8.3\%\pm1.7\%$}
\newcommand{\varnlttrans}{104}
\newcommand{\vardensityl}{$(4.41\pm0.33)\times10^{-3}$}
\newcommand{\vardensityt}{$(5.45\pm0.40)\times10^{-3}$}
\newcommand{\varcompletedist}{16}
\newcommand{\varcompletedistsig}{20}
\newcommand{\varcompletenum}{132}
\newcommand{\varvmaxfull}{0.41}
\newcommand{\varvmaxfullerr}{0.03}
\newcommand{\varcompletefull}{$83\%\pm5\%$}
\newcommand{\varcompletel}{$90\%\pm8\%$}
\newcommand{\varcompleteearly}{$92\%\pm7\%$}
\newcommand{\varcompletelate}{$69\%\pm8\%$}
\newcommand{\varvmaxearlyl}{$0.46\pm0.05$}
\newcommand{\varvmaxlatel}{$0.43\pm0.05$}
\newcommand{\varvmaxearlyt}{$0.50\pm0.07$}
\newcommand{\varnnewphot}{934}
\newcommand{\varnnewphotvollim}{238}
\newcommand{\varnnewphotother}{696}

\shorttitle{An L/T Transition Gap}
\shortauthors{Best, W. M. J. et al}

\received{October 30, 2019}
\revised{August 28, 2020}
\accepted{October 27, 2020}

\begin{document}

{\large Accepted by {\it The Astronomical Journal}, October 27, 2020}

\title{A Volume-Limited Sample of Ultracool Dwarfs. I. Construction, Space Density,
  and a Gap in the L/T Transition}

\correspondingauthor{William M. J. Best}
\email{wbest@utexas.edu}

\author[0000-0003-0562-1511]{William M. J. Best}
\affil{University of Texas at Austin, Department of Astronomy, 2515 Speedway C1400, Austin, TX 78712, USA}
\affil{Institute for Astronomy, University of Hawaii, 2680 Woodlawn Drive, Honolulu, HI 96822, USA}

\author[0000-0003-2232-7664]{Michael C. Liu}
\affil{Institute for Astronomy, University of Hawaii, 2680 Woodlawn Drive, Honolulu, HI 96822, USA}

\author[0000-0002-7965-2815]{Eugene A. Magnier}
\affil{Institute for Astronomy, University of Hawaii, 2680 Woodlawn Drive, Honolulu, HI 96822, USA}

\author[0000-0001-9823-1445]{Trent J. Dupuy}
\affil{Gemini Observatory, 670 N. A`ohoku Place, Hilo, HI, 96720, USA}
\affil{Institute for Astronomy, University of Edinburgh, Blackford Hill, Edinburgh, EH9 3HJ, United Kingdom}

\begin{abstract}

  We present a new volume-limited sample of L0--T8 dwarfs out to 25~pc defined
  entirely by parallaxes,
  using our recent measurements from UKIRT/WFCAM along with \gaiat\ and
  literature parallaxes.
  With \varnvollim\ members, our sample is the largest parallax-defined
  volume-limited sample of L and T dwarfs to date,
  yielding the most precise space densities for such objects.
  We find the local L0--T8~dwarf population includes {\varyoungfraccorr} young
  objects ($\lesssim$200~Myr) and {\varsubdfraccorr} subdwarfs,
  as expected from recent studies favoring representative ages $\lesssim$4~Gyr
  for the ultracool field population.
  This is also the first volume-limited sample to comprehensively map the
  transition from L to T~dwarfs (spectral types $\approx$L8--T4).
  After removing binaries, we identify a previously unrecognized, statistically
  significant ($>4.4\sigma$) gap $\approx$0.5~mag wide in \jkmko\ colors in the
  L/T transition,
  i.e., a lack of such objects in our volume-limited sample,
  implying a rapid phase of atmospheric evolution.
  In contrast, the most successful models of the L/T transition to date --- the
  ``hybrid'' models of Saumon \& Marley --- predict a pileup of objects at
  the same colors where we find a deficit,
  demonstrating the challenge of modeling the atmospheres of cooling brown
  dwarfs.
  Our sample illustrates the insights to come from even larger parallax-selected
  samples from the upcoming Legacy Survey of Space and Time (LSST) by the Vera
  Rubin Obsevatory.

\end{abstract}


\section{Introduction}
\label{intro}

Brown dwarfs have masses $\lesssim$70~\mjup\
\citep[e.g.,][]{Chabrier:2000hq,Dupuy:2017ke}, insufficient to sustain hydrogen
fusion and achieve the steady-state luminosity that defines main-sequence stars.
Brown dwarfs therefore cool as they age, causing their atmospheres to undergo
complex chemical transformations over time.  This is particularly true in the
L/T transition (spectral types $\approx$L8--T4), where evolutionary and
atmospheric models have difficulty reproducing observed magnitudes,
luminosities, and effective temperatures for L/T objects with known masses
and/or ages
\citep[e.g.,][]{Barman:2011fe,Bowler:2010ft,Dupuy:2009ga,Dupuy:2015gl,Leggett:2008kq,Naud:2014jx}.
Likewise, a fully physical model that matches observations of the L/T transition
in color-magnitude diagrams has yet to be developed
\citep[e.g.,][]{Burrows:2006ia,Saumon:2008im,Marley:2010kx,Tremblin:2016hi}.  Up
to this point, the small number of accurate parallax-based luminosities measured
for L/T transition dwarfs \citep[$<$30;][]{Best:2018kw} has hindered the
development of an accurate map for this distinctive evolutionary phase.

Volume-limited samples are ideal for population studies, as they minimize the
selection biases intrinsic to the more common magnitude-limited samples.
Parallaxes provide the most direct measures of distance to nearby objects and
are therefore preferred for establishing volume-limited samples in the solar
neighborhood.  Parallax-defined volume-limited samples of nearby brown dwarfs
enable the best estimates of the underlying mass and age distributions of the
local substellar population.  The most complete previous sample encompassing all
brown dwarf spectral types is the full-sky 8~pc sample of
\citet{Kirkpatrick:2012ha}, which contains only 33 L, T, and Y dwarfs.
\citet[hereinafter K19]{Kirkpatrick:2019kt} have recently assembled a
volume-limited sample of 278~objects out to 20~pc, defined using K19's
{\spitzer} parallaxes along with \gaiat\ and literature values.  However, their
sample comprises L0--L5 dwarfs (complete at 20~pc) and T6~and later-type dwarfs
(complete at $<$20~pc due to the faintness of these objects), entirely skipping
L6--T5 dwarfs, and therefore cannot be used to study the L/T~transition.  The
largest volume-limited samples of ultracool dwarfs (spectral types M6 and later)
published to date are those of \citet[196~M7--T2.5 dwarfs out to 20~pc, but
incomplete for types $>$L6 and selected primarily using photometric
distances]{Reid:2008fz} and \citet[hereinafter BG19; 410~M7--L5~dwarfs out to
25~pc]{BardalezGagliuffi:2019gn}. The latter study presents the most
comprehensive analysis to date of the warmest ultracool dwarfs down to the
hydrogen-burning limit \citep[spectral type $\approx$L4;][]{Dupuy:2017ke}, but
does not include the L/T~transition or cooler T~dwarfs.

In this paper, we present a new volume-limited sample of L and T~dwarfs out to
25~pc containing \varnvollim\ members, spanning spectral types L0--T8 and chosen
entirely by parallaxes, the largest such sample to date.  Using a near-infrared
(NIR) color-magnitude diagram for this sample, we identify a gap in $J-K$ color
in the L/T transition that is $\approx$0.5~mag wide.  We describe our
volume-limited sample in Section~\ref{vollim.sample}, present the L/T gap in
Section~\ref{lt.gap}, and discuss its implications in Section~\ref{discussion}.
We summarize our findings in Section~\ref{summary}.

\section{Volume-limited Sample}
\label{vollim.sample}

\subsection{Construction}
\label{vollim.construction}
\citet[hereinafter Best20]{Best:2020jr} used \linebreak UKIRT/WFCAM to obtain
parallaxes for \varngoodplx\ L0--T8~dwarfs with declinations
$-30^\circ\le\delta\le60^\circ$ and photometric distances \dponesig~pc (i.e., no
further beyond 25~pc than the distance uncertainty) based on AllWISE $W2$-band
photometry \citep{Cutri:2014wx}, with the goal of completing a volume-limited
25~pc sample for this portion of the sky (68.3\% of the full sky).
Additionally, we searched the literature for all spectroscopically confirmed
objects in Best20's spectral type and declination ranges that had parallax
measurements with errors $<20\%$, including parallaxes from \gaiat\
\citep{GaiaCollaboration:2018io}.  As in Best20, in all instances where an
object has both an optical and an NIR spectral type, we adopted the optical
types for L~dwarfs and the NIR types for T~dwarfs.  After merging the Best20 and
literature parallax lists and choosing the most precise parallax available for
each object, we removed objects with \hbox{parallaxes $<40$ mas} (i.e.,
\hbox{distances $>25$ pc}) to define our volume-limited sample.

\subsection{The Sample}
\begin{figure}
\begin{center}
    \includegraphics[width=1\columnwidth]{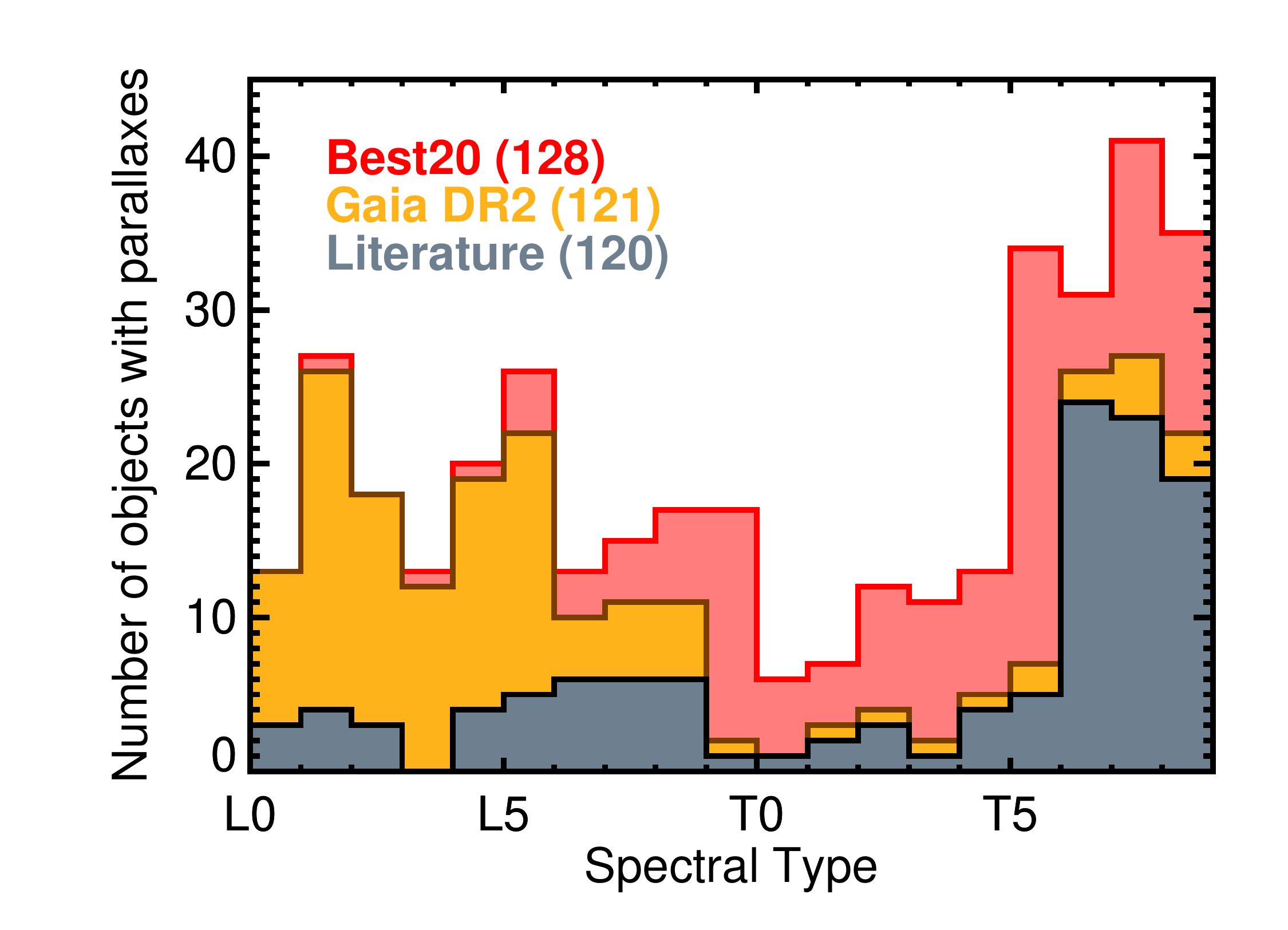}
    \caption{Distribution of spectral types for our volume-limited sample,
      indicating the sources of the parallaxes for sample members: Best20 (red),
      \gaiat\ (orange), and other literature sources (gray).  For objects with
      more than one parallax, we use the most precise measurement. (Most of the
      T dwarf parallaxes from \gaiat\ are actually parallaxes for main-sequence
      stars with T dwarf companions.)  The relative scarcity of L/T transition
      objects (especially types T0--T4) reflects the shorter timescale for this
      phase of brown dwarf cooling.}
  \label{fig.spt.source}
\end{center}
\end{figure}

We present our volume-limited sample in Table~\ref{tbl.sample}.  The sample
includes \varnvollim\ L0--T8 dwarfs and is constructed from \varnfromukirt\
Best20 parallaxes, \varnfromgaia\ from \gaiat, and \varnfromlit\ from other
literature sources.  Figure~\ref{fig.spt.source} shows the spectral type
distribution of our sample, featuring a clear deficit at early-T spectral types.
This deficit was predicted by evolutionary models \citep[e.g.,][hereinafter
SM08]{Saumon:2008im} and seen in previous photometrically selected samples
\citep{Burgasser:2007fl,Metchev:2008gx,Reid:2008fz,Marocco:2015iz,Best:2018kw},
and is now confirmed in our volume-limited sample, indicating that brown dwarfs
cool through these spectral types
\citep[$\teff\approx1400-1100$;][]{Luhman:2007fu,Cushing:2008kb,Stephens:2009cc,King:2010iu,Deacon:2012eg,Deacon:2012gf,Deacon:2017kd,Reyle:2014hz,Filippazzo:2015dv,Dupuy:2017ke}
in a relatively short time.  We also note an uneven distribution of
L0--L5~dwarfs whose origin is unclear given that our sample is $\gtrsim$90\%
complete at these types (Section~\ref{vollim.completeness}).  One possible
contributing factor is that the spectral types are a mixture of optical and NIR
assignments by many different astronomers using several methods.  At the L0
boundary of our sample in particular, this lack of consistency in typing could
potentially have impacted the choice of objects we included in our sample (we
included L0 and later types, but not M9.5 and earlier types).  However, the
relative lack of L0~dwarfs compared with L1~dwarfs was seen previously in the
more homogenously typed 20~pc samples of \citet{Cruz:2007kb},
\citet{Reid:2008fz}, and BG19, which also included late-M dwarfs, suggesting
that the L0 deficit is not a selection effect at the spectral type boundary of
our sample.

\begin{longrotatetable}
\pagestyle{empty}
\begin{deluxetable*}{lCCccCCCCr}
\centering
\tablecaption{Our Volume-limited 25 pc Sample of L0--T8 Dwarfs \label{tbl.sample}}
\tabletypesize{\scriptsize}
\setlength{\tabcolsep}{0.04in}
\tablewidth{0pt}
\rotate
\tablehead{   
  \colhead{Object} &
  \colhead{Distance} &
  \colhead{Parallax} &
  \colhead{Spectral Type\tablenotemark{a}} &
  \colhead{Flag\tablenotemark{b}} &
  \colhead{\ymko} &
  \colhead{\jmko} &
  \colhead{\hmko} &
  \colhead{\kmko} &
  \colhead{References} \\
  \colhead{} &
  \colhead{(pc)} &
  \colhead{(mas)} &
  \colhead{(Optical/NIR)} &
  \colhead{} &
  \colhead{(mag)} &
  \colhead{(mag)} &
  \colhead{(mag)} &
  \colhead{(mag)} &
  \colhead{(Disc; \plx; SpT; Flag; Phot)}
}
\startdata
SDSS J000013.54+255418.6 & $14.1\pm0.4$ & $70.8\pm1.9$ & T5/T4.5 & \nodata & $15.80\pm0.06$ & $14.73\pm0.03$ & $14.74\pm0.03$ & $14.82\pm0.03$ & 108; 64; 153,29; --; 64,108 \\
2MASS J00132229$-$1143006 & $24.8\pm1.9$ & $40.3\pm3.1$ & .../T3pec & \nodata & \nodata & $16.05\pm0.02$ & $[[15.74\pm0.22]]$ & $[[15.76\pm0.22]]$ & 97; 9; 97; --; 1,9,51 \\
2MASSW J0015447+351603 & $17.06\pm0.11$ & $58.6\pm0.4$ & L2/L1.0 & \nodata & $[14.95\pm0.05]$ & $13.74\pm0.02$ & $[12.96\pm0.02]$ & $[12.25\pm0.02]$ & 101; 77; 101,5; --; 1,9 \\
PSO J004.6359+56.8370 & $21.5\pm1.8$ & $46.5\pm3.9$ & .../T4.5 & \nodata & \nodata & $16.22\pm0.02$ & \nodata & \nodata & 125; 9; 125; --; 9 \\
2MASS J00282091+2249050 & $24.2\pm1.2$ & $41.3\pm2.0$ & .../L7: & \nodata & $[16.82\pm0.05]$ & $15.49\pm0.02$ & $[14.55\pm0.06]$ & $[13.77\pm0.06]$ & 37; 77; 37; --; 1,9 \\
WISE J003110.04+574936.3 & $14.1\pm1.0$ & $71.0\pm5.0$ & .../L9 & B & $[15.92\pm0.05]$ & $14.796\pm0.012$ & $13.862\pm0.014$ & $[13.21\pm0.03]$ & 185; 9; 10; 12; 1,10 \\
PSO J007.9194+33.5961 & $22.0\pm1.8$ & $45.4\pm3.8$ & .../L9 & \nodata & $[17.48\pm0.06]$ & $16.38\pm0.02$ & $[15.46\pm0.05]$ & $[14.67\pm0.05]$ & 11; 9; 11; --; 1,9 \\
2MASS J00320509+0219017 & $24.4\pm0.3$ & $41.0\pm0.4$ & L1.5/M9 & \nodata & $15.443\pm0.004$ & $14.220\pm0.002$ & $13.446\pm0.002$ & $12.797\pm0.002$ & 164; 77; 164,193; --; 111 \\
\enddata
\end{deluxetable*}
\end{longrotatetable}

\begin{figure}
  \centering
  \includegraphics[width=1\columnwidth, trim = 20mm 0 10mm 0]{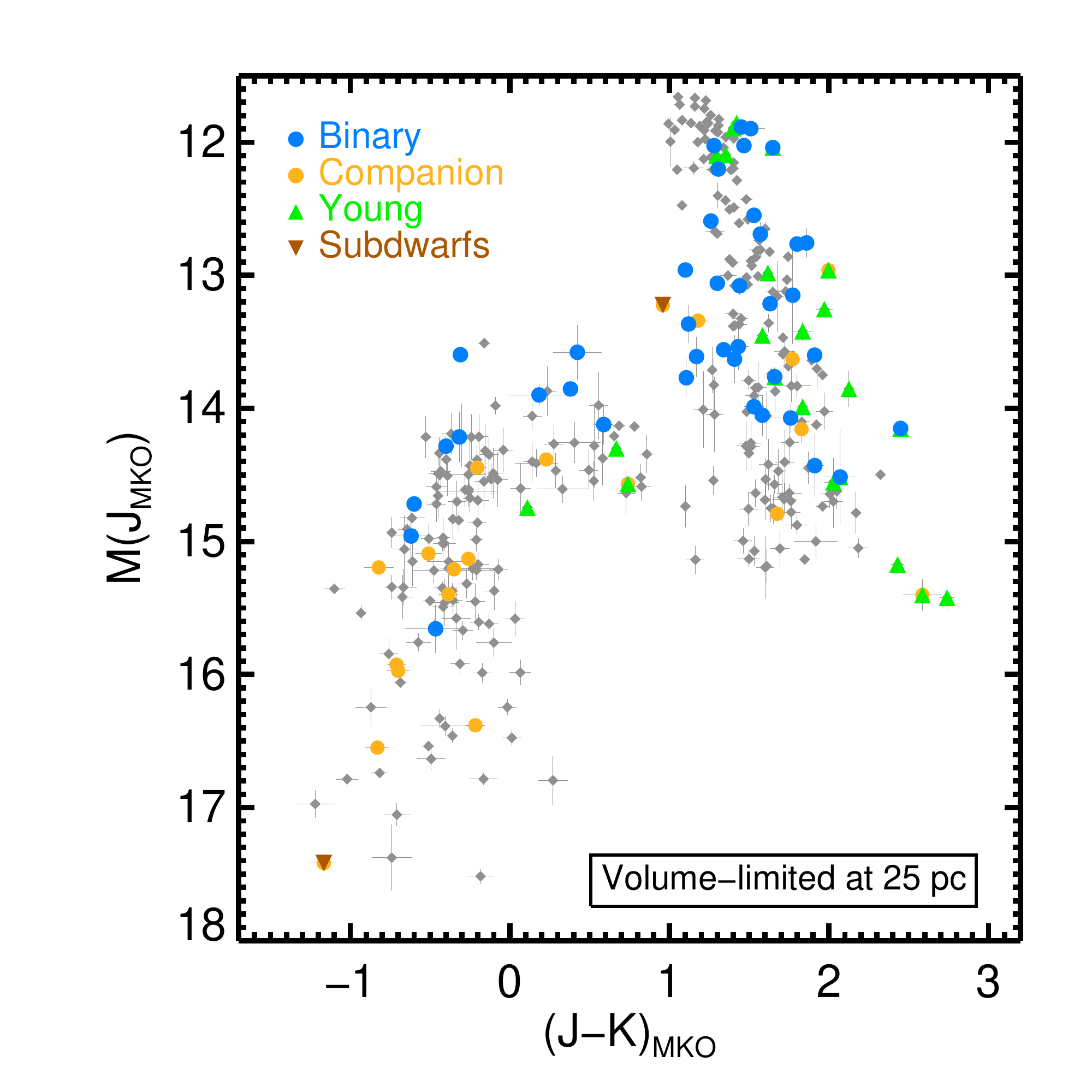}
  \caption{$M_J$ vs. $J-K$ (Maunakea Observatories system) color-magnitude
    diagram (CMD) for our volume-limited sample of L0--T8~dwarfs, highlighting
    binaries (blue circles), companions (orange circles), young objects
    ($\lesssim$200~Myr; green triangles), and subdwarfs (brown triangles), with
    other members plotted as gray diamonds.  As previously noted
    \citep[e.g.,][]{Liu:2013gy,Liu:2016co,Faherty:2016fx}, young L~dwarfs
    ($J-K>1$~mag) typically have redder colors than field-age L~dwarfs.
    Binaries (shown here with integrated-light photometry) tend to sit higher on
    the CMD than single brown dwarfs, as expected for intrinsically brighter
    objects.  The companions are not distinguishable from single objects in this
    CMD.  The gap in the L/T transition at $(J-K, M_J)\approx(1, 14.5)$~mag has
    not been identified previously and is discussed in Sections \ref{lt.gap}
    and~\ref{discussion}.}
  \label{fig.cmd.jjk.binyoung}
\end{figure}

Figure~\ref{fig.cmd.jjk.binyoung} shows our volume-limited sample in an NIR
color-magnitude diagram (CMD), highlighting the young ($\lesssim$200~Myr)
objects and binary systems (Sections \ref{vollim.young}
and~\ref{vollim.binaries} respectively).  We have removed objects with parallax
uncertainties $\ge$20\% of the parallax and objects with {\jkmko} uncertainties
$\ge$0.2~mag from the figure.  The CMD clearly shows the established
evolutionary sequence for brown dwarfs \citep[e.g.,][hereinafter
DL12]{Dahn:2002fu,Knapp:2004ji,Dupuy:2012bp}: L~dwarfs ($J-K>1$~mag) moving down
the right-hand sequence as they cool and their luminosities decline; the L/T
transition (types $\approx$L8--T4) where brown dwarfs become $\approx$2~mag
bluer in $J-K$, moving right to left on the CMD, while brightening by
$\approx$0.5~mag in $J$~band; followed by a drop in NIR absolute magnitudes and
widening spread of $J-K$ colors for cooling late-T dwarfs.  We note that types
$\lesssim$L4 are mostly hydrogen-burning stars \citep{Dupuy:2017ke} that will
remain at $M_J\lesssim12.5$~mag throughout their main-sequence liftetimes.  Our
volume-limited CMD exhibits a clear, previously unidentified gap in the L/T
transition at $\jkmko\approx0.9$--1.4~mag which we discuss in detail in Sections
\ref{lt.gap} and~\ref{discussion}.

\subsection{Completeness}
\label{vollim.completeness}
We assessed the completeness of our volume-limited sample using the \vmax\
statistic \citep{Schmidt:1968jc}.  Here $V$ is the volume of space enclosed at
the distance to a given object in the sample, and $V_{\rm max}$ is the volume of
space enclosed by the outer boundary of the sample.  \vmax\ thus quantifies the
position of a given object within the sample with a value between 0 and 1;
objects in the inner half of the sample's volume have $\vmax<0.5$, while objects
in the outer half have $\vmax>0.5$. For a sample with uniform spatial
distribution, the expectation value is therefore $\exvmax=0.5$.  Significant
differences from $\exvmax=0.5$ indicate that a sample is not uniformly
distributed in its volume.  Our 25~pc volume resides near the midplane of the
Galaxy and contains no clusters, so uniform spatial distribution is a reasonable
assumption for our sample, and thus deviations from $\exvmax=0.5$ would imply
incompleteness.  Because more distant objects in samples are fainter and more
difficult to observe, samples centered on the Sun tend to be less complete in
their outer portions.  Determining \exvmax\ for a sample over a series of
distances can reveal the extent to which a sample becomes incomplete approaching
its outer boundary.

\begin{figure*}
  \centering
  \includegraphics[width=2\columnwidth]{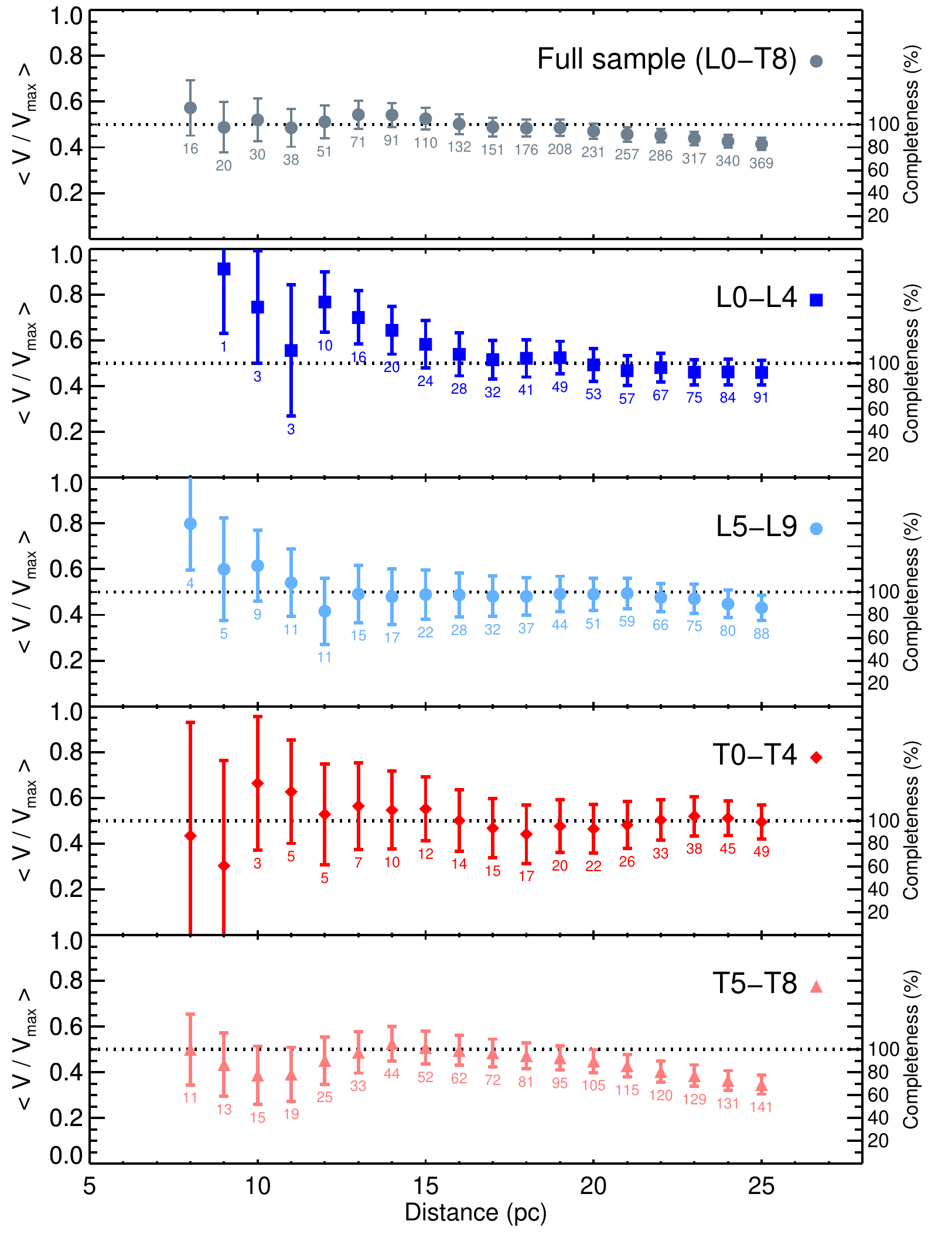}
  \caption{\exvmax\ as a function of limiting distance for our volume-limited
    sample, for all spectral types in our full sample (L0--T8, top panel) and
    for four spectral type bins ({\it other panels}).  The right-hand axes
    indicate the completeness values corresponding to \exvmax\ values less than
    0.5.  Our full sample is consistent with completeness (i.e., {\exvmax} is
    within $1\sigma$ of 0.5) out to 20~pc, with the declining trend of {\exvmax}
    implying that the sample starts to become incomplete beyond 16~pc.  Our
    sample is complete at 22~pc for L0--T4.5 dwarfs, i.e., through the
    L/T~transition.  The completeness at 25~pc is {\varcompleteearly} for
    L0--T4.5~dwarfs, and {\varcompletelate} for T5--T8~dwarfs.  The lower
    completeness for later-T~dwarfs is expected because T7--T8 dwarfs near 25~pc
    are too faint to be discovered by most large sky surveys.}
  \label{fig.vmax}
\end{figure*}

Figure~\ref{fig.vmax} shows \exvmax\ as a function of boundary distance for 
our volume-limited sample.  We estimated uncertainties for our {\exvmax}
calculations using the method described in Appendix~\ref{appendix.uncert.vmax}.
Our full volume-limited sample has $\exvmax=0.50\pm0.05$ at \varcompletedist~pc
(\varcompletenum\ objects), indicating completeness at this distance.  {\exvmax}
is within $1\sigma$ of 0.5 (implying consistency with completeness) out to
\varcompletedistsig~pc, and the steady decrease of {\exvmax} beyond 16~pc
implies that the sample is becoming less complete.  At 25~pc,
$\exvmax=\varvmaxfull\pm\varvmaxfullerr$, indicating {\varcompletefull}
completeness for our full sample.

Figure~\ref{fig.vmax} also breaks our sample into four spectral type bins. The
{\exvmax} trends imply completeness out to $\approx$22~pc for L~dwarfs and the
full 25~pc for T0--T4.5~dwarfs, suggesting that our sample is complete at 22~pc
through the L/T~transition.  The sample is complete at $\approx$16~pc for
T5-T8~dwarfs.  At the 25~pc limit of our sample, it is {\varcompletel} complete
for L~dwarfs and {\varcompletelate} for T5--T8 dwarfs.  We expected our sample
to be less complete for spectral types later than T6 due to the limiting
magnitude of the WISE survey \citep{Wright:2010in}, the primary source of
late-T~dwarf discoveries in our sample.  We present our {\exvmax} values for our
sample and several subsets thereof, including the four spectral type bins shown
in Figure~\ref{fig.vmax} as well as individual spectral subtypes, in
Table~\ref{tbl.space.density}.

\floattable
\begin{deluxetable}{lCCCCCC}
\centering
\tablecaption{Space Density and {\exvmax} for Our 25 pc Sample of L0--T8 Dwarfs\label{tbl.space.density}}
\tablewidth{0pt}
\tablehead{   
  \colhead{} &
  \colhead{} &
  \colhead{} &
  \colhead{} &
  \multicolumn{3}{c}{Space Density} \\
  \colhead{} &
  \colhead{} &
  \colhead{} &
  \colhead{} &
  \multicolumn{3}{c}{(10$^{-3}$ objects pc$^{-3}$)} \\
  \cline{5-7}
  \colhead{Objects} &
  \colhead{Number} &
  \colhead{\exvmax\tablenotemark{a}} &
  \colhead{Corrected Number\tablenotemark{a}} &
  \colhead{Value} &
  \colhead{$\sigma_{\rm binomial}$} &
  \colhead{$\sigma_{\rm Poisson}$}
}
\startdata
L0 $\le{\rm SpT}<$ L1 & 13 & $0.47\pm0.14$ & $14.2\pm2.1$ & 0.32 & 0.05 & 0.09 \\
L1 $\le{\rm SpT}<$ L2 & 27 & $0.50\pm0.10$ & $28.0\pm3.0$ & 0.63 & 0.07 & 0.12 \\
L2 $\le{\rm SpT}<$ L3 & 18 & $0.49\pm0.12$ & $19.6\pm2.5$ & 0.44 & 0.06 & 0.10 \\
L3 $\le{\rm SpT}<$ L4 & 13 & $0.34\pm0.13$ & $20.4\pm3.1$ & 0.46 & 0.07 & 0.13 \\
L4 $\le{\rm SpT}<$ L5 & 20 & $0.46\pm0.12$ & $21.8\pm2.7$ & 0.49 & 0.06 & 0.11 \\
L5 $\le{\rm SpT}<$ L6 & 26 & $0.42\pm0.10$ & $30.7\pm3.5$ & 0.69 & 0.08 & 0.14 \\
L6 $\le{\rm SpT}<$ L7 & 13 & $0.42\pm0.15$ & $15.4\pm2.6$ & 0.34 & 0.06 & 0.10 \\
L7 $\le{\rm SpT}<$ L8 & 15 & $0.47\pm0.14$ & $15.8\pm2.3$ & 0.35 & 0.05 & 0.09 \\
L8 $\le{\rm SpT}<$ L9 & 17 & $0.40\pm0.12$ & $22.2\pm3.0$ & 0.50 & 0.07 & 0.12 \\
L9 $\le{\rm SpT}<$ T0 & 17 & $0.45\pm0.13$ & $17.9\pm2.6$ & 0.40 & 0.06 & 0.10 \\
T0 $\le{\rm SpT}<$ T1 & 6 & $0.60\pm0.22$ & $5.8\pm1.4$ & 0.13 & 0.03 & 0.05 \\
T1 $\le{\rm SpT}<$ T2 & 7 & $0.45\pm0.21$ & $8.2\pm1.8$ & 0.18 & 0.04 & 0.07 \\
T2 $\le{\rm SpT}<$ T3 & 12 & $0.41\pm0.15$ & $15.7\pm2.6$ & 0.35 & 0.06 & 0.10 \\
T3 $\le{\rm SpT}<$ T4 & 11 & $0.62\pm0.17$ & $8.4\pm1.6$ & 0.19 & 0.04 & 0.06 \\
T4 $\le{\rm SpT}<$ T5 & 13 & $0.45\pm0.14$ & $16.2\pm2.5$ & 0.36 & 0.06 & 0.10 \\
T5 $\le{\rm SpT}<$ T6 & 34 & $0.41\pm0.09$ & $40.7\pm4.2$ & 0.91 & 0.09 & 0.16 \\
T6 $\le{\rm SpT}<$ T7 & 31 & $0.35\pm0.09$ & $43.4\pm4.6$ & 0.97 & 0.10 & 0.18 \\
T7 $\le{\rm SpT}<$ T8 & 41 & $0.33\pm0.07$ & $62.4\pm5.8$ & 1.40 & 0.13 & 0.22 \\
T8 $\le{\rm SpT}<$ T8.5 & 35 & $0.29\pm0.08$ & $60.0\pm6.3$ & 1.34 & 0.14 & 0.23 \\
\hline
L0 $\le{\rm SpT}<$ L5 & 91 & $0.46\pm0.05$ & $99\pm6$ & 2.22 & 0.13 & 0.23 \\
L5 $\le{\rm SpT}<$ T0 & 88 & $0.43\pm0.05$ & $98\pm6$ & 2.20 & 0.14 & 0.24 \\
T0 $\le{\rm SpT}<$ T5 & 49 & $0.50\pm0.07$ & $50\pm4$ & 1.12 & 0.10 & 0.16 \\
T5 $\le{\rm SpT}\le$ T8 & 141 & $0.35\pm0.04$ & $200\pm10$ & 4.48 & 0.23 & 0.39 \\
\hline
L0 $\le{\rm SpT}<$ T0 & 179 & $0.45\pm0.04$ & $197\pm9$ & 4.41 & 0.19 & 0.33 \\
T0 $\le{\rm SpT}\le$ T8 & 190 & $0.39\pm0.04$ & $243\pm11$ & 5.45 & 0.24 & 0.40 \\
\hline
Single & 301 & $0.41\pm0.03$ & $363\pm13$ & 8.13 & 0.28 & 0.48 \\
Binary/triple\tablenotemark{b} & 44 & $0.50\pm0.08$ & $43\pm4$ & 0.97 & 0.09 & 0.15 \\
Companion\tablenotemark{c} & 27 & $0.37\pm0.03$ & $36\pm4$ & 0.81 & 0.09 & 0.16 \\
Young & 22 & $0.45\pm0.11$ & $24\pm3$ & 0.54 & 0.07 & 0.12 \\
\hline
All & 369 & $0.42\pm0.03$ & $438\pm13$ & 9.80 & 0.30 & 0.52 \\
\enddata
\tablecomments{
  \emph{Number}: number of objects in our volume-limited sample.
  \exvmax: calculated for our volume-limited sample. A sample with uniform
  spatial distribution will have $\exvmax=0.5$. Twice the {\exvmax} gives an
  estimate of the volume-completeness of each sample bin (1 = complete).
  \emph{Corrected Number}: number of objects in each bin, corrected for
  incompleteness (i.e., divided by $2\times\exvmax$).  The numbers in smaller
  bins may not add up to the numbers in larger bins because the Monte Carlo
  trials were run separately for each bin.
  \emph{Space Density}: corrected number for each bin divided by the volume of
  our sample (44703.031~pc$^3$). $\sigma_{\rm binomial}$ describes how precisely
  our space density measurements represent the full 25~pc volume around the
  Sun. $\sigma_{\rm Poisson}$ describes how precisely our space density
  measurements represent brown dwarfs in our general neighborhood of the Galaxy.
  The calculation of $\sigma_{\rm binomial}$ and $\sigma_{\rm Poisson}$ is
  described in Appendix~\ref{appendix.uncert.density}.
}
\tablenotetext{a}{Mean and standard deviation from Monte Carlo trials that
  resample the parallaxes from their errors and incorporate binomial
  uncertainties to account for statistical fluctuations in our sample
  (Appendix~\ref{appendix.uncert.density}).}
\tablenotetext{b}{Close binaries and triples are counted as single objects with
  unresolved spectral types.}
\tablenotetext{c}{Three companions are themselves binaries (see the text for
  details) and are also included in the binary/triple bin.}
\end{deluxetable}

The fact that the L~dwarfs in our sample appear to be complete out to a smaller
distance than the T0--T4.5~dwarfs is somewhat surprising, since later-type
objects are overall less luminous and should in principle be more difficult to
detect.  Given that the {\exvmax} values for the L0--L4~dwarfs (\varvmaxearlyl),
L5--L9~dwarfs (\varvmaxlatel), and T0--T4~dwarfs (\varvmaxearlyt) are all within
$1\sigma$ of each other out to 25~pc, the greater completeness of the early-Ts
could simply be a random statistical feature arising from our smaller spectral
type subsamples.  However, it could also arise from selection effects that may
have impacted our volume-limited sample, which we consider briefly here.  The
objects in our sample were discovered by multiple searches using different
telescopes and methods, but nearly all relied on photometry from large sky
surveys, in particular 2MASS \citep{Skrutskie:2006hl}, the {\PS} $3\pi$ Survey
\citep[PS1;][]{Chambers:2020vk}, WISE, SDSS \citep{York:2000gn}, and UKIDSS
\citep{Lawrence:2007hu}.  Searches using PS1 \citep[optical;
e.g.,][]{Best:2015em}, 2MASS \citep[NIR;
e.g.,][]{Kirkpatrick:2000gi,Reid:2008fz}, and WISE \citep[mid-infrared (MIR);
e.g.,][]{Mace:2013jh,Thompson:2013kv} cover the entire area of our
volume-limited sample in 12 bands spanning the full spectral energy distribution
of L0--T8 dwarfs, and each alone is sensitive enough to detect all of the
spectral types in our sample out to 25~pc, except for the more distant T6--T8
dwarfs (detected only by WISE).  SDSS \citep[optical;
e.g.,][]{Chiu:2006jd,Schmidt:2010ex} and UKIDSS \citep[NIR;
e.g.,][]{Burningham:2013gt,Marocco:2015iz} discovered many objects in smaller
regions of the sky, with the depth of UKIDSS also contributing to the discovery
of numerous late-T dwarfs.  While each search used different surveys and
criteria, in aggregate they have enabled detection of all late-L and early-T
spectral types to well beyond 25~pc.  In particular, the targeted search for
L6--T4.5~dwarfs by \citet{Best:2015em}, which discovered 24\% of the objects
with those spectral types in our volume-limited sample, also discovered objects
beyond 30~pc.  Many of the searches avoided the Galactic plane
($|b|\lesssim20^\circ$), and many looked only for objects with clear proper
motion, so it is possible that the majority of the undiscovered objects are
concentrated in the Galactic plane and/or are slow-moving, although there is no
reason why these factors would lead to the discovery of T~dwarfs at greater
distances than L~dwarfs.  We therefore conclude that if the early-T dwarfs in
our volume-limited sample are truly complete to a greater depth than the
L~dwarfs, it is most likely a statistical fluctuation arising from the smaller
sizes ($<$100~objects) of these spectral type bins.

\subsubsection{Anisotropy}
\label{vollim.completeness.anisotropy}
Our primary tool for analyzing the completeness of our sample, the {\vmax}
statistic, is unable to account for anisotropies over the sky.  We consider two
possible cases of anisotropy by other means.  One is the claim by
\citet{Bihain:2016ht} of a highly non-uniform distribution of brown dwarfs
within 6.5~pc, whereby 21 out of 26 objects lie ahead of the Sun in its Galactic
orbit.  K19 found no such anisotropy in their complete samples of early-L dwarfs
(out to 20~pc) and late-T dwarfs (out to 12.5~pc), describing the distribution
of the smaller \citet{Bihain:2016ht} sample as a random statistical effect.  Our
larger 25~pc volume-limited sample spanning L0--T8 spectral types can provide a
robust assessment of this claim, but we must account for the declination limits
in our sample that exclude more of the space trailing the Sun than leading.  In
a model population of isotropically distributed objects filling our sample's
volume, we find 59\% of the objects are ahead of the Sun and 41\% are behind.
For our volume-limited sample, 60\% of the objects are ahead of the Sun and 40\%
behind.  We thus find no evidence for anisotropy in our sample relative to the
Sun's Galactic orbit other than the cuts imposed by our declination limits,
supporting K19's explanation.

\begin{figure}
  \centering
  \includegraphics[width=1\columnwidth]{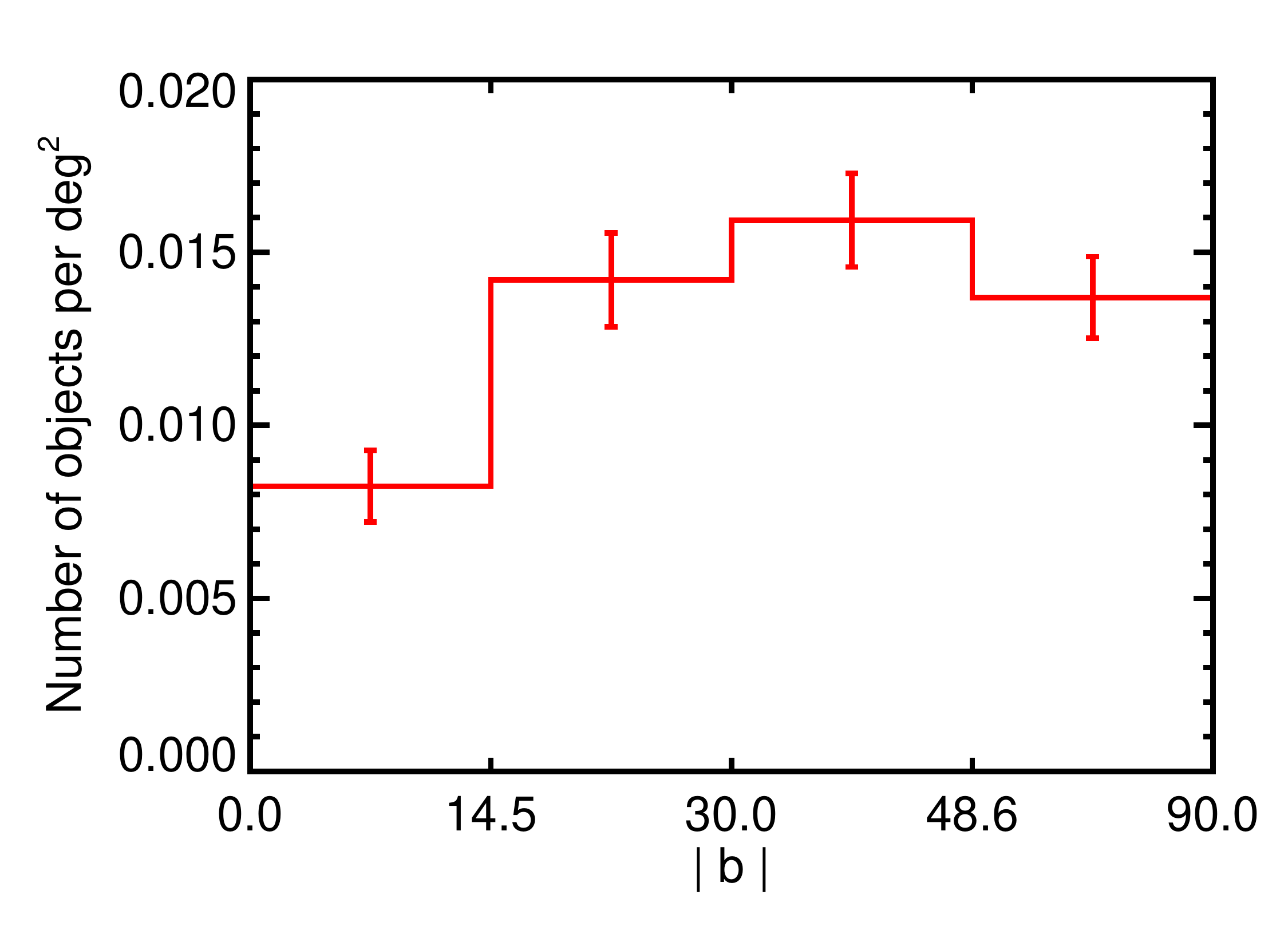}
  \caption{Distribution of objects in our volume-limited sample as a function of
    Galactic latitude, divided into four bins of $|b|$ each corresponding to a
    quarter of the full sky.  To account for the differing fractions of each bin
    that are actually within the declination limits of our sample, we normalized
    the number of objects in each bin to the area of sky it covers.  There is a
    clear deficit of objects near the Galactic plane, likely an artifact of the
    choice to avoid low Galactic latitudes made by many searches for brown
    dwarfs.  Objects missing from our volume-limited sample mostly reside at
    these low Galactic latitudes.}
  \label{fig.glat}
\end{figure}

The other anisotropy to consider is motivated by the fact that many searches for
ultracool dwarfs eschewed the Galactic plane ($|b|\lesssim20^\circ$), and by
K19's finding that their own 20~pc sample was deficient at low Galactic
latitudes.  To assess whether our volume-limited sample is similarly deficient,
we take the same approach as K19, dividing the sky into eight equal-area slices
at Galactic latitudes $b=0^{\circ}$, $\pm14.47^{\circ}$, $\pm30.00^{\circ}$, and
$\pm48.49^{\circ}$, except that we combine slices that have the same absolute
Galactic latitudes to obtain four bins, each covering 25\% of the sky.  We then
count the number of objects in each bin, calculating uncertainties from the
binomial distribution as described in Appendix~\ref{appendix.uncert.density}
(Equation~\ref{sig.binomial.num}).  Since our sample does not cover the full
sky, we must also account for the fact that the declination limits of our sample
include different fractions of the Galactic latitude bins.  Specifically, our
sample covers 62.4\% of the $0^{\circ}\le|b|<14.47^{\circ}$ slices, 62.8\% of
the $14.47^{\circ}\le|b|<30.00^{\circ}$ slices, 70.0\% of the
$30.00^{\circ}\le|b|<48.49^{\circ}$ slices, and 77.2\% of the
$48.49^{\circ}\le|b|\le90^{\circ}$ slices.  We use the coverage fractions to
convert the number of objects in each bin into a number of objects per square
degree.  We show the resulting distribution of objects as a function of absolute
Galactic latitude in Figure~\ref{fig.glat}.  There is a clear deficit at low
Galactic latitudes, echoing what K19 saw in their 20~pc sample, and confirming
the combined impact of many brown dwarf searches that avoided the Galactic
plane.  If we assume the sample is complete for $|b|\ge14.47^{\circ}$, then the
deficit seen at lower Galactic latitudes would comprise $\approx$13\% of a
complete sample, consistent with our {\varcompletefull} completeness estimate.
This indicates that most objects missing from our volume-limited sample reside
at $|b|<14.47^{\circ}$.

\subsection{Space Density}
\label{vollim.density}
We calculated the space density of our 25~pc volume-limited sample and several
subsets thereof, including young objects (Section~\ref{vollim.young}); binaries,
triples, and companions (Section~\ref{vollim.binaries}); and single objects.  We
divided the number of objects in each sample or subset by twice the
corresponding {\exvmax} value to obtain an estimate of the true number of
objects in our 25~pc volume, including objects missing from our sample.  We then
divided these corrected numbers of objects by our sample volume
(44703.031~pc$^3$) to obtain space densities.  We present our results in
Table~\ref{tbl.space.density}, including two different sets of uncertainties
($\sigma_{\rm binomial}$ and $\sigma_{\rm Poisson}$) for the space densities
whose calculation and purpose we describe in
Appendix~\ref{appendix.uncert.density}.  Briefly, the binomial uncertainties
$\sigma_{\rm binomial}$ are appropriate for estimating how well our
volume-limited sample's space density represents the full 25~pc volume, of which
our sample covers 68.3\%.  The Poisson uncertainties $\sigma_{\rm Poisson}$ are
appropriate for estimating how well our volume-limited sample represents L0--T8
dwarfs in the much larger local neighborhood of our Galaxy.  Both of these
uncertainties also incorporate the parallax uncertainties of individual objects
in our sample, whose impact is small compared to the statistical fluctuations
described by $\sigma_{\rm binomial}$ and $\sigma_{\rm Poisson}$.  For our
discussion here, we adopt the Poisson uncertainties in order to comment on the
space density of L0--T8 dwarfs in general and compare with previous estimates
(which adopt Poisson uncertainties).  Like previous studies, we do not account
in this work for uncertainties in the spectral types themselves.

The uncertainties on our space densities for L and T0--T8 dwarfs are $<$10\%,
making ours the most precise estimates to date spanning these spectral type
ranges.  (BG19 quote a similar precision for their M7--L5 sample.)
\setcitestyle{notesep={; }} Our \vardensityl~pc$^{-3}$ space density for
L~dwarfs and subsets thereof are consistent with most previous estimates
\citep[K19]{Cruz:2007kb,Reyle:2010gq,Marocco:2015iz}, but are a factor of 1.8
smaller for L0--L5~dwarfs than the estimate of BG19.  \setcitestyle{notesep={,
  }} The latter work attempts to correct for sample selection effects as well as
incompleteness in their sample.  BG19 identify their completeness correction as
larger than their correction for selection effects, so the most straightforward
explantion for their much larger space density is that their adopted sample
completeness is underestimated.
 
For T~dwarfs, our space densities are consistent with most previous estimates
\citep{Kirkpatrick:2012ha,Burningham:2013gt,Marocco:2015iz}.  Our overall
T0--T8~dwarf space density of \vardensityt~pc$^{-3}$ is lower than the
$\approx(7\pm3)\times10^{-3}$~pc$^{-3}$ estimates of \citet{Metchev:2008gx} and
\citet{Reyle:2010gq}, with much of the discrepancy coming at later T6--T8
dwarfs, although our estimate differs by less than 1$\sigma$ due to the large
uncertainties presented in those studies.  K19 present their space densities
using bins of {\teff}, which do not necessarily map to specific spectral types,
but assuming that their bins spanning 600--1050~K correspond approximately to
spectral types T6--T8, their estimate for those types appears also to be
$\approx$40\% higher than ours.  If this discrepancy is real, its source is
unclear, as our sample includes all of the
$\mathrm{T6}\le\mathrm{SpT}\le\mathrm{T8}$~dwarfs in K19's 20~pc sample, and the
incompleteness of our sample beyond 20~pc for these spectral types is addressed
by our {\exvmax} analysis (Figure~\ref{fig.vmax}) and consequent correction for
completeness.

\subsection{Photometry}
\label{vollim.photometry}
Table~\ref{tbl.sample} reports $YJHK$ photometry on the Maunakea Observatories
system \citep[MKO;][]{Simons:2002hh,Tokunaga:2002ex} for our volume-limited
sample.  Best20 used the $J$ band for parallax observations, providing us with
{\jmko} photometry for those objects.  For other objects, we use {\jmko}
photometry from the literature where available, and for all objects we use
{\ymko}, {\hmko}, and {\kmko} photometry from the literature where available.

When MKO photometry was not available, we calculated synthetic MKO photometry
using SpeX prism spectra \citep{Rayner:2003hf} from the SpeX Prism Library
\citep{Burgasser:2014jm} or from the literature.  We calibrated our synthetic
photometry with 2MASS photometry in the same band when available, or MKO
photometry from another NIR bandpass.  We calculated uncertainties in a Monte
Carlo fashion using the measurement uncertainties of the spectrum, adding in
quadrature the uncertainties from the calibrating photometry.  Following the
analysis of DL12, we added an additional 0.05~mag uncertainty in quadrature for
all synthesized colors between bands within a photometry system (e.g., \jhmko),
but added no additional uncertainty for conversions between systems in a single
band (e.g., ${\jmko}-{\jtwo}$).

When no prism spectra were available, we converted 2MASS magnitudes into the MKO
system using \kabstwo\ and the polynomials of
\citet[Appendix~A.2]{Dupuy:2017ke}.

\subsection{Young and Old Objects}
\label{vollim.young}
Our volume-limited sample contains \varnyoung\ young objects
($\lesssim$200~Myr), identified as such via spectroscopic indications of low
surface gravity, lithium absorption, or H$\alpha$ emission, and/or kinematic
association with a young moving group.  With completeness corrections, our
sample implies a population that is {\varyoungfraccorr} young (assuming Poisson
statistics), more than the expected $\approx$2\% ($7\pm3$~objects in our sample)
for a commonly assumed uniform age distribution in a 10~Gyr old galaxy.  This
suggests that the local population of brown dwarfs is not uniform and skews
toward younger ages.  BG19 similarly identify 33 out of 410 (8\%) of the M7--L5
dwarfs in their sample as young, and \citet{Kirkpatrick:2008ec} found that
$7.6\%\pm1.6\%$ of a sample of 303 L~dwarfs are younger than 100~Myr.  These
young-leaning samples echo the 0.4--4~Gyr distribution (median age 1.3~Gyr)
derived by \citet{Dupuy:2017ke} from evolutionary models using individual
luminosities and dynamical masses of 20 resolved L and T~dwarfs in binaries, as
well as the 0.5--2~Gyr age distribution found by \citet{ZapateroOsorio:2007cg}
in the kinematics of nearby 21 L and T~dwarfs, but are less consistent with the
statistical 3--8~Gyr kinematic age found by \citet{Faherty:2009kg} from the
proper motions of 184~L0--T8 dwarfs.  \citet{Dupuy:2017ke} point out that a
young-leaning age distribution such as ours is consistent with a constant star
formation rate coupled with dynamical heating, which tends to scatter older
objects out of the Galactic plane where our Sun resides.  Congruently, our
volume-limited sample contains only three likely old objects
(completeness-corrected \varsubdfraccorr): the T8 subdwarf
WISE~J200520.38+542433.9 \citep{Mace:2013ku}, and the two components of the
comoving brown dwarf pair SDSS~J141624.08+134826.7
\citep[sdL6;][]{Bowler:2010gd,Kirkpatrick:2010dc,Schmidt:2010dz} and
ULAS~J141623.94+134836.3
\citep[(sd)T7.5;][]{Burgasser:2010bj,Burningham:2010du,Scholz:2010gm}.  More
analysis of our volume-limited sample, including its kinematics and comparison
to synthetic populations of differing ages, will yield better constraints on its
age distribution.

Figure~\ref{fig.cmd.jjk.binyoung} shows the positions of the young objects in
the $M_J$ versus $J-K$ CMD of our volume-limited sample.  All but three of these
young objects are L~dwarfs, which display the previously observed trends
\citep[e.g.,][]{Liu:2013gy,Liu:2016co,Faherty:2016fx} of redder colors and (for
late-L~types) lower $J$-band luminosities than field-age L~dwarfs.  In general
we would expect T~dwarfs to be older than L~dwarfs because brown dwarfs cool as
they age, but since there are currently no clear spectroscopic identifiers of
youth for $\ge$L8~dwarfs, it is possible that our sample contains T~dwarfs with
unrecognized young ages, which would make the fraction of young objects in our
sample even higher.  The three identified young T~dwarfs in our sample are the
T2.5~dwarf companion HN~Peg~B, age $300\pm200$~Myr \citep{Luhman:2007fu}, the
T2.5~dwarf SIMP~J013656.5+093347.3 \citep{Artigau:2006bh} in the $200\pm50$~Myr
old Carina-Near Moving Group \citep{Zuckerman:2006ke,Gagne:2017bl}, and the
T5.5~dwarf SDSS~J111010.01+011613.1 \citep[hereinafter
SDSS~J1110+0116;][]{Geballe:2002kw} in the $149^{+51}_{-19}$~Myr old AB Doradus
Moving Group \citep{Bell:2015gw,Gagne:2015kf}.  Atmospheric models predict a
wide range of $J$-band luminosities for the L/T transition as a function of
gravity \citep[e.g., SM08;][]{Charnay:2018jn}, though \citet{Liu:2016co} do not
find such a wide range in their parallax sample.  Consistently with
\citet{Liu:2016co}, we see an $\approx$1~mag spread of \mjmko\ in the L/T
transition of our volume-limited sample, which could be an indication of a
variety of gravities.  All three of the young T~dwarfs in our sample are part of
the group of young benchmark T~dwarfs that \citet{Zhang:2020jn} describe as
marginally ($\lesssim$0.5~mag) fainter in {\jmko} than older objects of the same
spectral type.  We note that none of these T~dwarfs has ages less than 100~Myr,
so they are not as young as many of the young L~dwarfs in our sample.

Figure~\ref{fig.cmd.jjk.binyoung} also shows the positions of two \linebreak of
the subdwarfs (SDSS~J141624.08+134826.7 and ULAS~J141623.94+134836.3) on our
volume-limited CMD, both of which are among the bluest objects in their
respective regions of the brown dwarf evolutionary sequence.  The third subdwarf
in our sample, WISE~J200520.38+542433.9, does not appear on our CMD plot because
it lacks \kmko\ photometry.

\subsection{Binaries and Companions}
\label{vollim.binaries}
Binaries and multiples require special consideration in population studies.
These systems can be considered separately from single objects, or the
components can be treated as individual objects; both of these approaches
require identification of the binaries and multiples in the sample.
Alternatively, the impact of unresolved binaries on the photometry, luminosity,
and mass of the sample can be accounted for statistically
\citep[e.g.,][]{Metchev:2008gx,DayJones:2013hm}.  We searched the literature to
identify binaries in our volume-limited sample detected via high-angular
resolution imaging, radial velocity measurements, or astrometric signatures.
Contemporaneously with observations for the Best20 parallaxes, we also conducted
our own high-angular resolution imaging survey of candidate members of the
volume-limited sample lacking previous such observations across all spectral
types, using laser guide star adaptive optics (LGSAO) on Keck~II/NIRC2 at
$\approx$0{\farcs}05--0{\farcs}10 resolution (W. Best et al., in preparation).
The overwhelming majority of our sample (88\%) has now been observed with
high-angular resolution, without preference for particular spectral types,
meaning that our full L0--T8 evolutionary sequence has been explored for
binaries.  (The remaining objects were largely not observed due to the lack of
tip-tilt guide stars needed for LGSAO, which is also independent of spectral
type.)

We identified \varnbinary\ binaries in our volume-limited sample, from which we
calculate a {\varbinfraccorr} binary fraction for L0--T8~dwarfs, correcting for
spatial incompleteness (Section~\ref{vollim.completeness}) and assuming Poisson
statistics.\footnote{We do not treat the components of L and T dwarf binaries as
  separate objects in this discussion.  We also do not include L and T dwarfs
  that are in close binaries with earlier-type primaries in our sample.  Known L
  and T dwarf secondaries not included in our sample are LHS 1070C
  \citep{Leinert:1994ti,Rajpurohit:2012ef}, 2MASSW J0320284-044636B
  \citep{Blake:2008co,Burgasser:2008cj}, WISE J072003.20-084651.2B
  \citep{Burgasser:2015cn}, LHS 2397aB \citep{Freed:2003js}, 2MASS
  J17072343-0558249B \citep{McElwain:2006ca}, and LSPM J1735+2634
  \citep{Law:2006dm}.}  (Two-thirds of these binaries have resolved $J$- and
$K$-band photometry.)  This includes two systems that are thought to include a
third component: DENIS~J020529.0$-$115925 \citep{Bouy:2005de} and
2MASS~J07003664+3157266 \citep{Dupuy:2017ke}; for simplicity, we include these
systems in the category of ``binaries'' for this work.
Figure~\ref{fig.cmd.jjk.binyoung} highlights the binaries (with unresolved
photometry) in the $M_J$ versus $J-K$ CMD of our volume-limited sample.
Unresolved binaries combine the light of two objects and are therefore more
luminous than single objects, so their tendency to sit higher (brighter $M_J$)
than single objects on the CMD is expected.

A total of {\varnbinldwarfs} of the binaries have L-dwarf primaries in our
volume-limited sample.  After correcting for spatial incompleteness
(Section~\ref{vollim.completeness}), we find a {\varbinlfraccorr} L-dwarf binary
fraction, consistent with previous estimates \citep{Gizis:2003fj,Reid:2008gr}
for binaries resolvable with high-angular resolution imaging ($\gtrsim$1.5~au),
but somewhat lower than the $\approx$24\% fraction estimate of
\citet{Reid:2006dy} that includes L-dwarf binaries with smaller separations
\citep[see also][]{BardalezGagliuffi:2015fd}.  It is therefore reasonable to
expect that our volume-limited sample contains $\approx$10 unresolved binaries
with L-dwarf primaries.  For T~dwarfs, \varnbintdwarfs\ (corrected
\varbintfraccorr) in our sample are binaries, a fraction consistent with the
recent comprehensive assessment of \citet{Fontanive:2018cq}.

Our volume-limited sample also contains \varnwidecomp\ L and T~dwarfs that are
companions to hotter objects (corrected \varwidecompfraccorr).  These L and
T~dwarfs were identified as resolved (typically $>$10'') common proper motion
companions mostly to main-sequence stars, but also include
VHS~J125601.92$-$125723.9 b \citep[hereinafter
VHS~J1256$-$1257b;][]{Gauza:2015fw}, whose primary is a young M7 binary with
component masses near the 
stellar/substellar boundary \citep{Stone:2016fz,Dupuy:2020kb}, and the wide
sdL7+(sd)T7.5 pair SDSS~J141624.08+134826.7 and ULAS~J141623.94+134836.3.  Three
of the companions --- Gl~337CD \citep{Wilson:2001db,Burgasser:2005jp}, Gl~417BC
\citep{Kirkpatrick:2000gi,Bouy:2003eg}, and Gl~564BC \citep{Potter:2002ie} ---
are themselves close binaries and are included in the preceding discussion of
binaries.  Figure~\ref{fig.cmd.jjk.binyoung} shows the positions of our
non-binary companions in the $M_J$ versus $J-K$ CMD.  The companions include a
smaller fraction of early-L dwarfs and a larger fraction of late-T dwarfs than
the full volume-limited sample, but otherwise appear to follow the same sequence
on the CMD as the single L and T~dwarfs.

Unresolved binaries blend the light of two objects and therefore do not
accurately represent the evolution of individual objects on a CMD.  The
components of binaries and the resolved companions, on the other hand, are as
much individual ultracool dwarfs as the isolated ones.  However, these
components and companions may have different distributions of masses from the
single objects, in which case they could be distributed differently along the L
and T dwarf sequence in Figure~\ref{fig.cmd.jjk.binyoung}.  The overabundance of
late-T type companions, for example, is suggestive of a different underlying
distribution.  Previous efforts to constrain the mass distribution of companions
have found results consistent with most substellar initial mass function (IMF)
constraints \citep[within large error bars;
e.g.,][]{Brandt:2014cw,Bowler:2016jk,Baron:2019fg,Nielsen:2019cb}.  However,
those studies have primarily focused on planetary-mass companions found by
adaptive optics surveys, and have not encompassed the full range of brown dwarf
masses and separations that characterize the mostly wide companions in our
volume-limited sample.  There is also evidence that the mass distribution of the
components of binaries is different from single objects.
\setcitestyle{notesep={; }} The mass ratio distribution of brown dwarf binaries
skews strongly toward equal masses \citep{Burgasser:2006hd}, whereas the IMF for
brown dwarfs is thought to be roughly flat: for a power-law IMF of the form
($\frac{dN}{dM}\propto M^{-\alpha}$), most studies have found
$-0.5\lesssim\alpha\lesssim0.5$
\citep[e.g.,][K19]{Allen:2005jf,Metchev:2008gx,DayJones:2013hm}.
\setcitestyle{notesep={, }} In addition, to the extent that selection effects
are present in our sample, they may be different for companions and binaries as
a result of different goals and methods used for past searches (compared with
field objects).  In short, since the singles, companions, and components of
close binaries in our sample may all have different mass distributions, it is
appropriate to consider those groups separately, especially when studying their
population properties.  Therefore, we have removed the known binaries and
companions from our volume-limited sample for the remainder of this paper.  This
allows us to present a clean picture of the NIR photometric evolution of
\emph{single} brown dwarfs, uncluttered by the blended photometry of unresolved
binaries and free of any bias from possible differences in the mass
distributions of binary components or companions.

\section{A Gap in the L/T Transition}
\label{lt.gap}

\begin{figure*}
  \centering
  \begin{minipage}[t]{0.48\textwidth}
    \includegraphics[width=1\columnwidth, trim = 12mm 0 10mm 0]{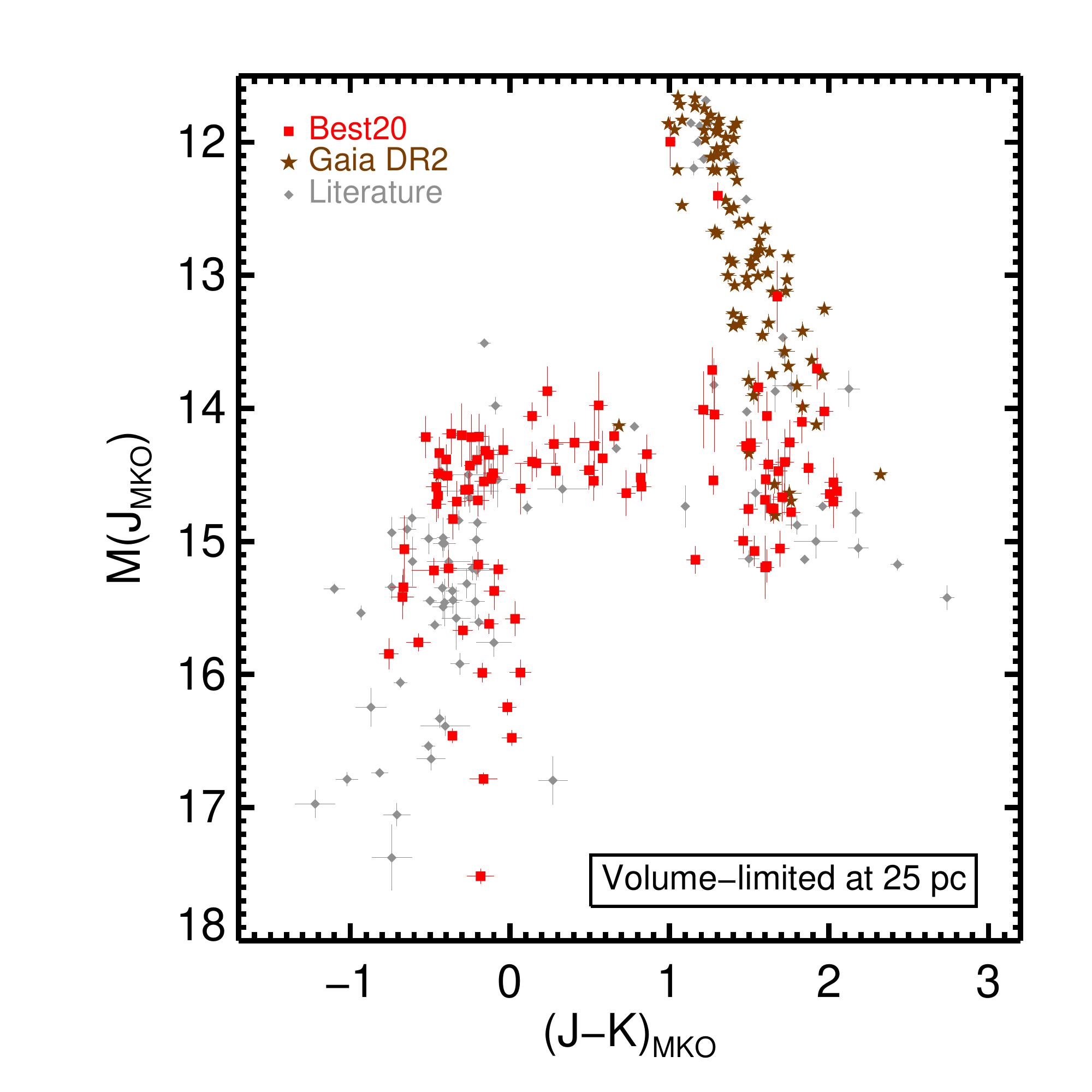}
  \end{minipage}
  \hfill
  \begin{minipage}[t]{0.48\textwidth}
    \includegraphics[width=1\columnwidth, trim = 12mm 0 10mm 0]{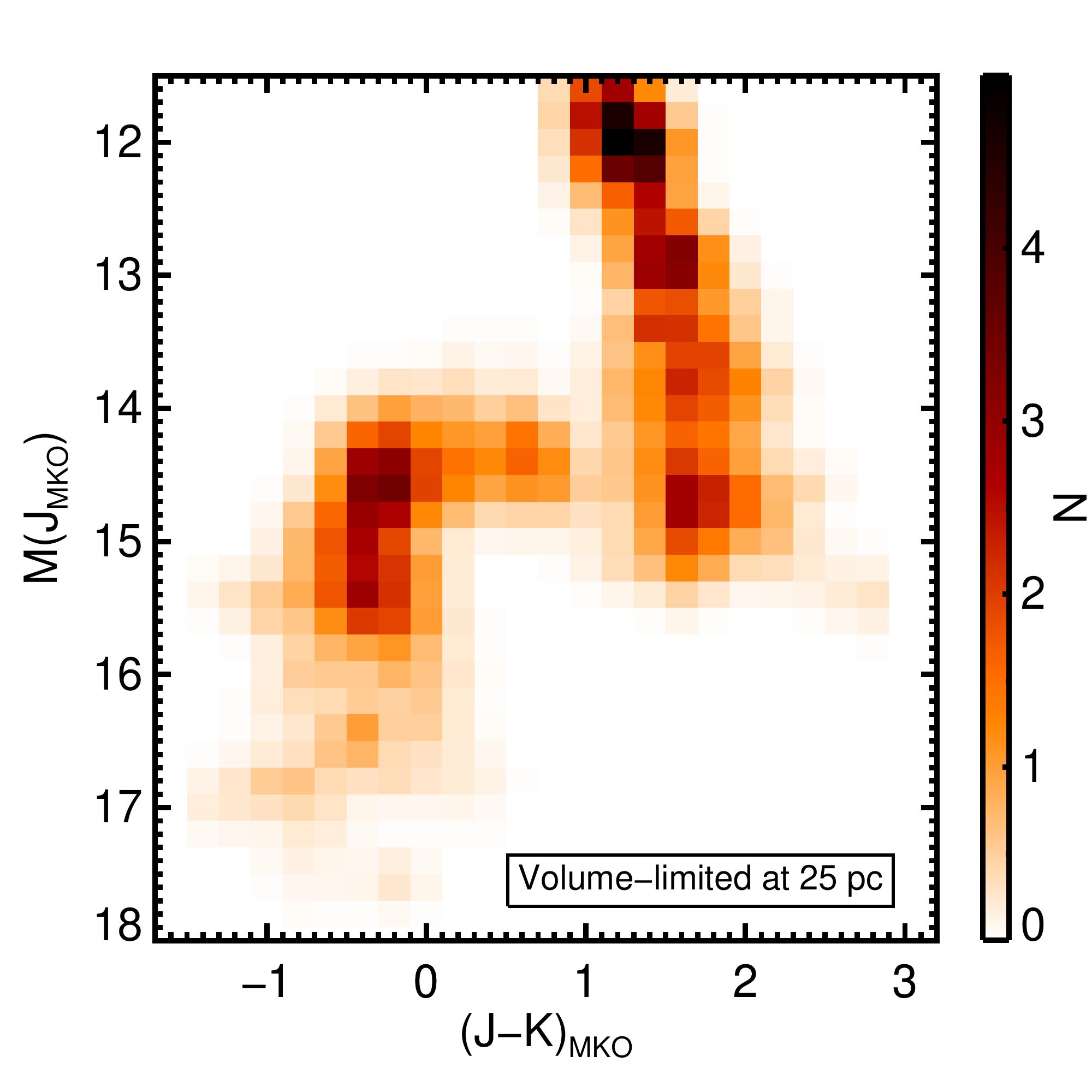}
  \end{minipage}
  \caption{Left: $M_J$ vs. $J-K$ (MKO) color-magnitude diagram for our
    volume-limited sample of {\varnsingle} single objects (binaries and
    companions removed). Symbols indicate the source of the parallax
    measurement: \gaiat\ (brown stars), literature sources (gray diamonds), and
    Best20 (red squares).  For objects with more than one parallax measurement,
    we use the most precise one.  Right: density map of the same CMD, obtained
    by smoothing the mean number of objects in $0.2 \times 0.2~{\rm mag}$~cells
    from Monte Carlo trials.  The color bar at right indicates the number of
    objects in the smoothed cells.  The large gap (underdensity) at
    $J-K\approx1$~mag, $M_J\approx14.5$~mag is clearly identified for the first
    time by our volume-limited sample.  The gap occurs early in the L/T
    transition and implies a phase of rapid atmospheric evolution as brown
    dwarfs cool through $\teff\approx1300$~K \citep[e.g.,][]{Dupuy:2017ke}.}
  \label{fig.cmd.jjk}
\end{figure*}

Figure~\ref{fig.cmd.jjk} shows the $M_J$ versus $J-K$ (MKO) CMD for our
volume-limited sample of {\varnsingle} single objects after removing the
{\varnbinary} binaries and {\varnwidecomp} companions (three of which are
themselves binaries; Section~\ref{vollim.binaries}).  Figure~\ref{fig.cmd.jjk}
highlights the sources of the parallaxes that define our sample.  The CMD
reveals a significant gap at $\jkmko\approx0.9$--1.4~mag, early in the evolution
of L/T transition dwarfs from redder to bluer colors, implying a phase of rapid
atmospheric evolution as brown dwarfs cool through $\teff\approx1300$~K
\citep[e.g.,][]{Dupuy:2017ke}.  The gap was not identifiable in CMDs using
previous samples requiring parallaxes because those samples (e.g., DL12)
contained too few L/T transition dwarfs to securely identify trends in the
population.  The Best20 parallaxes now allow us to present a volume-limited CMD
that reveals this L/T transition gap for the first time.

Figure~\ref{fig.cmd.jjk} also shows the $M_J$ versus $J-K$ CMD as a density map,
obtained by Monte Carlo sampling of the objects (to account for observational
uncertainties), binning into $0.2 \times 0.2~{\rm mag}$~cells, and smoothing by
a two-dimensional Gaussian of FWHM 1.5~cells.  The smoothed cells contain 0--5
objects.  The gap in the L/T transition seen in Figure~\ref{fig.cmd.jjk} stands
out clearly as the most prominent underdensity in the brown dwarf sequence
through mid-T dwarfs ($\mjmko\lesssim15.5$~mag, after which our sample becomes
less complete).

\begin{figure}
  \centering
  \includegraphics[width=1\columnwidth, trim = 20mm 0 10mm 0]{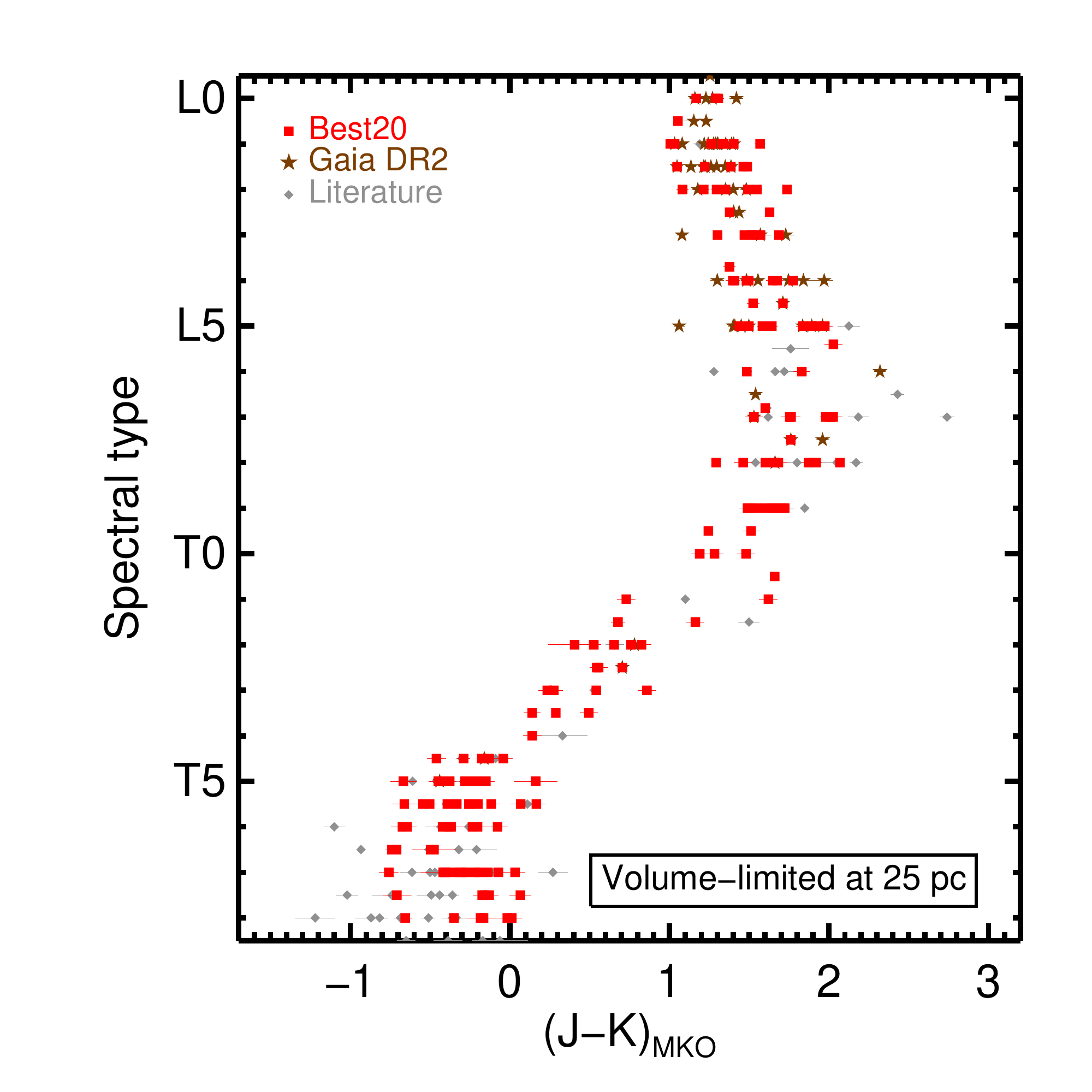}
  \caption{Spectral type as a function of \jkmko\ color for the single objects
    in our volume-limited sample.  The gap occurs at $J-K\approx0.9-1.4$~mag, in
    the vicinity of spectral types $\approx$T0--T3.}
  \label{fig.spt.jk}
\end{figure}

The $J-K$ color of L/T transition spectral subtypes varies as much as the width
of the gap \citep{Best:2018kw}, and the spectral types in our sample have not
been homogeneously assigned, so we cannot claim that the gap occurs at a
specific spectral type.  For reference, however, we show spectral type as a
function of \jkmko\ for our sample in Figure~\ref{fig.spt.jk}, and note that the
gap occurs in the vicinity of spectral types $\approx$T0--T3.  We note also that
Figure~\ref{fig.vmax} shows our sample to be $\approx$100\% complete for
T0--T4~dwarfs, the spectral type range that likely encompasses the gap, so the
gap is not an incompleteness artifact.

\begin{figure*}
  \centering
  \includegraphics[width=2\columnwidth]{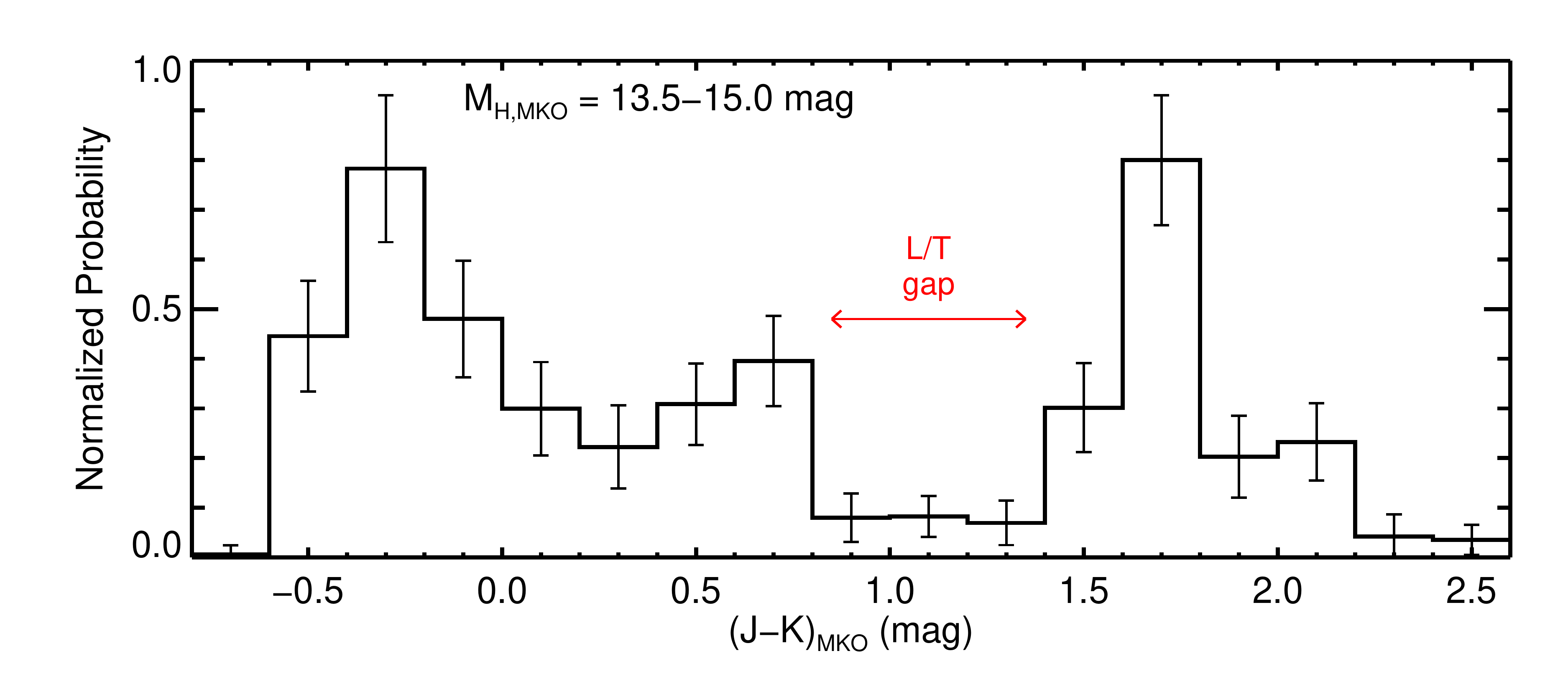}
  \caption{Distribution of \jkmko\ colors for single L/T transition dwarfs in
    our volume-limited sample.  We computed the histogram from objects having
    $13.5\le\mhmko\le15.0$~mag (we used \mhmko\ because it is essentially
    constant across the L/T transition; DL12), sampling their color
    uncertainties in a Monte Carlo fashion.  The error bars are the RMS for each
    color bin from our Monte Carlo trials added in quadrature to the binomial
    uncertainties for a sample covering 68.3\% of the sky.  The gap seen in
    Figure~\ref{fig.cmd.jjk} is labeled in red.  A gap seen previously at
    $\jkmko\approx0.0-0.5$~mag in smaller, less complete samples by DL12 and
    \citet{Best:2015em} appears as a shallow deficit in our volume-limited
    sample, suggesting another slightly accelerated evolutionary phase.}
  \label{fig.hist.jk}
\end{figure*}

Figure~\ref{fig.hist.jk} shows the $J-K$~color distribution of single L/T
transition dwarfs in our volume-limited sample.  We used Monte Carlo trials to
incorporate the uncertainties in the colors.  To account for random variations
due to the limited size and incomplete sky coverage of our sample, we calculated
uncertainties from the binomial distribution
(Appendix~\ref{appendix.uncert.density}, Equation~\ref{sig.binomial.num}) and
added these in quadrature.  To select L/T transition objects in an objective
fashion, we chose objects with $13.5\le\mhmko\le15.0$~mag, using \mhmko\ because
it is essentially constant across the L/T transition (DL12). This selection
includes \varnlttrans\ objects.  The gap seen in the CMD stands out clearly in
the color distribution.

How significant is this gap in the L/T transition?  We assessed this in two
stages.  We first addressed the statistical significance of our measurement of a
gap by evaluating its depth.  The normalized histogram in
Figure~\ref{fig.hist.jk} shows a sudden drop in probability from the 0.2~mag
wide bin centered at $J-K=0.7$~mag to the bin centered at 0.9~mag.  Adding the
uncertainties for these bins in quadrature, we found the difference between
these bins to be 3.1 times the combined uncertainty, i.e., a
$3.1\sigma$~measurement.  At the red edge of the gap, the rise from the 1.3~mag
bin to the 1.5~mag bin is a $2.3\sigma$~measurement.  Treating the gap as a
single 0.6~mag wide bin centered at $J-K=1.1$~mag, in comparison to adjacent
0.6~mag wide bins centered at 0.5 and 1.7~mag, the gap's significance is more
pronounced: the drop at the blue side is $4.6\sigma$, while the rise at the red
side is $6.6\sigma$.  Clearly the gap is a statistically significant feature in
our sample.

We then considered the possibility that the gap is simply a random fluctuation
in the color distribution of L/T transition dwarfs of our particular
volume-limited sample.  Without prior knowledge of the true {\jkmko}
distribution of L/T transition dwarfs, one cannot directly determine the extent
to which the gap is a deviation from a specific physical scenario.  We can,
however, estimate the likelihood of a gap like the one in our volume-limited
sample randomly appearing when the underlying distribution has no gaps.  Again
referring to Figure 7 and considering the gap as one 0.6~mag wide bin centered
at $J-K=1.1$~mag, that gap contains on average 4.99 objects in our Monte Carlo
trials, out of a total of \varnlttrans\ objects.  The 0.6~mag bin centered at
$J-K=0.5$~mag (adjacent to the gap) contains an average of 19.94 objects, 14.95
more than the gap does.  For comparison, we drew random samples of \varnlttrans\
objects with {\jkmko} colors between $-0.5$ and 2.1~mag from a uniform
distribution.  In these draws, a 0.6~mag wide bin anywhere in the interval with
both adjacent bins having at least 15 more objects occurs 1.9\% of the time.  In
our volume-limited sample, the 0.6~mag wide bin to the red side of the gap is
even taller than the bin on the blue side, so a gap like ours is actually less
likely to occur.  In addition, a 0.6~mag wide bin with five or fewer objects
anywhere in the interval occurred less than 0.001\% of the time in our random
draws, i.e., it is a $>$4.4$\sigma$ event.  We therefore conclude that the gap
is extremely unlikely to occur if the true $J-K$ distribution across the L/T
transition is uniform.

Could our removal of binaries and companions have biased our sample in such a
way as to create the gap? Figure~\ref{fig.cmd.jjk.binyoung} shows there are no
binaries or companions in the gap, partly answering the question.  We must also
consider that the individual components of a binary can have different photomety
than the binary's integrated-light photometry, and thus a resolved component
could lie in the gap even when the blended light of the unresolved binary does
not. For the addition of binary components to change the appearance of the CMD,
such components would need to be highly concentrated in the gap, which seems
implausible given that binaries account for only 10\% of our entire sample,
spanning all spectral types. More generally, binary components (as well as
resolved companions) are themselves ultracool dwarfs with the same spectral and
evolutionary properties as single objects, so it is extremely unlikely that the
components would preferentially congregate in the single-object gap (e.g., given
the previous paragraph's estimate of a $<$0.001\% chance of a uniform color
distribution across the L/T transition).  In other words, removing binary
components cannot create a phase of evolution that is not seen in single
objects.

Using a much smaller sample of 36~objects defined by parallaxes but not
volume-limited, DL12 tentatively identified a gap at a bluer
$\jkmko\approx0.1$--0.5~mag color, along with a pileup at
$\jkmko\approx0.5$--0.8~mag, similar to the predictions of the ``hybrid''
evolutionary models of SM08 (see the discussion in
Section~\ref{discussion.models}).  \citet{Best:2015em} found further evidence
for this bluer gap in a sample of 70~objects, volume-limited at 25~pc but
incomplete and selected in part using photometric distances.  Our
parallax-defined 25~pc sample now reveals this tentative gap to in fact be a
minor underdensity (Figure~\ref{fig.hist.jk}), possibly another phase of mildly
accelerated evolution.  The samples of DL12 and \citet{Best:2015em} do show
hints of the gap we now clearly identify at $\jkmko\approx1$~mag, but its
significance is unapparent without the large number of early-T dwarf parallaxes
now provided by Best20.  These discrepancies with previous tentative results and
models demonstrate how essential a complete volume-limited sample is for
accurate population analysis.

\section{Discussion}
\label{discussion}

\subsection{L/T Transition Colors in Multiple Bands}
\label{discussion.ltphot}

\setcitestyle{notesep={; }} L/T transition dwarfs simultaneously brighten in the
$J$ band while dimming in the $K$ band as they cool through the transition
\citep[e.g.,][DL12]{Tinney:2003eg}. \setcitestyle{notesep={, }} This behavior is
thought to be caused by a depletion of condensate clouds
\citep[e.g.,][]{Ackerman:2001gk,Burrows:2006ia} or the evolution of
thermochemical instabilities \citep[cf.
\citealt{Leconte:2018ft}]{Tremblin:2016hi}.  The gap we identify suggests that
the distinctive blueward evolution of the L/T transition is occurring much more
quickly at $\jkmko\approx1$~mag than in subsequent parts of the transition.

In Figure~\ref{fig.cmd.others} we present CMDs of single objects in our
volume-limited sample using eight different colors spanning \PS\ \yps\
(0.96~\um) to WISE $W2$ (4.6~\um).  We highlight the objects inside, above, and
at the edges of the \jkmko\ L/T transition gap to illustrate the location (or
lack) of the gap in other colors.  The gap appears widest in \jkmko, but is also
clearly visible in \jhmko\ and \jmkowa, underscoring the importance of $J$-band
flux in the L/T transition evolution.  The gap is also apparent in similar
colors that replace \jmko\ with \yps\, including \ykm\ and \ywa, and is also
visible (although narrower) in \yjm.  The gap is not visible at all in \jmkowb\
and \wawb, indicating a more gradual evolution in these colors.

\begin{figure*}
  \centering
  \begin{minipage}[t]{0.48\textwidth}
    \includegraphics[width=1\columnwidth, trim = 20mm 0 10mm 0]{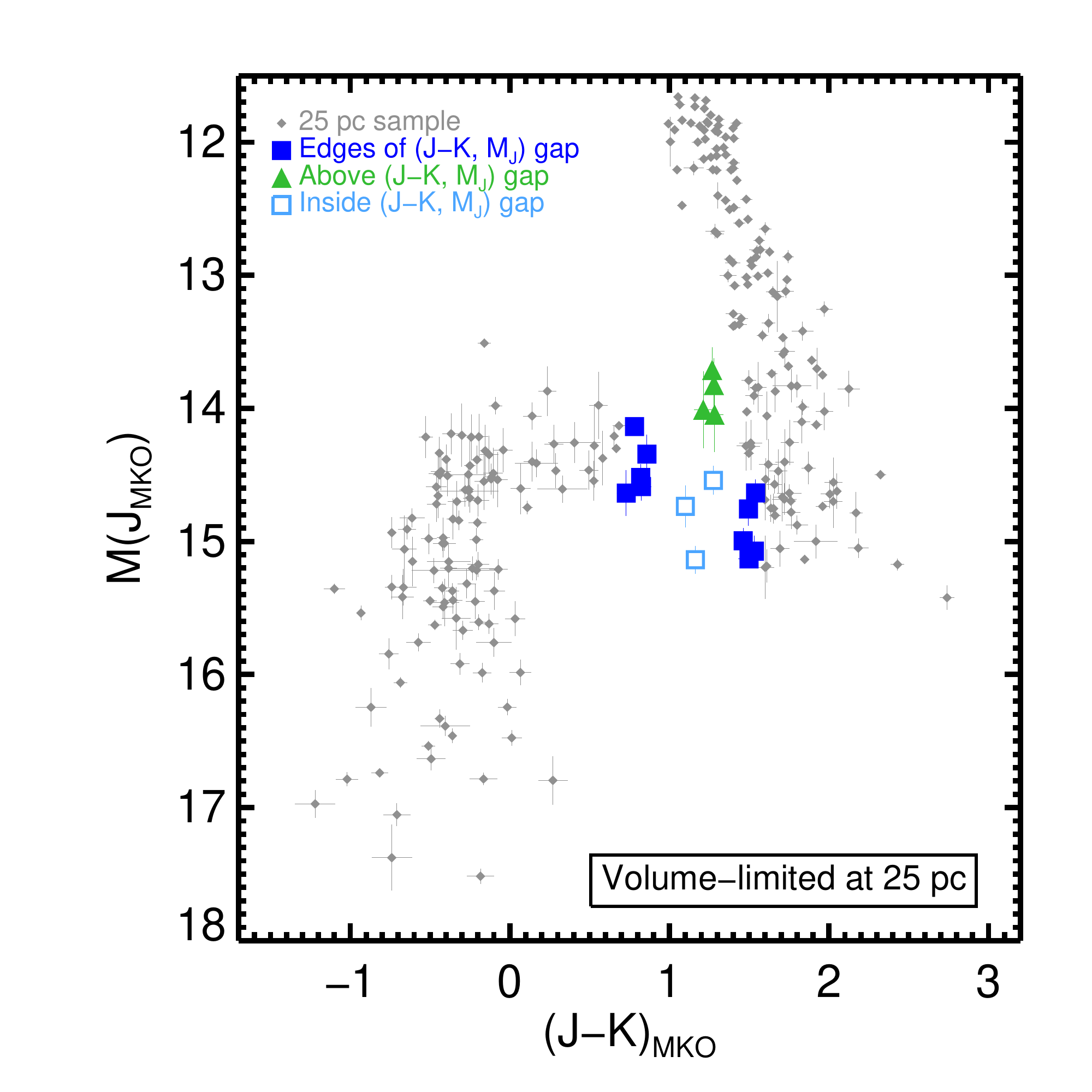}
  \end{minipage}
  \hfill
  \begin{minipage}[t]{0.48\textwidth}
    \includegraphics[width=1\columnwidth, trim = 20mm 0 10mm 0]{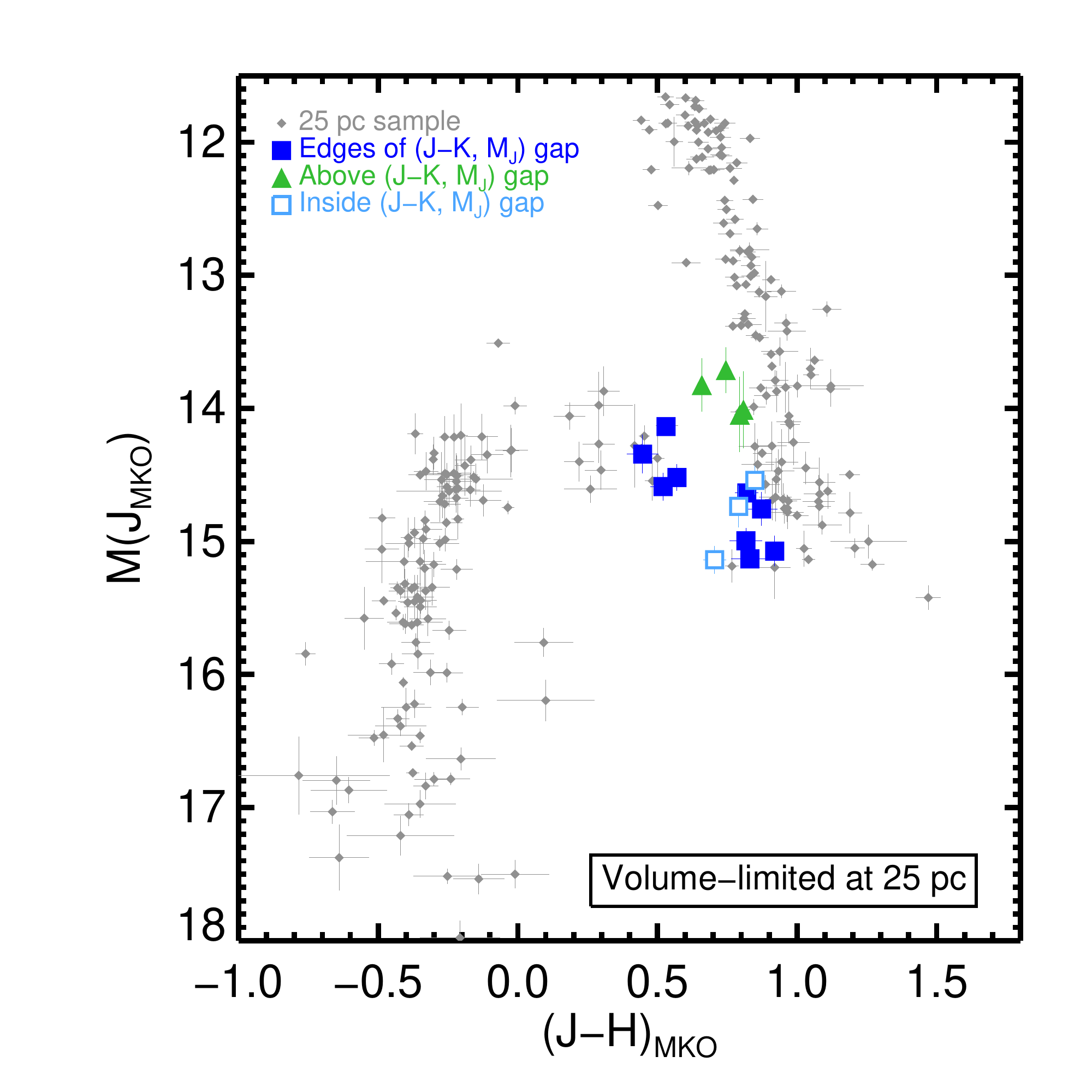}    
  \end{minipage}
  \begin{minipage}[t]{0.48\textwidth}
    \includegraphics[width=1\columnwidth, trim = 20mm 0 10mm 0]{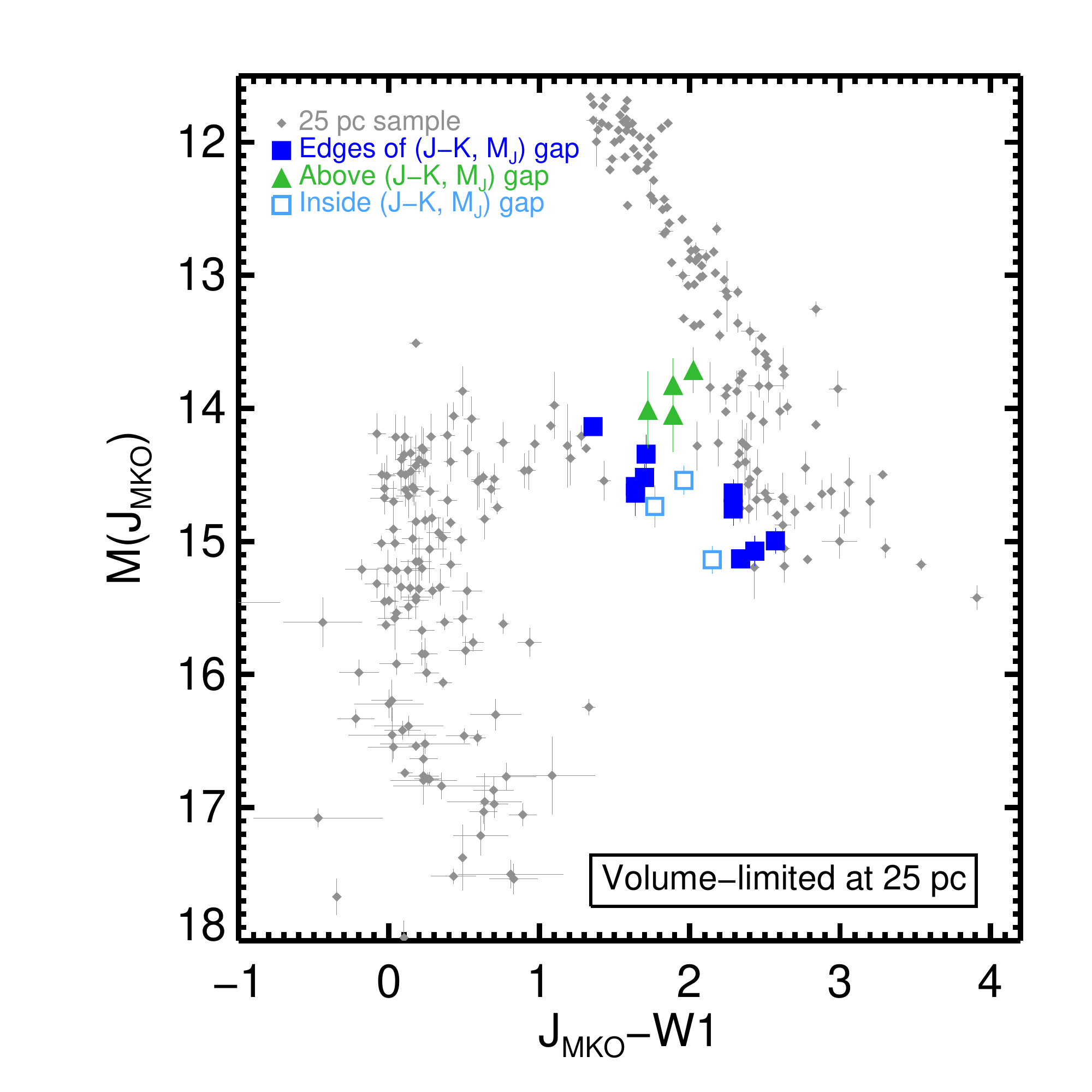} 
  \end{minipage}
  \hfill
  \begin{minipage}[t]{0.48\textwidth}
    \includegraphics[width=1\columnwidth, trim = 20mm 0 10mm 0]{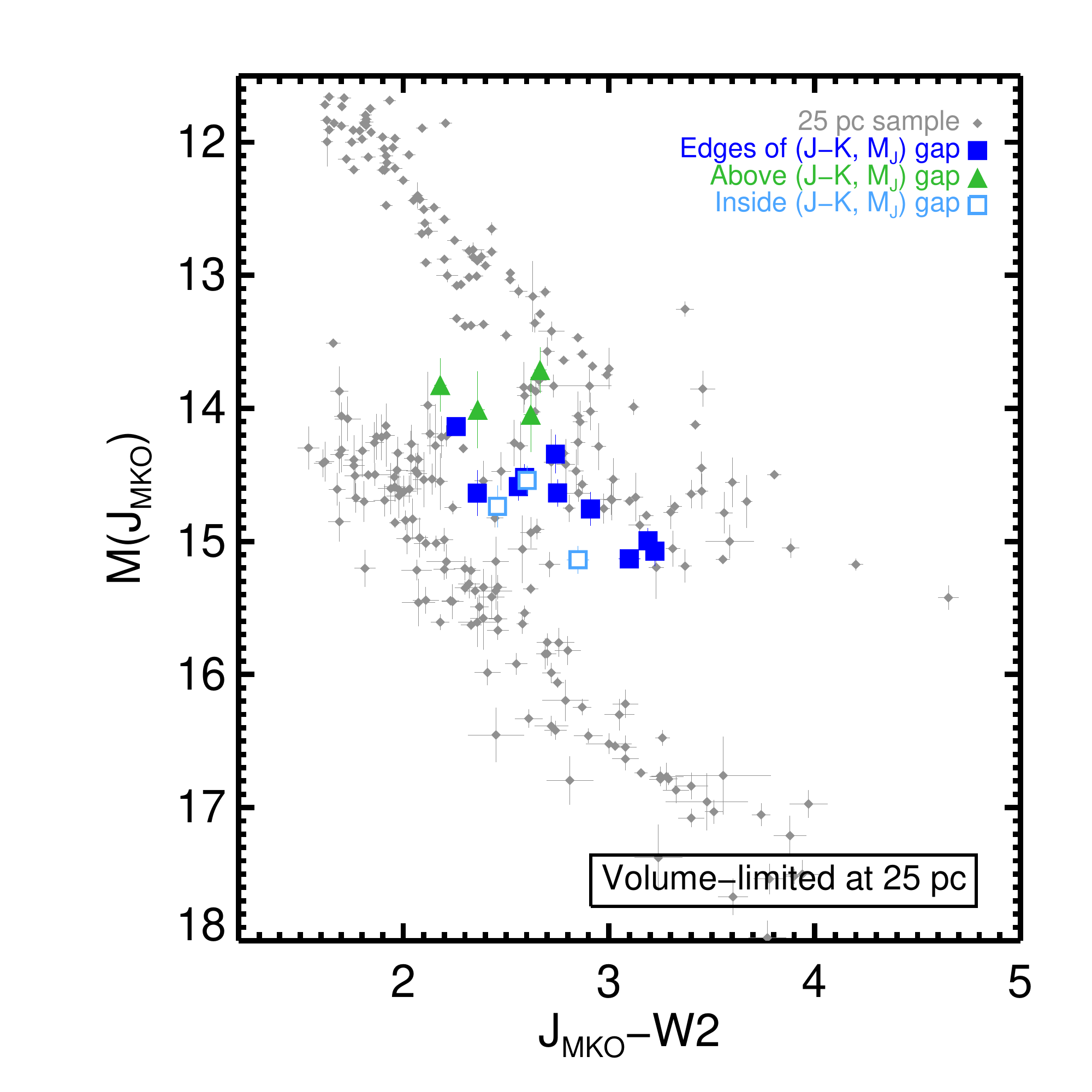}    
  \end{minipage}
  \caption{CMDs for single objects in our volume-limited sample, tracing objects
    in or near the \jkmko\ L/T transition gap through eight different colors.
    The top left CMD reproduces Figure~\ref{fig.cmd.jjk} but highlights only
    objects inside the $M_J$ vs. $J-K$ (MKO) gap (light blue open squares), on
    the edges of the gap (dark blue filled squares), or directly above the gap
    (green triangles). The same objects are highlighted in the same colors in
    the other diagrams.  The gap remains a prominent feature in \jhmko\ (top
    right) and \jmkowa\ (bottom left) but is not visible in \jmkowb\ (bottom
    right).}
  \figurenum{fig.cmd.others.1}
\end{figure*}

\begin{figure*}
  \centering
  \begin{minipage}[t]{0.48\textwidth}
    \includegraphics[width=1\columnwidth, trim = 20mm 0 10mm 0]{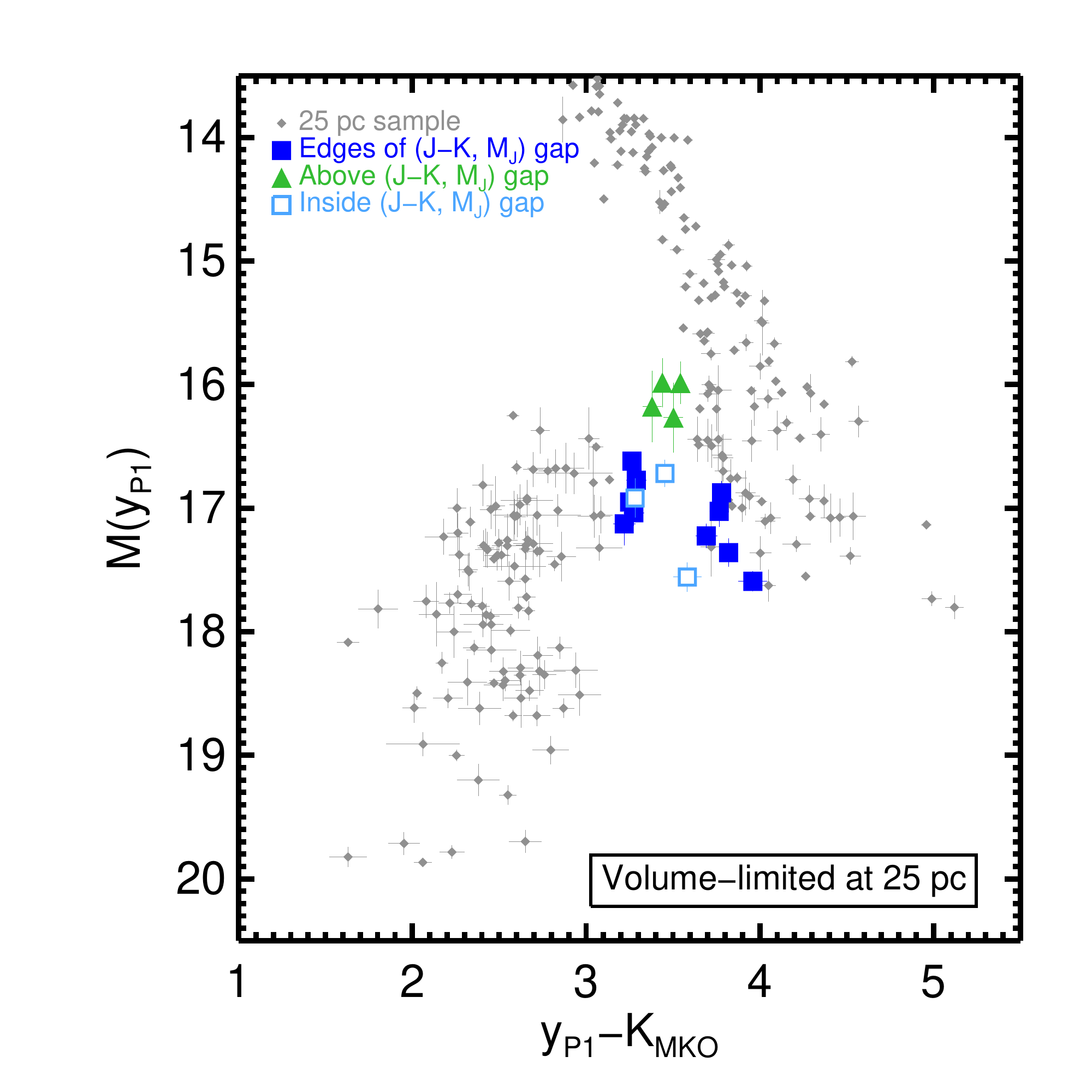} 
  \end{minipage}
  \hfill
  \begin{minipage}[t]{0.48\textwidth}
    \includegraphics[width=1\columnwidth, trim = 20mm 0 10mm 0]{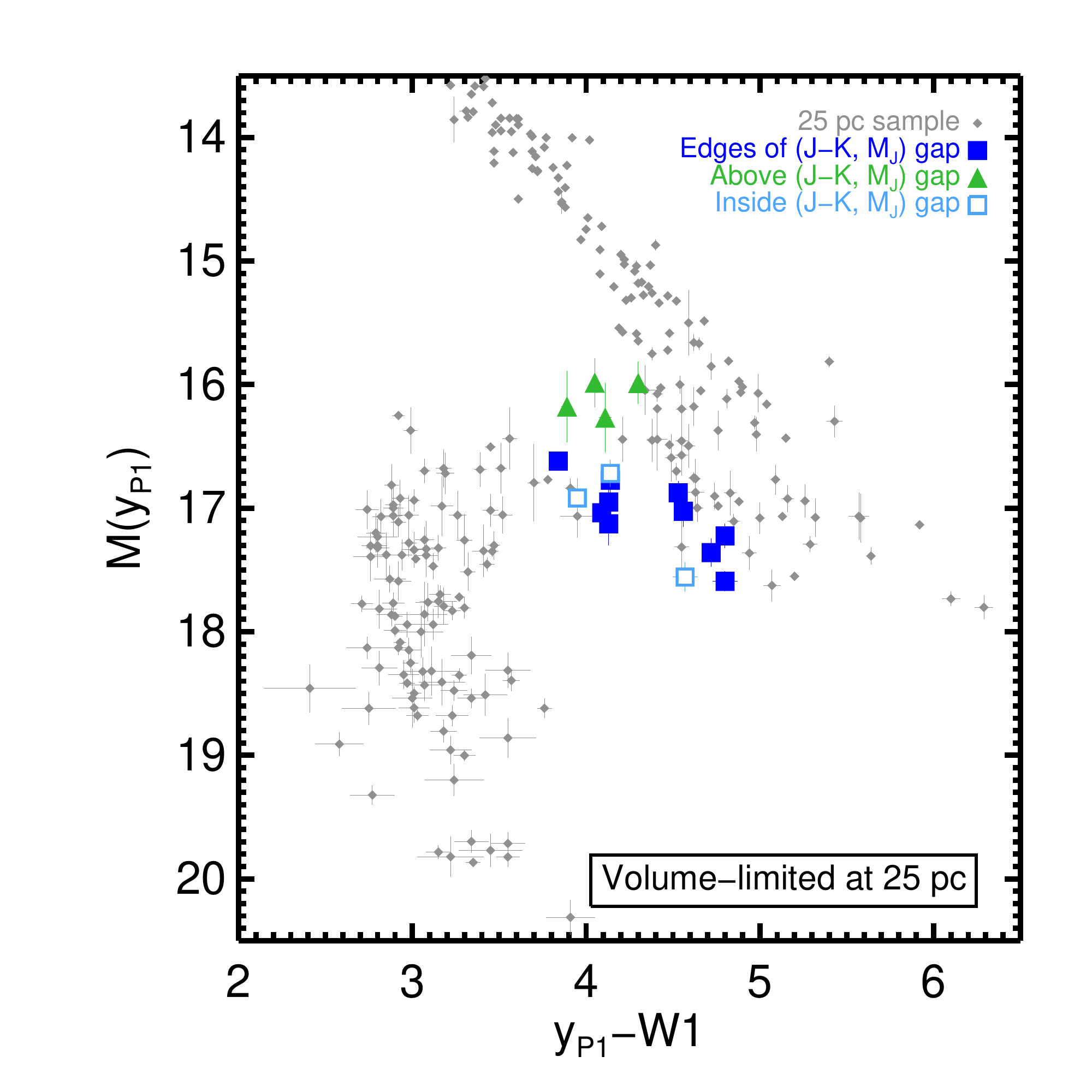} 
  \end{minipage}
  \begin{minipage}[t]{0.48\textwidth}
    \includegraphics[width=1\columnwidth, trim = 20mm 0 10mm 0]{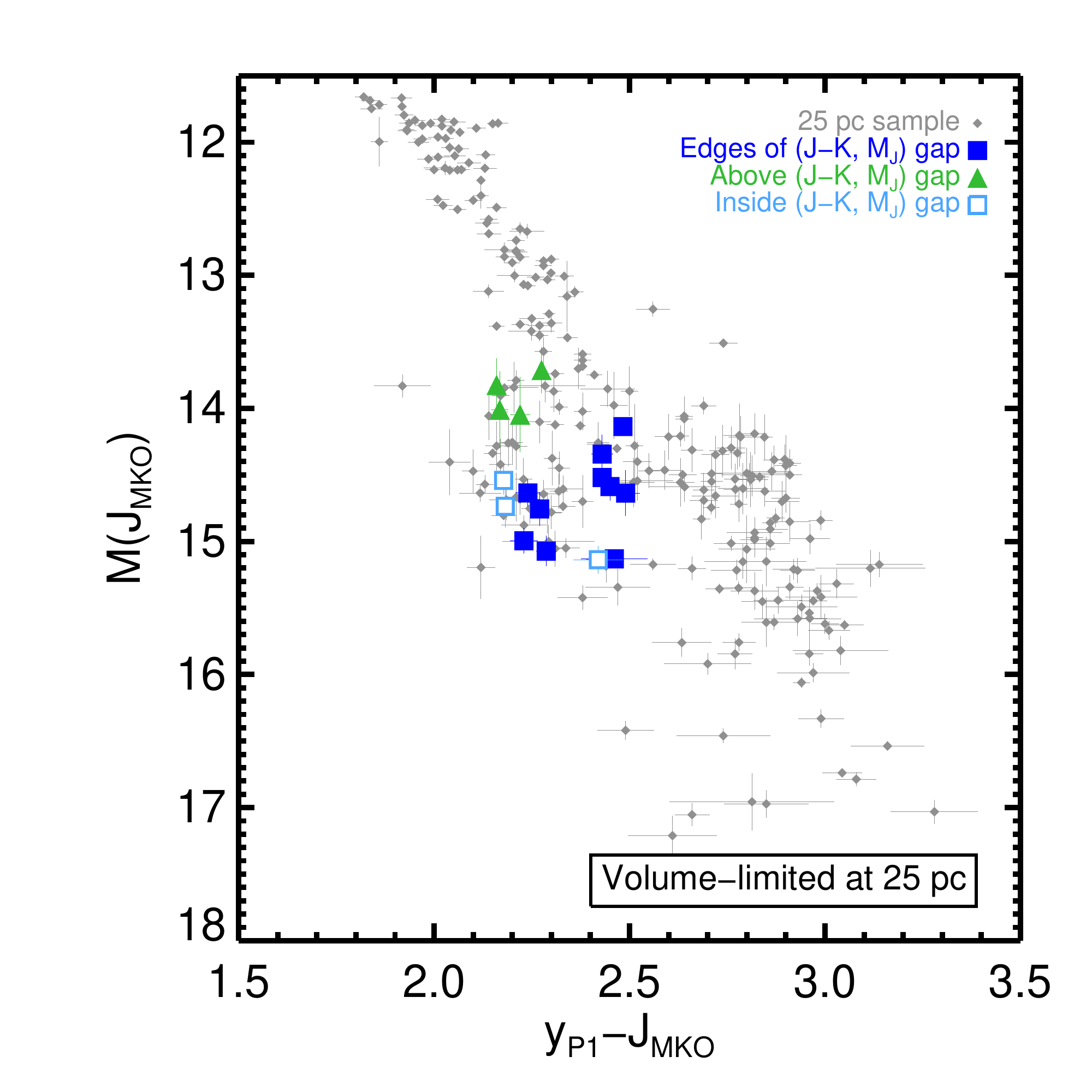}   
  \end{minipage}
  \hfill
  \begin{minipage}[t]{0.48\textwidth}
    \includegraphics[width=1\columnwidth, trim = 20mm 0 10mm 0]{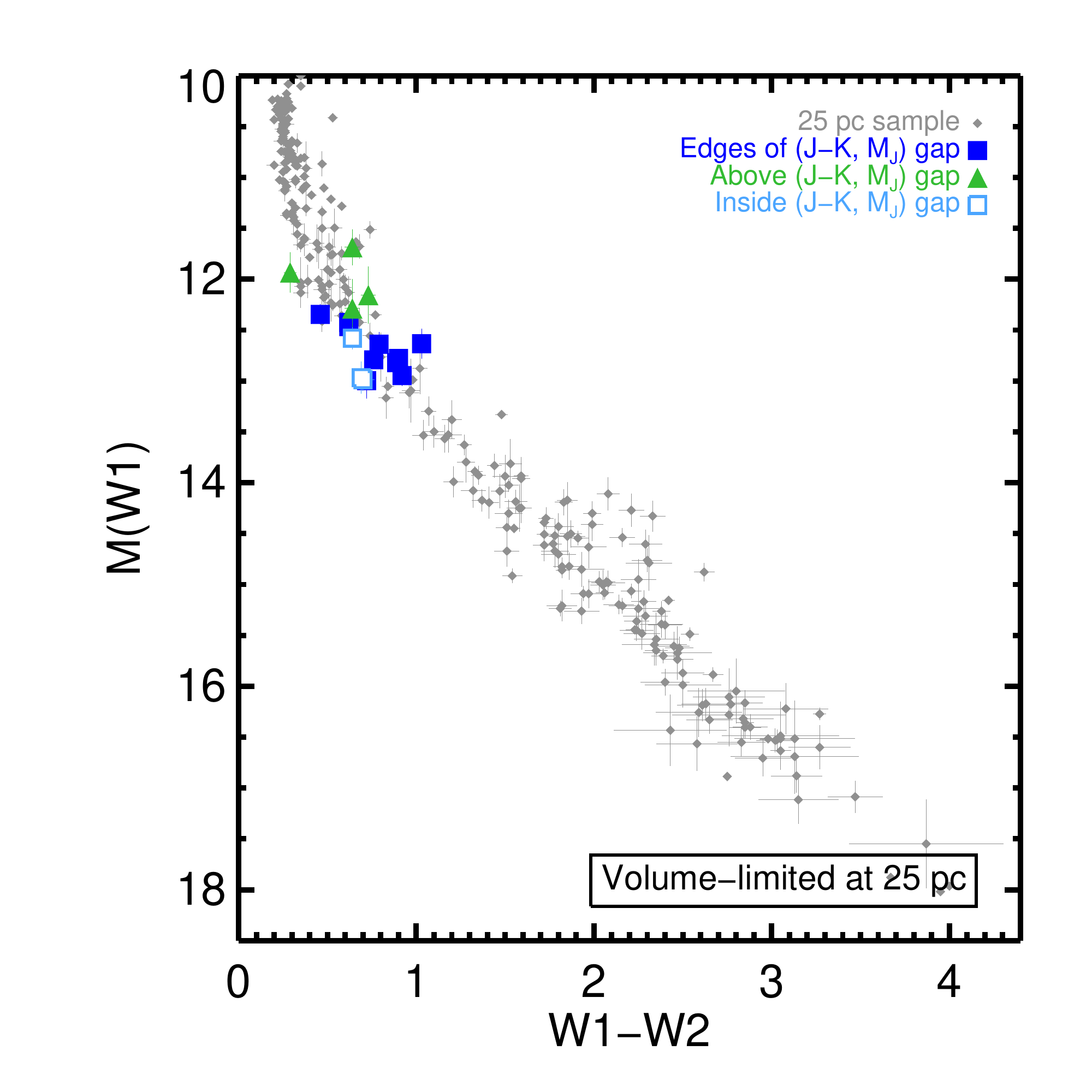} 
  \end{minipage}
  \caption{continued. The L/T transition gap is clearly visible in \ykm\ (top
    left) and \ywa\ (top right), but is contaminated with other objects in \yjm\
    (bottom left) and not at all present in \wawb\ (bottom right).}
  \label{fig.cmd.others}
\end{figure*}

\subsection{Objects in or near the L/T Transition Gap}
\label{discussion.binaries}

\begin{figure}
  \centering
  \includegraphics[width=1\columnwidth, trim = 20mm 0 10mm 0]{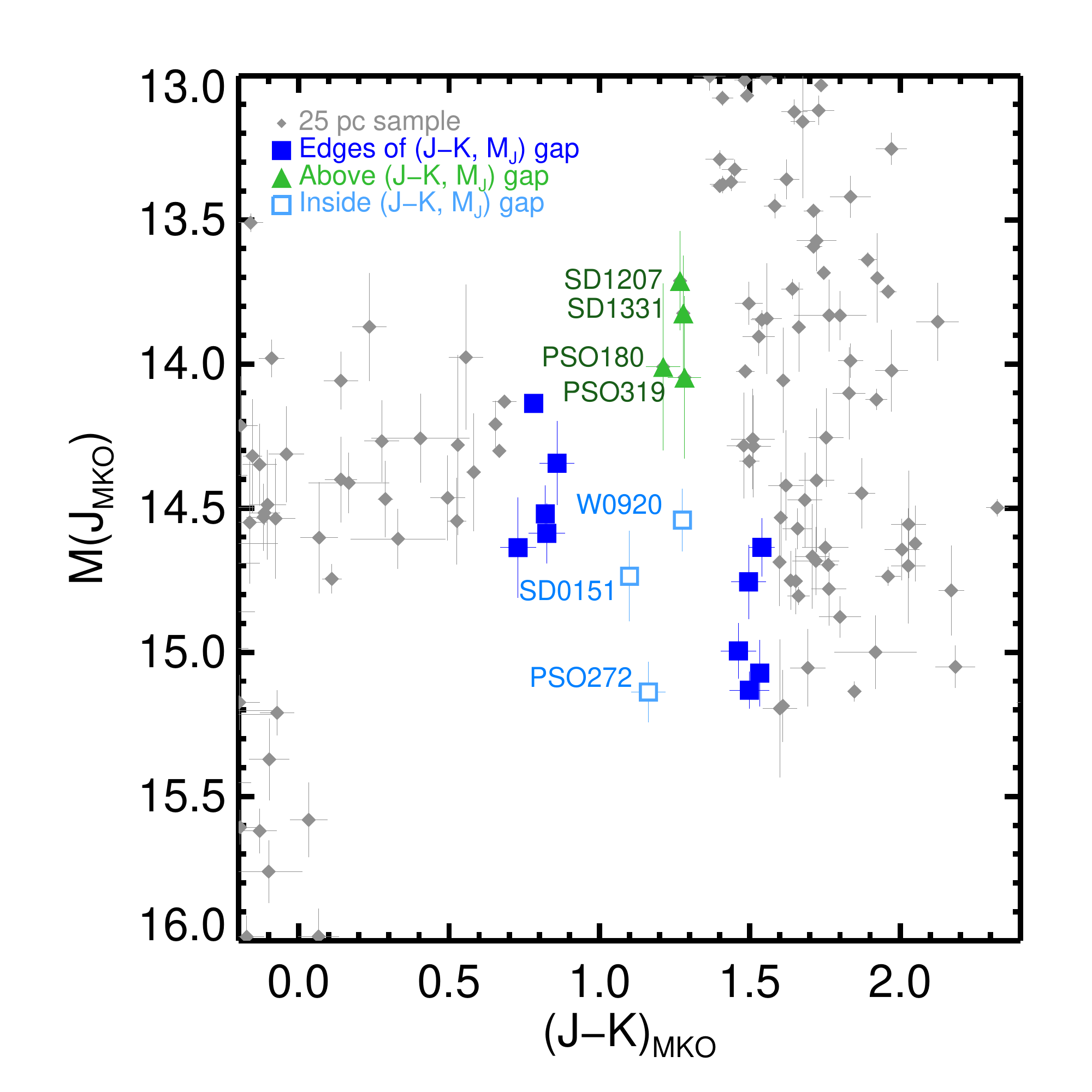}
  \caption{$M_J$ vs. $J-K$ (MKO) color-magnitude diagram for single objects in
    our volume-limited sample, using the same color scheme as
    Figure~\ref{fig.cmd.others} but showing only the central portion of the
    first panel of Figure~\ref{fig.cmd.others} in order to highlight the objects
    above and inside the L/T transition gap.  These objects are labeled here and
    discussed in Section~\ref{discussion.binaries}.}
  \label{fig.cmd.gap.labels}
\end{figure}

Assuming the L/T transition gap represents a rapid phase of brown dwarf
evolution, single objects with gap photometry ($J-K\approx1$~mag,
$M_J\approx14.5$~mag) should be relatively rare.  The gap in
Figure~\ref{fig.cmd.jjk} is not completely empty, featuring three objects within
the gap and four directly above it (highlighted in the first panel of
Figure~\ref{fig.cmd.others} and labeled in Figure~\ref{fig.cmd.gap.labels}),
suggesting that L/T transition objects may have gap colors for a brief but
observable period of time.

\begin{figure*}
  \centering
  \includegraphics[width=1\columnwidth, trim = 0mm 0 10mm 0]{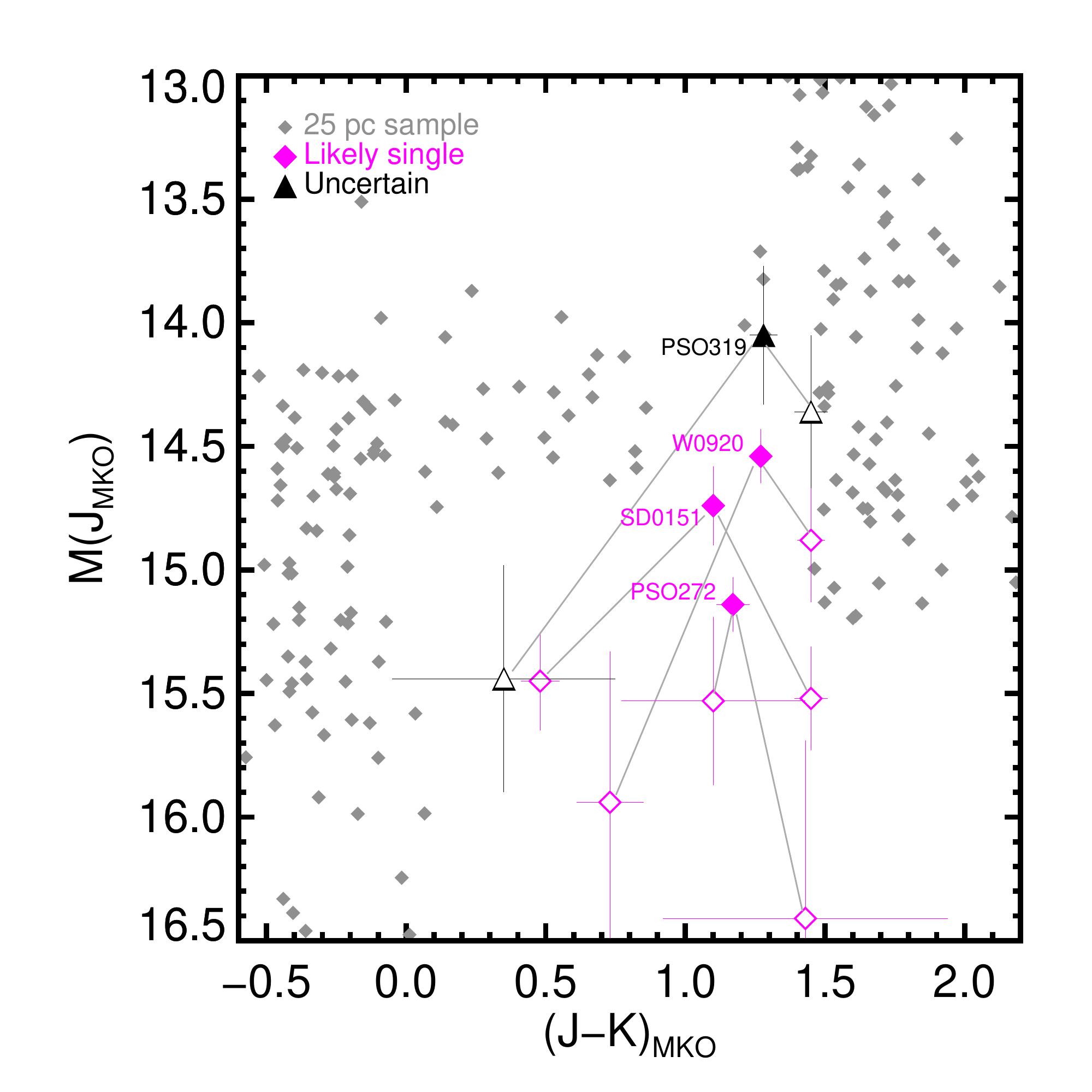}
  \includegraphics[width=1\columnwidth, trim = 0mm 0 10mm 0]{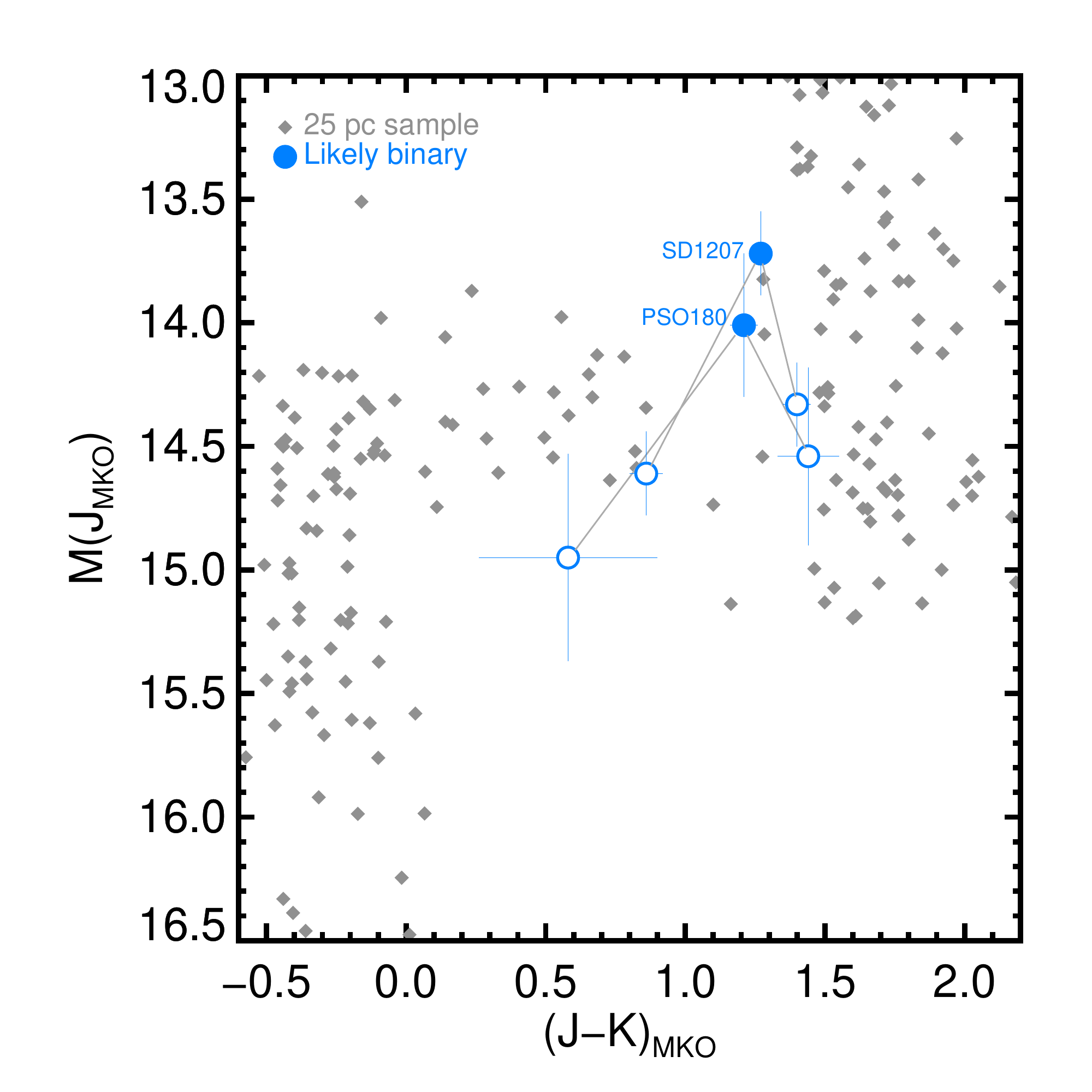}
  \includegraphics[width=1\columnwidth, trim = 0mm 0 10mm 0]{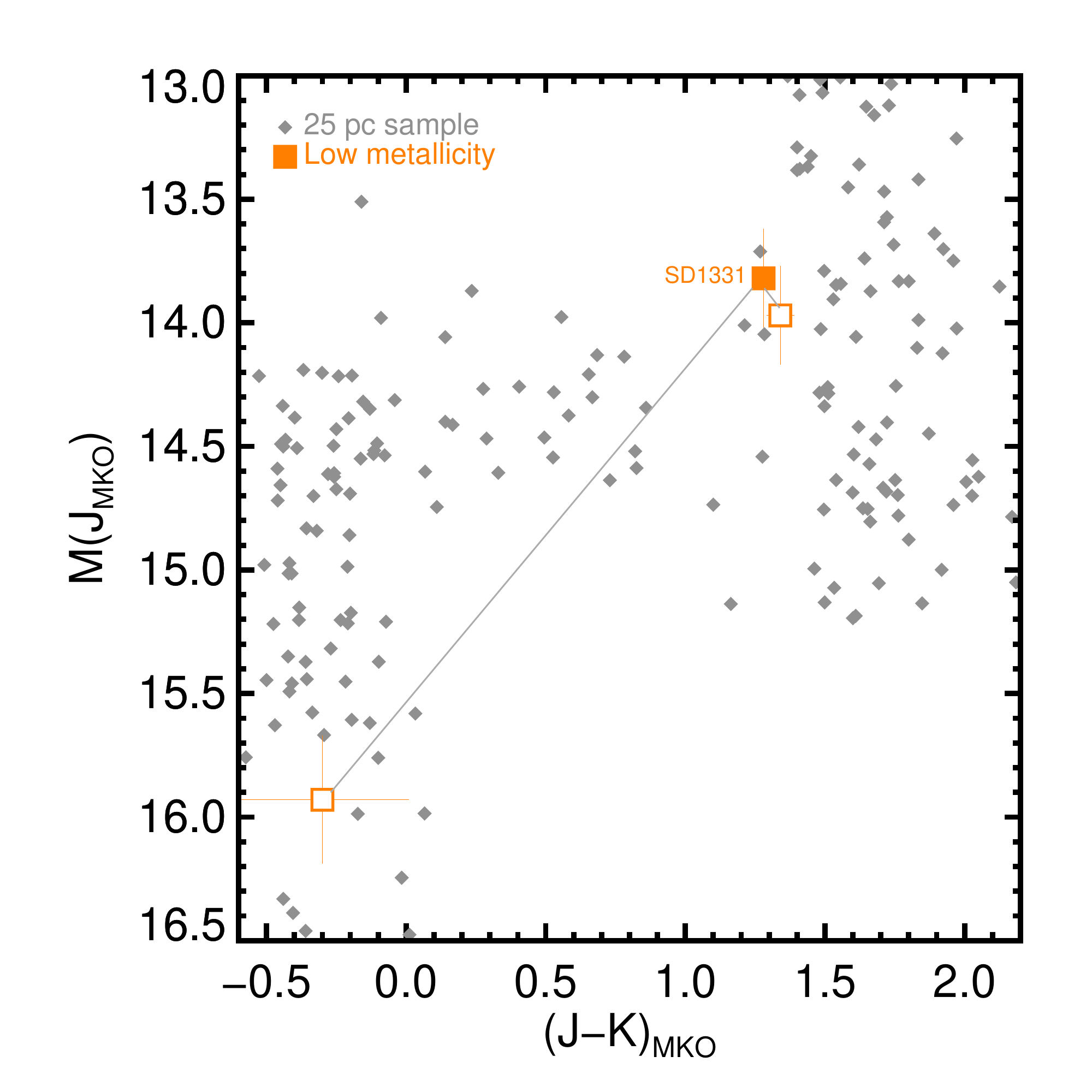}
  \caption{$M_J$ vs. $J-K$ (MKO) CMDs for our volume-limited sample, enlarging a
    portion centered on the L/T transition.  All three panels show the same
    portion of the CMD; we use three separate panels for clarity. The filled
    colored symbols show the objects in and above the L/T transition gap that
    are labeled in Figure~\ref{fig.cmd.gap.labels}, connected by light gray
    lines to the best-fit components from our binary template matching (open
    symbols).  Other (single) objects in the sample are plotted as light gray
    diamonds. Top left: for all three objects in the gap (SDSS~J0151+1244,
    WISE~J0920+4538, and PSO~J272.0$-$04.9; pink diamonds) as well as
    PSO~J319.3$-$29.6 (black triangles), at least one of the best-fit binary
    components is implausibly faint for its color and spectral type, suggesting
    that the objects are single. Top right: PSO~J180.1$-$28.6 and
    SDSS~J1207+0244 (blue circles) have plausible binary decompositions. Bottom:
    SDSS~J1331$-$0116 (orange square) is most likely a low-metallicity
    mid-L~dwarf \citep{Knapp:2004ji,Marocco:2013kv}.}
  \label{fig.cmd.decomp}
\end{figure*}

\begin{figure}
  \centering
  \includegraphics[width=1\columnwidth]{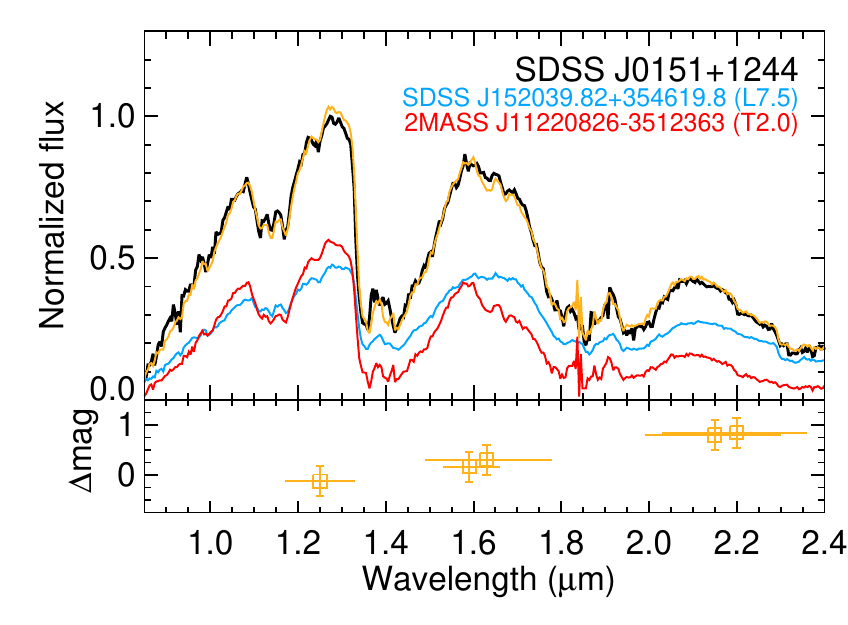}
  \includegraphics[width=1\columnwidth]{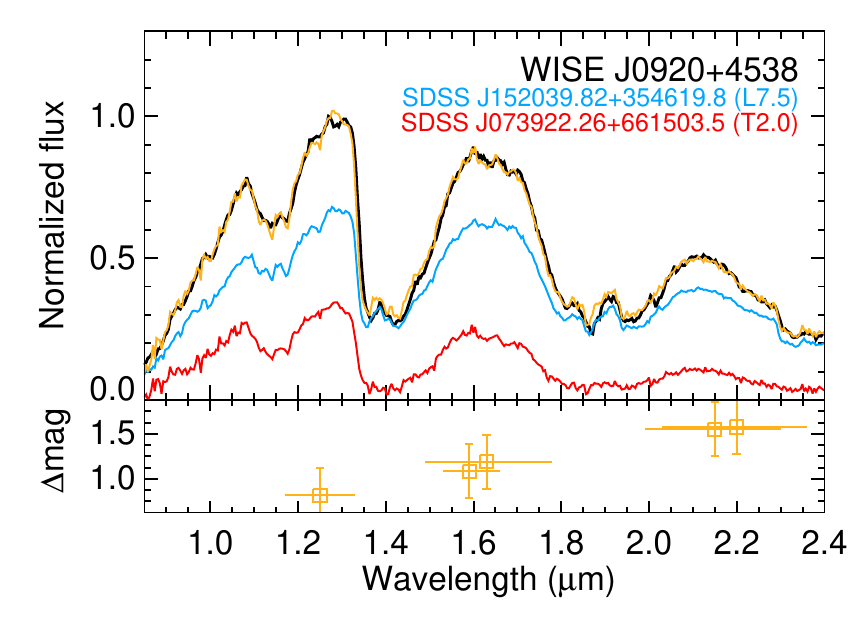}
  \includegraphics[width=1\columnwidth]{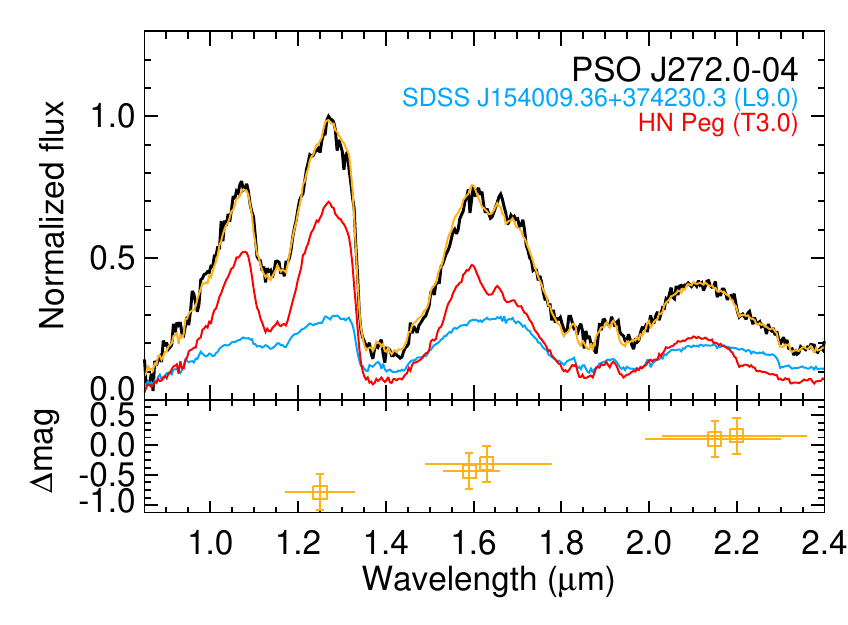}
  \caption{Best-matching template pairings for the three objects that lie inside
    the L/T transition gap (Figures \ref{fig.cmd.gap.labels}
    and~\ref{fig.cmd.decomp}).  Observed spectra are shown in black, individual
    component templates in blue and red, and convolved templates (i.e., template
    blended spectra) in orange.  The lower subpanels show the resulting flux
    ratios over standard NIR bandpasses computed from the best-matching template
    pairs (open orange squares) which are listed in Table~\ref{tbl.decomp}. The
    convolved templates provide good matches to the observed spectra but often
    require unrealistic flux ratios; see the object descriptions in
    Section~\ref{discussion.binaries} for details.}
  \label{fig.decomp.inside}
\end{figure}

\begin{figure*}
  \centering
  \begin{minipage}[t]{0.48\textwidth}
    \includegraphics[width=1\columnwidth]{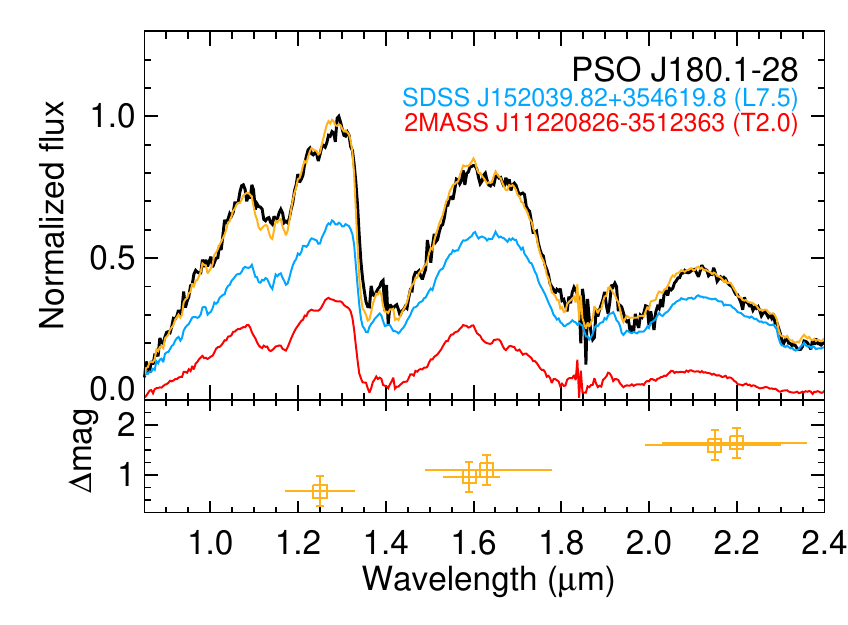}
  \end{minipage}
  \hfill
  \begin{minipage}[t]{0.48\textwidth}
    \includegraphics[width=1\columnwidth]{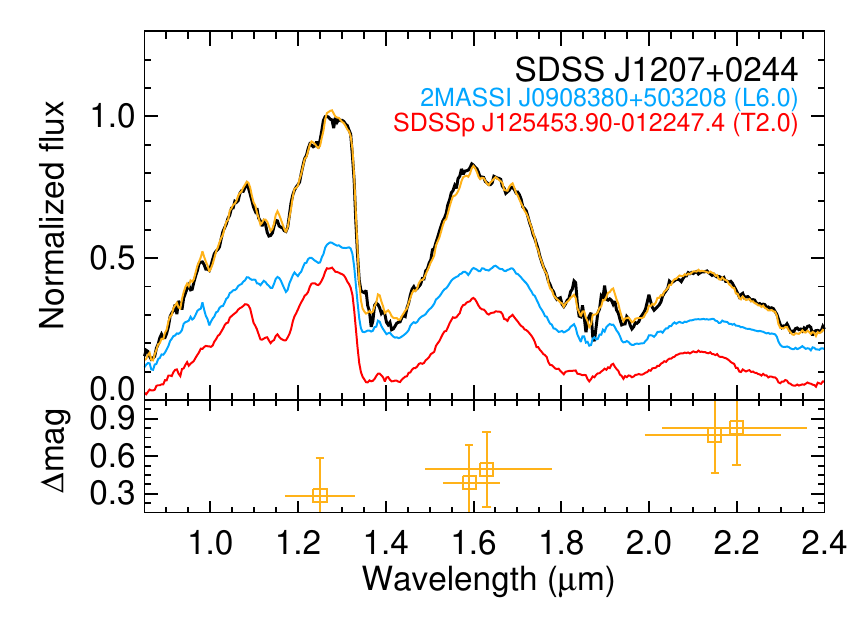}
  \end{minipage}
  \begin{minipage}[t]{0.48\textwidth}
    \includegraphics[width=1\columnwidth]{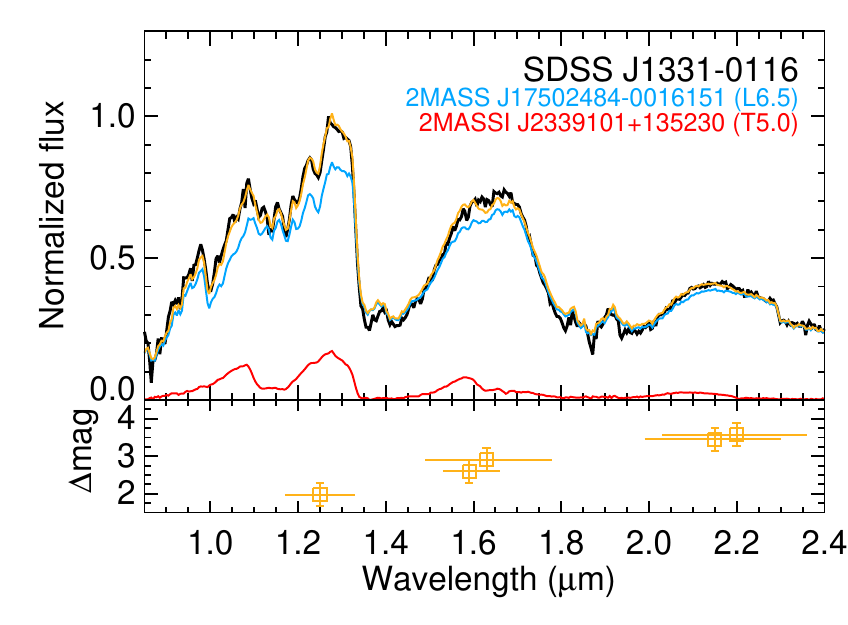}
  \end{minipage}
  \hfill
  \begin{minipage}[t]{0.48\textwidth}
    \includegraphics[width=1\columnwidth]{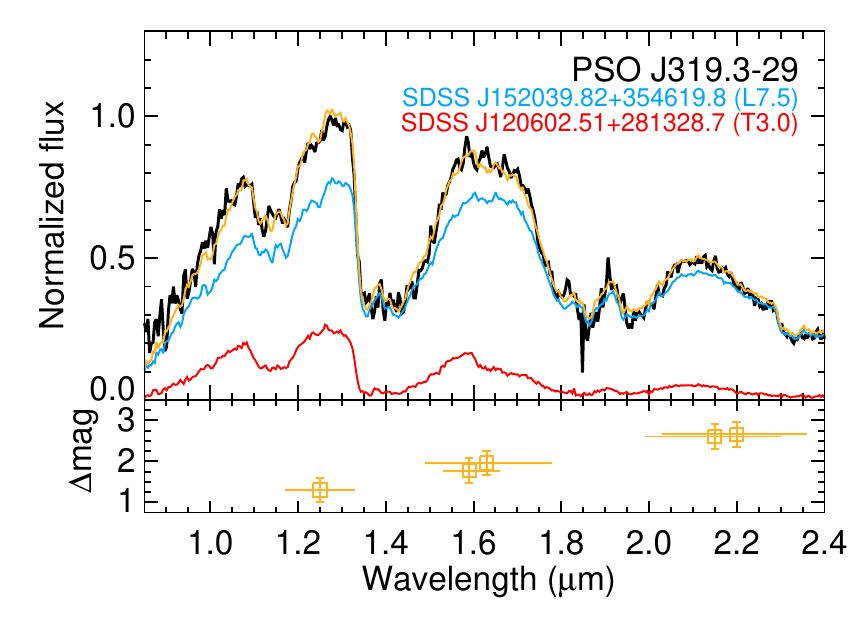}
  \end{minipage}
  \caption{Same as Figure~\ref{fig.decomp.inside}, but for objects lying above
    the L/T transition gap in Figures \ref{fig.cmd.gap.labels}
    and~\ref{fig.cmd.decomp}.}
  \label{fig.decomp.above}
\end{figure*}

However, another type of object could appear in or directly above the gap: an
unresolved binary with components that individually sit on either side of the
gap but whose blended color overlaps the gap.  To assess the possibility that
the seven objects in and above the gap are unrecognized binaries, we performed
spectral decomposition (e.g., \citealt{Burgasser:2005gj,Burgasser:2010df},
hereinafter B10; \citealt{Liu:2006ce}) of these objects following the method
described in DL12.  Briefly, we used the library of 178 IRTF/SpeX prism spectra
from B10 --- for which they determined uniform NIR spectral types and removed
binaries, young objects, and other unusual spectra --- as templates to create
blended spectra.  For each template pairing we determined the scale factor
needed to minimize the $\chi^2$ of the difference with the spectrum of a gap
object.  We then examined the resulting best pairing to determine the component
spectral types, taking into account spectral type uncertainties in the
best-match templates.  We estimated the flux ratios for the template pairings in
standard NIR bandpasses and $J-K$ colors using our $\chi^2$ values and the
weighting scheme described in B10.  For the $J-K$ colors, we added 0.05~mag in
quadrature to the uncertainties to account for systematic uncertainties we have
previously found in colors derived from low-resolution NIR spectra (DL12).  The
best template pairings for the gap objects are listed in Table~\ref{tbl.decomp},
placed on the $M_J$ versus $J-K$ color-magnitude diagram in
Figure~\ref{fig.cmd.decomp}, presented in Figures \ref{fig.decomp.inside}
and~\ref{fig.decomp.above}, and discussed below.  Unlike the approach used by
B10, our analysis does not make any prior assumptions about the flux ratios of
the components (e.g., based on spectral types), nor does it assess whether pairs
of templates are better matches to our observed spectra than single-object
templates.

We note that four of our seven decompositions identified the L7.5~dwarf
SDSS~J152039.82+354619.8 \citep{Chiu:2006jd} as the primary component template.
This brown dwarf has inconsistent NIR spectral types in the literature ---
T$0\pm1$ from \citet{Chiu:2006jd} and L7.5 from B10, and visually matches well
the NIR L9 standard DENIS-P~J025503.3$-$470049
\citep{Martin:1999er,Kirkpatrick:2010dc} --- but otherwise has no unusual
features. In the end, only one of the decompositions using
SDSS~J152039.82+354619.8 as the primary template (for PSO~J319.3102$-$29.6682)
is plausible, as discussed below.

\begin{deluxetable}{lccCCCCCCC}
\centering
\tablecaption{Spectral decomposition of objects in or above the L/T transition gap \label{tbl.decomp}}
\tabletypesize{\scriptsize}
\rotate
\tablewidth{0pt}
\tablehead{   
  \multicolumn{3}{c}{} &
  \multicolumn{7}{c}{MKO Photometry} \\
  \cline{4-10}
  \colhead{Object} &
  \colhead{Primary} &
  \colhead{Secondary} &
  \colhead{$M_J$ (combined)} &
  \colhead{$J-K$ (primary)} &
  \colhead{$J-K$ (secondary)} &
  \colhead{$\Delta J$} &
  \colhead{$\Delta H$} &
  \colhead{$\Delta {\rm CH_4short}$} &
  \colhead{$\Delta K$} \\
  \colhead{} &
  \colhead{(SpT)} &
  \colhead{(SpT)} &
  \colhead{(mag)} &
  \colhead{(mag)} &
  \colhead{(mag)} &
  \colhead{(mag)} &
  \colhead{(mag)} &
  \colhead{(mag)} &
  \colhead{(mag)}
}
\startdata
SDSS~J015141.69+124429.6   & L$7.5\pm2$ & T$2\pm1$ & 14.74\pm0.16 & 1.45\pm0.06 & 0.48\pm0.07  & -0.07\pm0.25 & 0.34\pm0.23 & 0.21\pm0.24 & 0.89\pm0.22 \\
WISE~J092055.40+453856.3   & L$7.5\pm1$ & T$2\pm2$ & 14.54\pm0.11 & 1.45\pm0.05 & 0.73\pm0.12  & 1.22\pm0.91  & 1.45\pm0.69 & 1.39\pm0.75 & 1.66\pm0.50 \\
PSO~J180.1475$-$28.6160    & L$7.5\pm1$ & T$2\pm1$ & 14.01\pm0.29 & 1.44\pm0.11 & 0.58\pm0.32  & 0.43\pm0.52  & 0.79\pm0.70 & 0.67\pm0.66 & 1.29\pm0.78 \\
SDSS~J120747.17+024424.8   & L$6\pm1  $ & T$2\pm1$ & 13.72\pm0.17 & 1.40\pm0.05 & 0.86\pm0.06  & 0.28\pm0.05  & 0.50\pm0.06 & 0.39\pm0.06 & 0.82\pm0.06 \\
SDSS~J133148.92$-$011651.4 & L$6.5\pm1$ & T$5\pm1$ & 13.82\pm0.20 & 1.34\pm0.05 & -0.30\pm0.31 & 1.97\pm0.18  & 2.86\pm0.37 & 2.55\pm0.31 & 3.61\pm0.48 \\
PSO~J272.0887$-$04.9943    & L$9\pm1  $ & T$3\pm1$ & 15.14\pm0.11 & 1.10\pm0.33 & 1.43\pm0.51  & 1.09\pm1.22  & 0.93\pm0.83 & 0.98\pm0.94 & 0.76\pm0.46 \\
PSO~J319.3102$-$29.6682    & L$7.5\pm1$ & T$3\pm1$ & 14.05\pm0.28 & 1.45\pm0.06 & 0.35\pm0.40  & 1.13\pm0.49  & 1.66\pm0.75 & 1.51\pm0.68 & 2.23\pm0.85 \\
\enddata
\end{deluxetable}

\subsubsection{Objects inside the Gap}
\label{discussion.binaries.inside}

The T1 dwarf SDSS~J015141.69+124429.6 \citep[hereinafter
SDSS~J0151+1244;][]{Geballe:2002kw} appears single at a resolution of 40~mas
\citep{Burgasser:2006hd} and has no previous spectral indication of binarity
(B10).  Our spectral decomposition does find a good match to an L7.5 + T2 blend,
and plausible matches for blends of spectral types L$8\pm2$ with T$2\pm1$.  For
the most likely match, the difference in \jmko\ between the two components is
$-0.07\pm0.25$~mag, consistent with equal-brightness components.  In this case,
both components would be $\approx$0.75~mag fainter than the combined absolute
magnitude of $14.74$~mag (Figure~\ref{fig.cmd.decomp}).  At $\mjmko=15.5$~mag,
both components would be implausibly faint in $J$-band --- fainter than any
known object with spectral type earlier than T7.  SDSS~J0151+1244 thus appears
to be a single T1~dwarf.

The L9.5 dwarf WISE~J092055.40+453856.3 \citep[hereinafter
WISE~J0920+4538;][]{Mace:2013jh,Best:2013bp} was characterized as a weak binary
candidate with L$7.5\pm1.5$ and T$1.5\pm1.5$ components by \citet{Mace:2013jh}.
Laser guide star adaptive optics imaging of this object using Keck II/NIRC2 with
an angular resolution of 95~mas has detected no sign of a companion (W. Best et
al, in preparation).  Our spectral decomposition identifies a possible blend
combination of L7.5 + T2, but the difference in $J$-band luminosity for the two
components would be $>$1~mag, inconsistent with an expected difference of
$\approx$0.3~mag in \jmko\ (DL12).  We therefore find it unlikely that
WISE~J0920+4538 is a binary.

The T1.5~pec~dwarf PSO~J272.0887$-$04.9943 \citep[hereinafter
PSO~J272.0$-$04.9;][]{Best:2015em}, which sits at the bottom of the gap
(Figure~\ref{fig.cmd.gap.labels}), is a candidate binary \citep{Best:2015em}
based on the appearance of its NIR spectrum and the spectral indices defined by
B10.  However, PSO~J272.0$-$04.9 is already the faintest early-T dwarf in our
sample, and its position at the bottom of the L/T transition sequence would
require potential binary components to be an unusually faint late-L dwarf and a
much fainter late-T dwarf.  Our spectral decomposition does not support this
pairing, finding instead an L9 + T3 pairing whose components are both
$\approx$1.5~mag fainter than expected for those spectral types
(Figure~\ref{fig.cmd.decomp}).  In addition, laser guide star adaptive optics
imaging of this object using Keck II/NIRC2 with an angular resolution of 80~mas
has detected no sign of a companion (W. Best et al, in preparation).
PSO~J272.0$-$04.9 therefore appears to be a single object, and an unusually
faint one for its spectral type.

SDSS~J0151+1244, WISE~J0920+4538, and \linebreak PSO~J272.0$-$04.9 thus all
appear to be single brown dwarfs caught crossing the L/T transition gap.  This
indicates that $\jkmko\approx0.9$--1.4~mag colors are rare ($\approx$5\% of L/T
transition dwarfs) but not forbidden.

\subsubsection{Objects above the Gap}
\label{discussion.binaries.above}

Unresolved binaries that combine the light of two objects will be brighter than
single objects with the same effective temperature.  We therefore expect to find
binaries sitting higher on a CMD than either of the single components.  Four
objects sit directly above the gap in Figure~\ref{fig.cmd.gap.labels}.

The T0~dwarf PSO~J180.1475$-$28.6160 \citep[hereinafter
PSO~J180.1$-$28.6;][]{Best:2015em} is a candidate binary based on its spectral
features.  Laser guide star adaptive optics imaging of this object using Keck
II/NIRC2 with an angular resolution of 110~mas has detected no sign of a
companion (W. Best et al, in preparation).  However, our spectral decomposition
finds a best match for types L7.5 + T2 with reasonable differences in component
fluxes, along with other plausible matches for L5/L6 + T1/T2 dwarfs.  At
$\mjmko=14.01\pm0.29$~mag, PSO~J180.1$-$28.6 could plausibly be either a bright
single T0~dwarf or a late-L + early-T blend.

SDSS~J120747.17+024424.8 \citep[SDSS~J1207+0244;][]{Hawley:2002jc}
is the NIR T0 spectral standard \citep{Burgasser:2006cf}, but has also been
identified as a candidate L6 + T3 spectral blend (B10).  It has not been
observed with high-resolution imaging.  SDSS~J1207+0244 has
$\mjmko=13.72\pm0.17$~mag, which would make it the brightest T~dwarf in our
sample if it is in fact single, almost a full magnitude above the center of the
gap.  On the other hand, our spectral decomposition finds a good match for L6 +
T2, similar to that of B10.  We therefore regard SDSS~J1207+0244 as a likely
binary.

SDSS~J133148.92$-$011651.4 (SDSS~J1331$-$0116) was discovered and
assigned a spectral type of L6 by \citet{Hawley:2002jc} based on an optical
spectrum.  Subsequent analyses of NIR photometry and spectra have noted the
object's unusually blue color and atypical spectral features, finding spectral
types of L$8\pm2.5$ \citep{Knapp:2004ji}, L1~pec \citep{Marocco:2013kv}, L6.5
\citep{BardalezGagliuffi:2014fl}, and L6 \citep{Marocco:2015iz}.  While this
degree of spectral-type discrepancy often points to a blend of binary components
with different spectral types, \citet{Knapp:2004ji} and \citet{Marocco:2013kv}
explain the spectral features as indicative of low metallicity, consistent with
the object's blue colors.  SDSS~J1331$-$0116's position on the CMD in
Figure~\ref{fig.cmd.gap.labels} is also consistent with that of an unusually
blue mid-L~dwarf.  Our spectral decomposition finds a best pairing of L6 + T5
with $\Delta J=1.97\pm0.18$~mag, much larger than expected from the observed
absolute magnitudes of these spectral types \citep[$\Delta J\approx0.6$~mag;
DL12;][]{Filippazzo:2015dv}, adding evidence that this object is not a spectral
binary.  SDSS~J1331$-$0116 has not been observed with high-resolution imaging.

The T0~dwarf PSO~J319.3102$-$29.6682 (hereinafter PSO~J319.3$-$29.6) is a
candidate spectral blend \citep{Best:2015em} as well as a candidate member of
the $\beta$ Pictoris Moving Group \citep{Best:2015em,Best:2020jr}.  The
best pair found by our spectral decomposition is L7.5 + T3, but for this
particular blend the T3~dwarf would need to be implausibly fainter ($<$1~mag) in the
$J$ and $H$ bands than the L~dwarf.  The $\mjmko=14.05$~mag of PSO~J319.3$-$29.6
is nevertheless consistent with the combined flux from $\approx$L8 and
$\approx$T2~dwarfs, as well as with the fluxes from brighter single early-T
dwarfs.  Laser guide star adaptive optics imaging of this object using Keck
II/NIRC2 with an angular resolution of 60~mas has detected no sign of a
companion (W. Best et al, in preparation).  We draw no conclusion about whether
PSO~J319.3$-$29.6 is a tighter unresolved binary.

\subsubsection{No Suggestion of Binarity in the Astrometric Solutions}
\label{discussion.binaries.chisq}
Astrometric solutions derived from observations of binaries can have larger
residuals and/or less accurate results if there is significant photocenter
motion due to the binary's orbital motion or if the binary is partially
resolved.  For five of the seven objects with gap colors described above, the
parallaxes come from the Best20 sample.  We examined the \rchi\ for their
astrometric solutions for indications of larger residuals and/or poor solutions.
For the five objects WISE~J0920+4538, PSO~J180.1$-$28.6, SDSS~J1207+0244,
PSO~J272.0$-$04.9, and \linebreak PSO~J319.3$-$29.6, the $\chi^2$/degrees of
freedom are 22.4/19, 3.6/7, 18.2/23, 12.9/17, and 8.9/11, respectively, all
within the range of typical values for the Best20 solutions.  We see no
indication of \rchi\ much greater than 1 or widely disparate values that could
indicate problems with the solutions due to partially resolved binaries.  We
also visually examined the solutions for these fits and found no structure in
the residuals.  The other two gap-color objects (SDSS~J0151+1244 and
SDSS~J1331$-$0116) do not have published \rchi\ for their astrometric solutions
\citep{Vrba:2004ee,Smart:2018en}.

\subsubsection{Objects outside the Volume-limited Sample}
\label{discussion.binaries.outside}

\begin{figure}
  \centering
  \includegraphics[width=1\columnwidth, trim = 12mm 0 10mm 0]{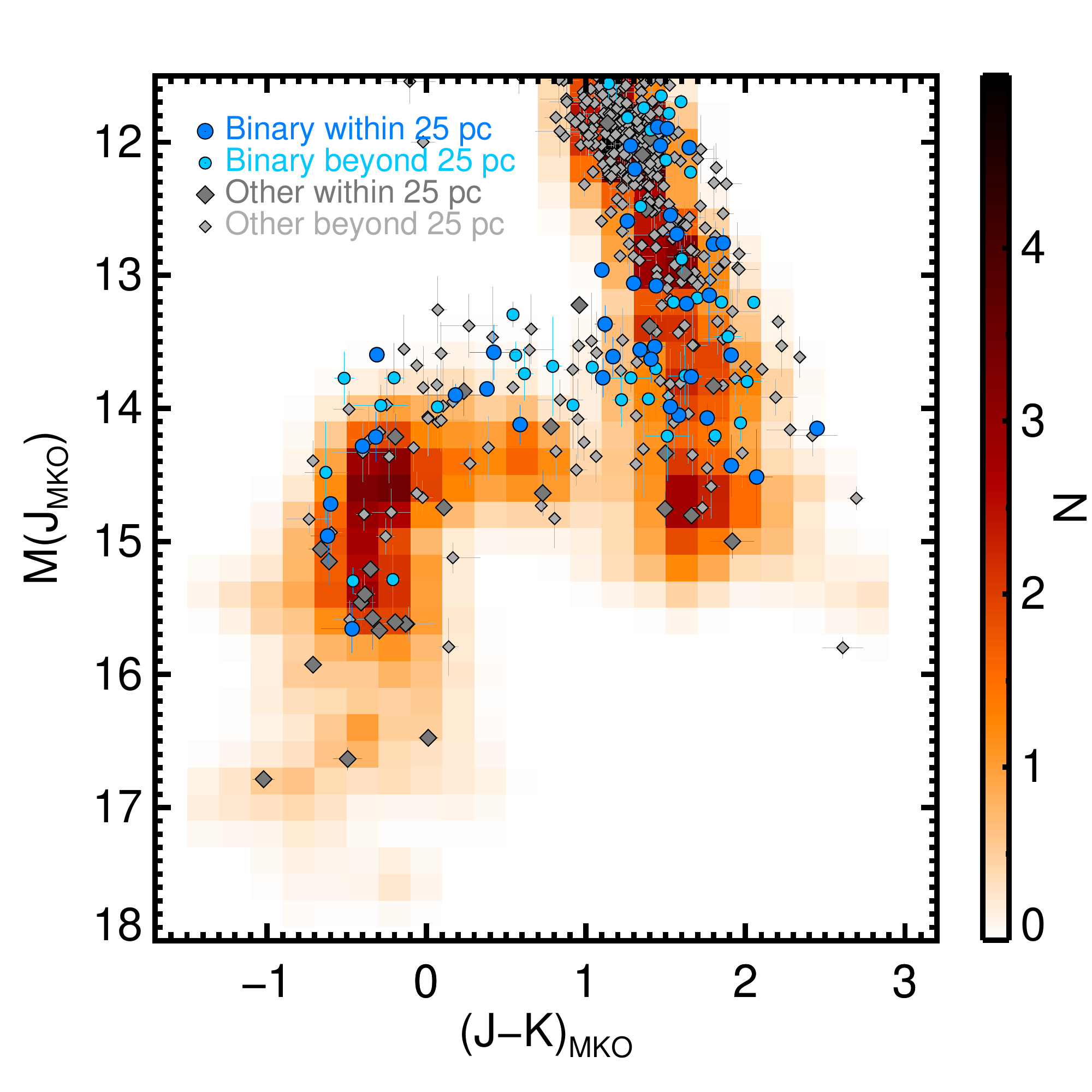}
  \caption{Density map of the CMD from Figure~\ref{fig.cmd.jjk}, overlaid with
    objects having parallaxes but excluded from our volume-limited sample of
    single objects, i.e., known binaries within 25~pc (larger dark blue circles)
    and beyond 25~pc (smaller light blue circles), and other apparently single
    objects (including companions) within 25~pc (larger dark gray diamonds) and
    beyond 25~pc (smaller light gray diamonds). The known binaries are clearly
    brighter as a population than the single objects, and the brightest
    apparently single objects in the L/T transition may actually be unrecognized
    binaries.}
  \label{fig.cmd.heat.excluded}
\end{figure}

Figure~\ref{fig.cmd.heat.excluded} shows the density map of the 25~pc
volume-limited sample from Figure~\ref{fig.cmd.jjk} overlaid with the L0--T8
objects having published parallaxes that are not part of our volume-limited
sample, including apparently single objects beyond 25~pc and known binaries at
any distance.  The known binaries are clearly
$\approx$$0.5-0.7$ brighter as a population than the single objects, indicating
that the brightest putatively single objects in the L/T transition may indeed be
unrecognized binaries.  Most of the objects directly above the gap are known
binaries, suggesting that their apparent colors may be a blend of two components
on either side of the gap similar to our decompositions for
PSO~J180.1$-$28.6 and SDSS~J1207+0244.  We note there are also several known
binaries in our full volume-limited sample (included in
Figure~\ref{fig.cmd.jjk.binyoung} and excluded from Figure~\ref{fig.cmd.jjk})
that lie in the same region directly above the gap.

\subsection{Do Models Predict the L/T Transition Gap?}
\label{discussion.models}

The ``hybrid'' models of SM08 are the only models that agree with the
mass-luminosity relationship of L/T transition dwarfs
\citep{Dupuy:2015gl,Dupuy:2017ke}.  Briefly, the hybrid models are a set of
evolutionary models coupled to cloudy atmospheres for objects with
$\teff>1400$~K (essentially L dwarfs), clear atmospheres for objects with
$\teff\le1200$~K (mid-T and later dwarfs), and a linear interpolation between
the two for $1200<\teff\le1400$~K (the L/T transition), interpolating the
surface boundary condition in \teff\ for each value of gravity.  We note that
this modeling of the L/T transition was developed to explore the ramifications
of cloud clearing in a notional sense and does not invoke a specific physical
explanation for the cloud clearing. The starting and ending {\teff} for the
transition were chosen to approximate the NIR colors of the ultracool dwarf
sequence.

\begin{figure*}
  \centering
  \begin{minipage}[t]{0.48\textwidth}
    \includegraphics[width=1\columnwidth, trim = 12mm 0 10mm 0]{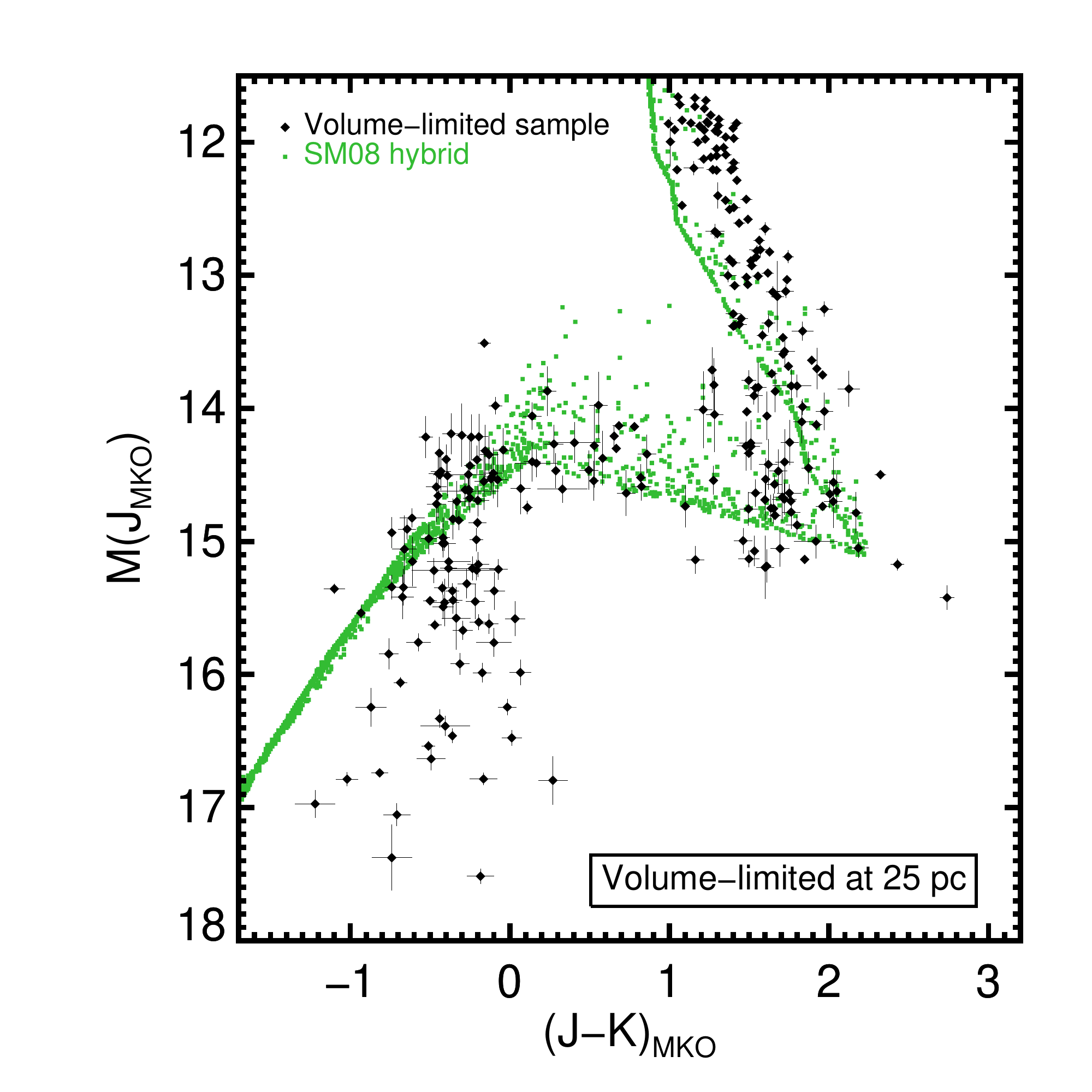}
  \end{minipage}
  \hfill
  \begin{minipage}[t]{0.48\textwidth}
    \includegraphics[width=1\columnwidth, trim = 12mm 0 10mm 0]{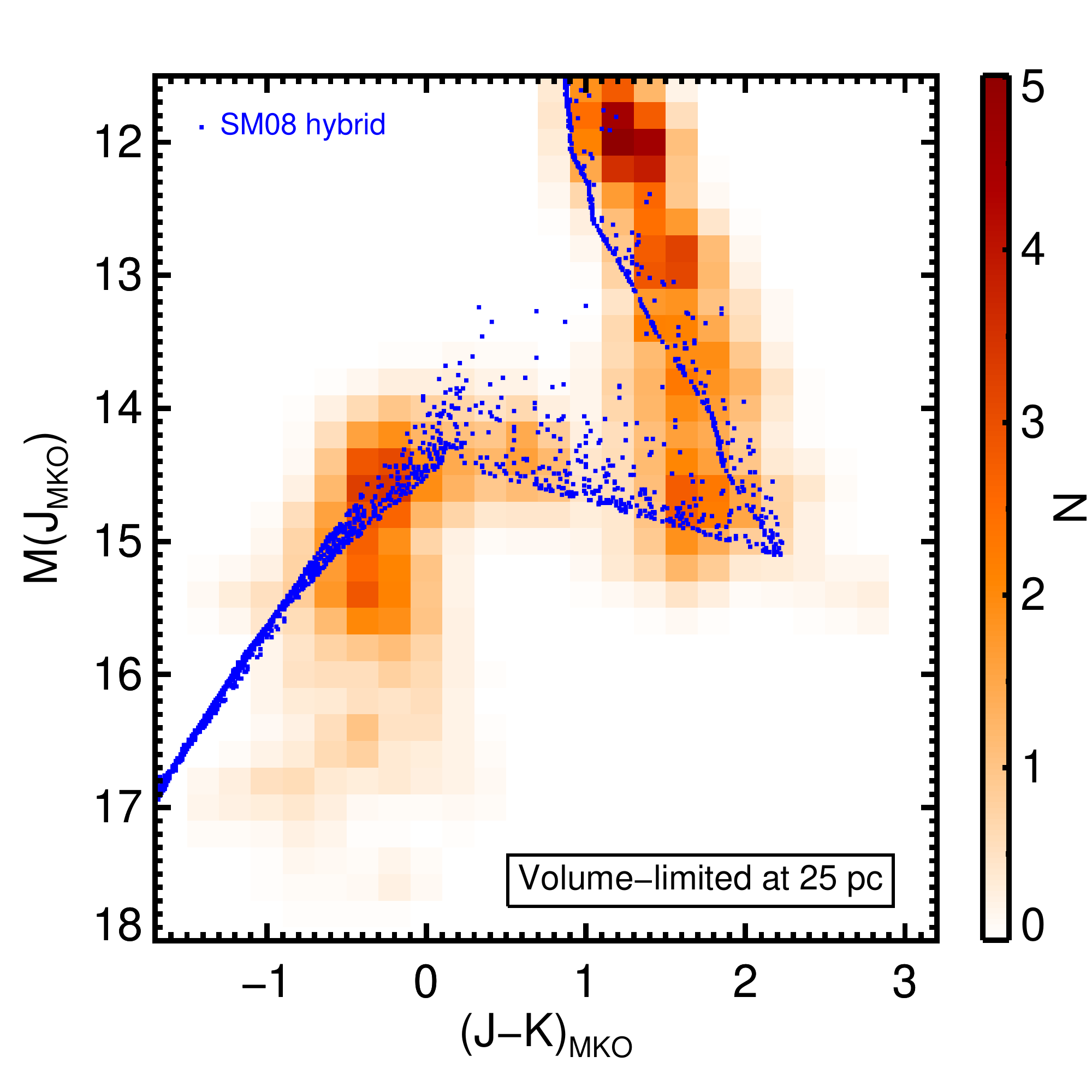}
  \end{minipage}
  \begin{minipage}[t]{0.48\textwidth}
    \includegraphics[width=1\columnwidth, trim = 12mm 0 10mm 0]{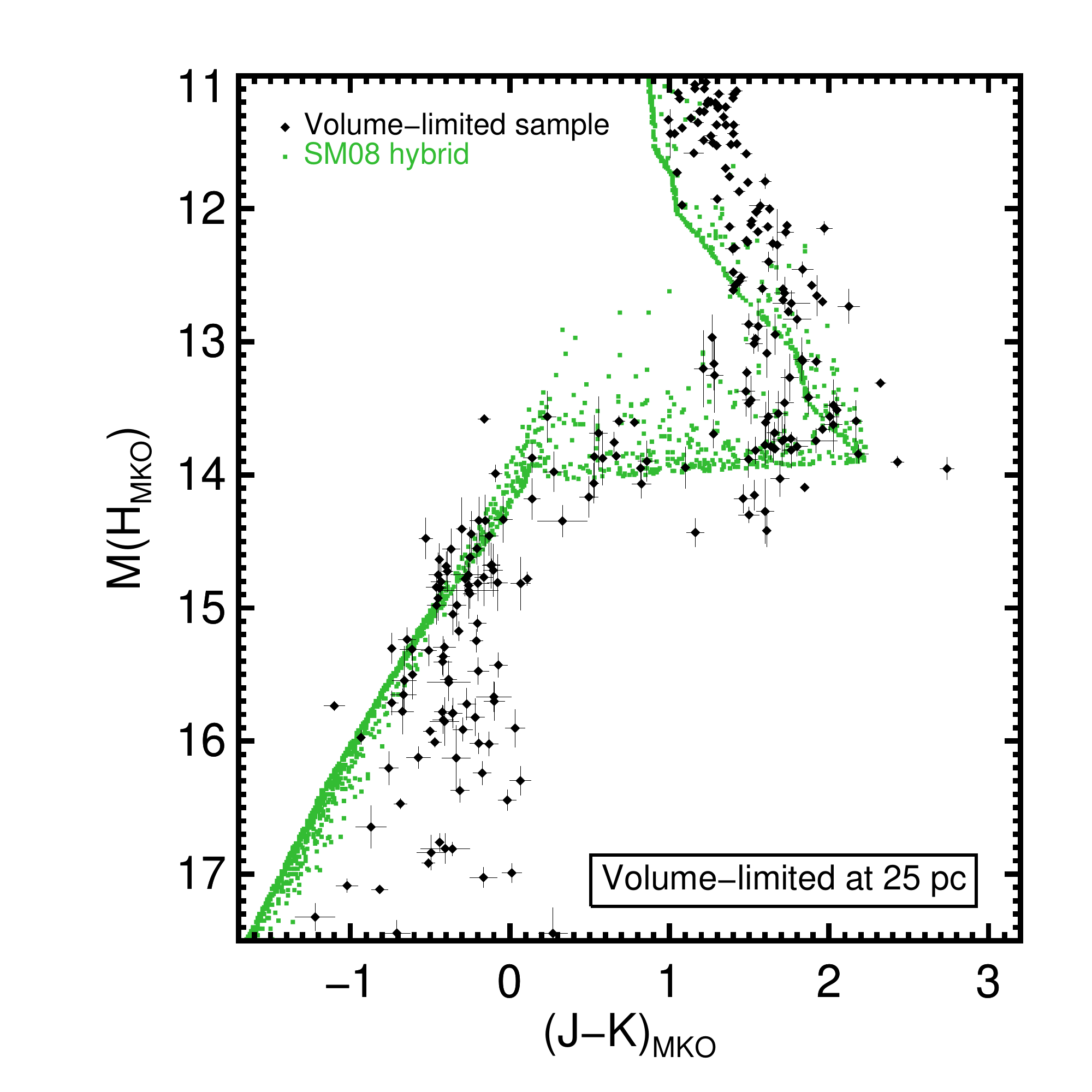}
  \end{minipage}
  \hfill
  \begin{minipage}[t]{0.48\textwidth}
    \includegraphics[width=1\columnwidth, trim = 12mm 0 10mm 0]{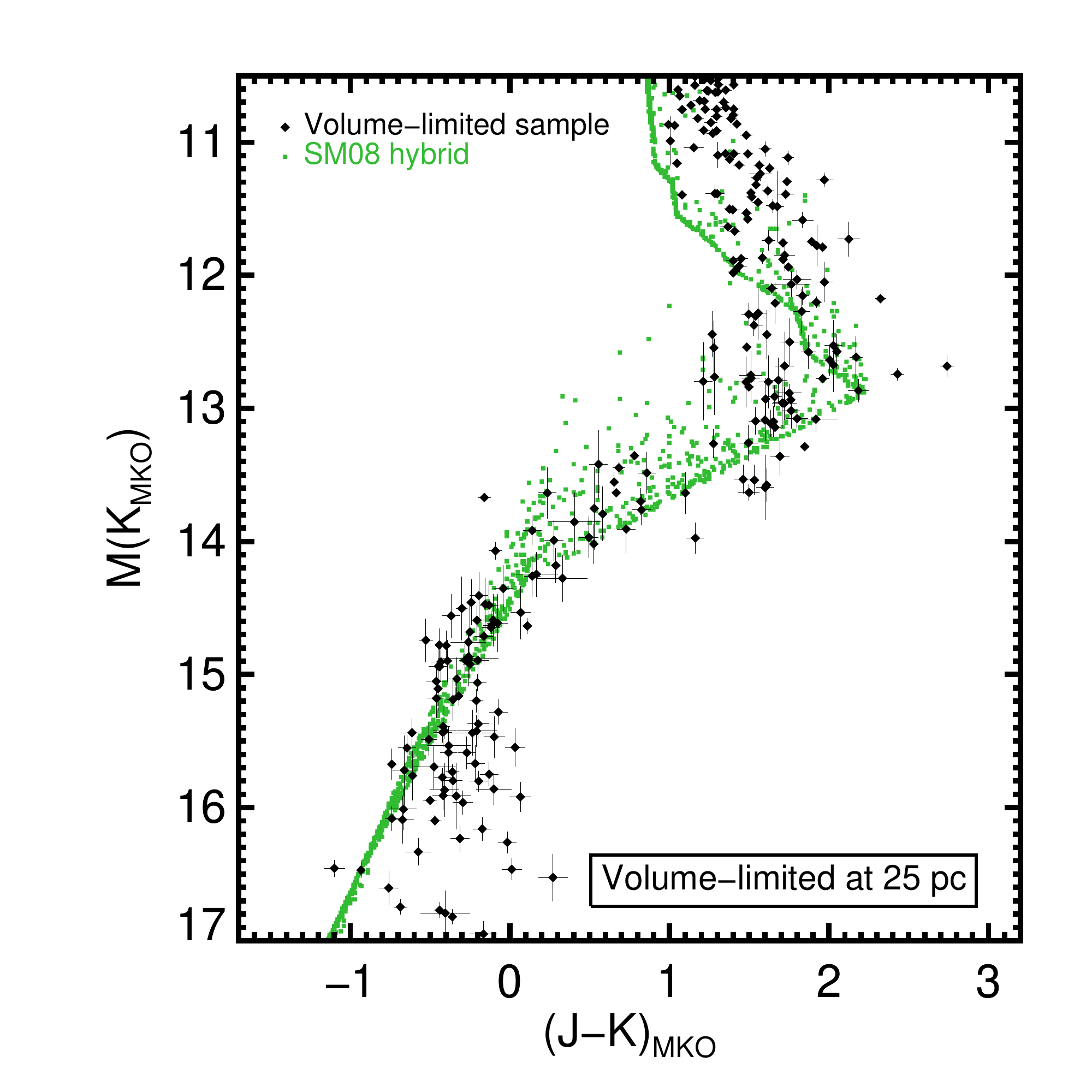}
  \end{minipage}
  \caption{$M_J$, $M_H$, and $M_K$ vs. $J-K$ (MKO) color-magnitude diagrams for
    single objects in our volume-limited sample (black), overlaid on the
    synthetic population derived by SM08 from their "hybrid" model (green,
    except in the top right panel where we use blue for visual clarity).  The
    top two panels reproduce the objects and density map from
    Figure~\ref{fig.cmd.jjk}.  The bottom right panel updates Figure~14 from
    SM08 with our much larger and volume-limited sample.  The hybrid models
    generally reproduce the shape of the L and T~dwarf evolutionary sequence,
    but predict significantly different \jkmko\ colors for most L~dwarfs and
    late-T~dwarfs.  In the L/T transition, the models show an uneven
    distribution of objects across the $J-K$~colors, but predict an overdensity
    of objects at the $J-K$~colors of our gap.}
  \label{fig.cmds.sm08}
\end{figure*}

We compare the volume-limited synthetic population of L and T~dwarfs generated
by SM08 from their hybrid models to our volume-limited sample of single objects
in Figure~\ref{fig.cmds.sm08}.  The hybrid models generally reproduce the
overall shape of the L and T~dwarf evolutionary sequence, but predict \jkmko\
colors that are too blue for early-L and late-T~dwarfs and too red for
late-L~dwarfs.  As described in SM08, this synthetic population did not attempt
to fully capture the observed scatter within the sequence, which is likely due
to variations in gravity, cloud properties, and/or metallicity, but SM08 also
generated alternative populations based on different assumptions to explore the
effects of these variations.  The original synthetic population shown in
Figure~\ref{fig.cmds.sm08} assumes a 0--10~Gyr uniform age distribution, and the
visible scatter in luminosities originates from this spread of ages, as younger
brown dwarfs have larger radii (lower gravity) and are therefore more luminous.
We note that this uniform age distribution is not consistent with the expected
younger age distribution of our sample (Section~\ref{vollim.young}).

For comparison, SM08 generated a younger population (0-5 Gyr) that spreads the
population somewhat toward brighter absolute magnitudes (lower gravity), better
matching the scatter in our volume-limited sample.  SM08 also generated an
old-skewing population (exponential decline in star-formation rate with
characteristic time 5~Gyr) which makes their sequence thinner and clearly a
worse match to our volume-limited sample, especially along the L~dwarf sequence
where the color offset is exaggerated for this population.

The SM08 synthetic population in Figure~\ref{fig.cmds.sm08} also assumes a
power-law IMF of the form ($\frac{dN}{dM}\propto M^{-\alpha}$) with $\alpha=1$;
\setcitestyle{notesep={; }} we note that while $\alpha$ is poorly constrained by
observations, most estimates have found $-0.5\lesssim\alpha\lesssim0.5$
\citep[e.g.,][K19]{Allen:2005jf,Metchev:2008gx,Burningham:2013gt}.
\setcitestyle{notesep={, }} Using a log-normal IMF, SM08 also generated a
population with fewer low-mass (and thus low-gravity and low-\teff) objects, so
this synthetic sequence is narrower toward the cooler end, which is again a
poorer match to our volume-limited sample, supporting their initial assumption
of a bottom-heavy IMF.  Finally, SM08 use cloudless models (appropriate for
later-T~dwarfs) with a moderate range of metallicities ($-0.9\le [M/H] \le 0.3$)
to generate a synthetic population that displays considerably more scatter in
its CMD sequence, suggesting that metallicity contributes significantly to the
scatter for the T~dwarfs in our volume-limited sample.

\begin{figure*}
  \centering
  \includegraphics[width=2\columnwidth]{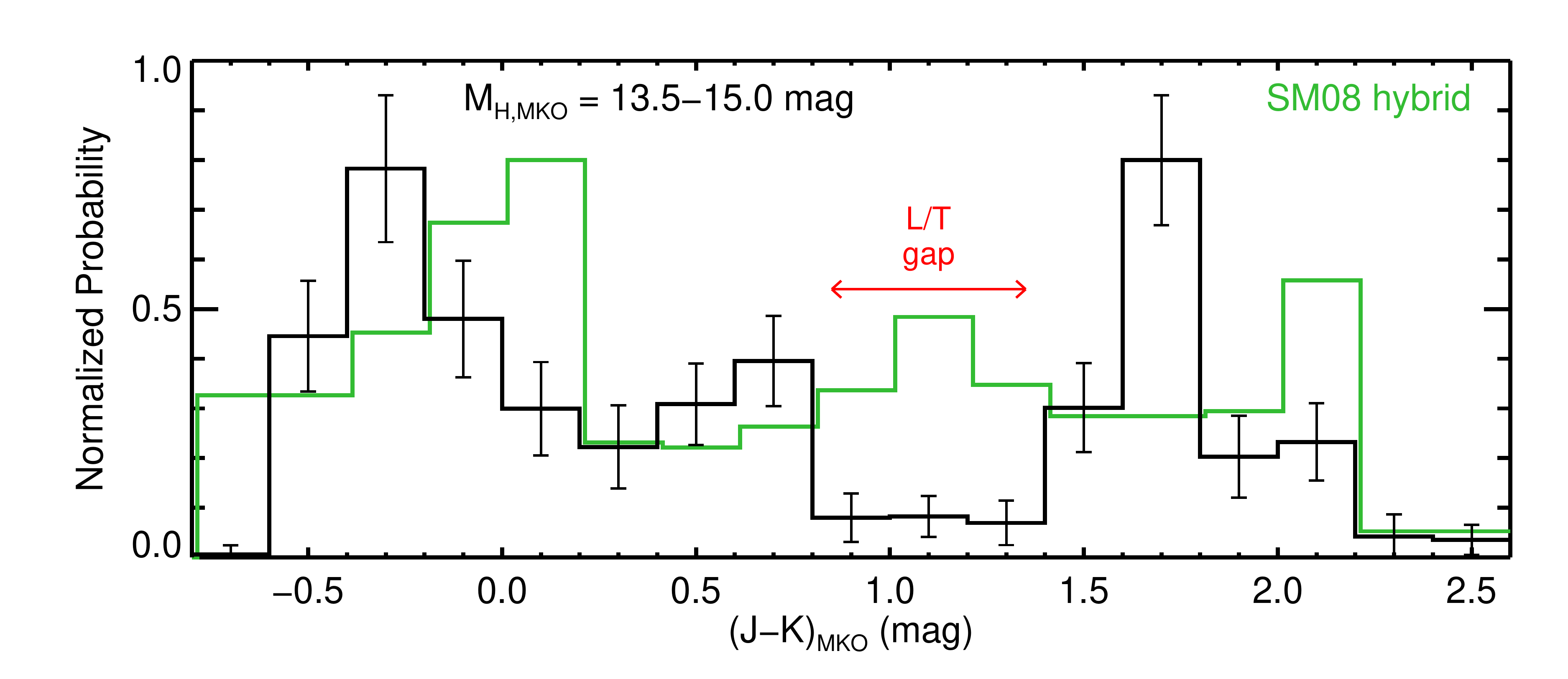}
  \caption{Same as Figure~\ref{fig.hist.jk}, but also including the distribution
    of \jkmko\ colors for the synthetic population generated by SM08 from their
    hybrid model (green, slightly offset horizontally for clarity).  As with the
    volume-limited sample, we computed the SM08 histogram for objects having
    $13.5\le\mhmko\le15.0$~mag.  The relative pileup of the synthetic
    population at the colors of our gap (labeled in red) is evident.  Shifting
    the synthetic colors blueward by $\approx$0.5~mag would appear to improve
    the agreement with our volume-limited sample (although without a significant
    gap), highlighting the need for a model that explains the cloud processes
    that lead to the L/T transition color changes.}
  \label{fig.hist.jk.sm08}
\end{figure*}

In the L/T transition, the SM08 hybrid models approximately replicate the slopes
of each of the NIR CMDs (Figure~\ref{fig.cmds.sm08}), and show an uneven
distribution of objects across the $J-K$~colors (Figure~\ref{fig.hist.jk.sm08})
that is reminiscent of the gap and clumps in our volume-limited sample.
However, the peaks of the model color distribution are inconsistent with our
sample; in particular, the models predict a ``pileup'' of objects at the
$J-K\approx1$~mag location of our gap, as well as a clear paucity of objects at
$J-K\approx0.3$--0.5~mag where our volume-limited sample shows only a marginal
underdensity.  Our volume-limited sample instead has clumps of objects at the
beginning ($J-K\approx1.5$~mag) and end ($J-K\approx-0.5$~mag) of the L/T
transition.  To quantitatively assess the degree of consistency of these $J-K$
distributions, we performed a two-sided Kolmogorov-Smirnov test and found a
probability of 0.06 that the two sets of L/T transition colors are drawn from
the same population.

\begin{figure*}
  \centering
  \includegraphics[width=2\columnwidth]{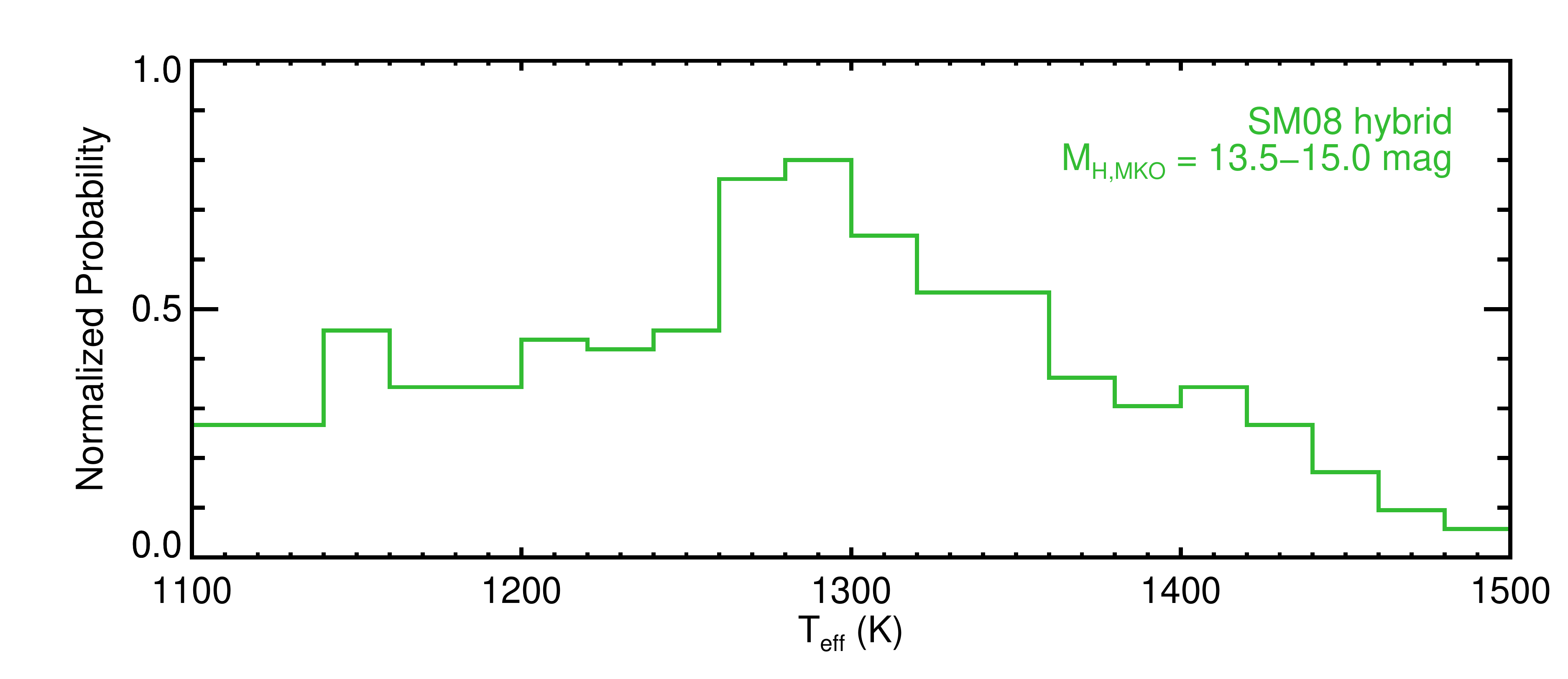}
  \includegraphics[width=1\columnwidth]{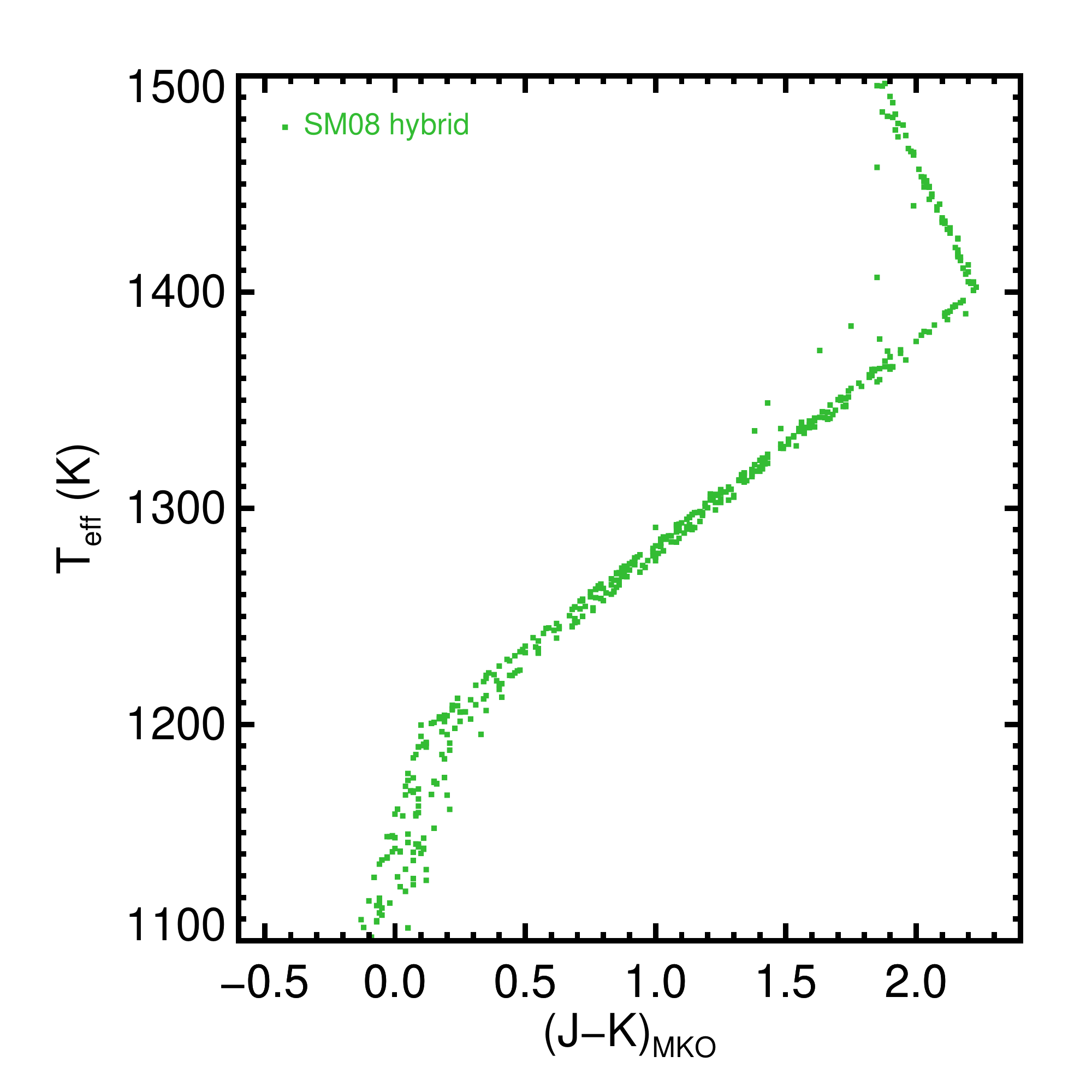}
  \caption{Top: distribution of model-derived \teff\ for the same synthetic
    population of single L/T~transition dwarfs from the SM08 hybrid model shown
    in Figure~\ref{fig.cmds.sm08}.  Bottom: \teff\ vs. $J-K$ (MKO) for this
    synthetic population.  The peak in the \teff\ distribution near 1300~K
    corresponds to the pileup at $J-K\approx1$~mag in
    Figure~\ref{fig.hist.jk.sm08}.  Changing the hybrid model's \teff\
    prescription for the L/T transition could shift the $J-K$ color of the
    pileup and improve agreement with the color distribution of our
    volume-limited sample.}
  \label{fig.teff.sm08}
\end{figure*}

SM08 attributed their pileup to a slowdown in the cooling of L/T~transition
dwarfs, as heat trapped by L-dwarf clouds takes time to radiate away when the
clouds begin to clear.  Figure~\ref{fig.teff.sm08} demonstrates that this pileup
in $J-K$ corresponds directly to a maximum in the \teff\ distribution at
$\approx$1300~K.  This suggests that the disagreement in color distribution with
our volume-limited sample could be mitigated by adjusting the SM08 temperature
prescription for the L/T transition to move their pileup to $\approx$0.5~mag
bluer colors (Figure~\ref{fig.hist.jk.sm08}).

\section{Summary}
\label{summary}
We present a volume-limited sample of \varnvollim\ L0--T8 dwarfs chosen entirely
by parallaxes, twice as large as any previous parallax-defined ultracool dwarf
sample and more than 10 times larger than the last sample covering our full
spectral type range.  Our sample spans 68.3\% of the sky ($\delta=-30^\circ$ to
$+60^\circ$), extends out to 25~pc, and combines parallaxes from Best20, \gaiat,
and the literature.  Using the {\exvmax} statistic, we determine that our sample
is {\varcompletefull} complete.  Breaking our sample into spectral type bins,
we find it is complete for T0--T4.5~dwarfs, {\varcompletel} complete for
L~dwarfs, and {\varcompletelate} complete for T5--T8~dwarfs, making ours the
first volume-limited sample to provide an unbiased picture of the L/T transition
(spectral types $\approx$L8--T4).  We find a completeness-corrected binary
fraction of {\varbinfraccorr} for our sample, and another {\varwidecompfraccorr}
are companions to main-sequence stars (except for one wide pair of brown dwarfs,
SDSS~J141624.08+134826.7 and ULAS~J141623.94+134836.3).  {\varyoungfraccorr} of
our sample is young objects ($\lesssim$200~Myr) but only {\varsubdfraccorr} is
subdwarfs, implying that the local brown dwarf population is younger than the
standard assumption of a 0--10~Gyr age distribution, in agreement with other
recent studies.

Our volume-limited sample reveals a previously unidentified gap at
$\jkmko\approx0.9$--1.4~mag (spectral types $\approx$T0--T3;
$\teff\approx1300$~K) in the L/T transition.  The gap's existence implies a
rapid blueward evolution in color resulting from changes in the atmospheres of
these cooling brown dwarfs.  The gap is apparent in several other NIR colors
spanning $y$ through $K$ bands, but is not present in the MIR \wawb\ color.  Two
objects that sit directly above the gap on the $M_J$ versus $J-K$ CMD are good
candidate binaries whose unresolved blends give them gap-like $J-K$ colors.  On
the other hand, the three objects that lie within the gap all appear to be
single objects in the process of crossing the gap.

The evolutionary and atmospheric models that to date have most accurately
matched the observed luminosities of L/T~transition dwarfs -- the ``hybrid''
models of SM08 -- also predict an L/T transition with an uneven distribution of
colors, reminiscent of the gap and comparative overdensities in our
volume-limited sample.  However, our gap is located at $J-K$ colors where the
hybrid models predict a pileup of objects, suggesting that a different
temperature prescription for the L/T transition is needed.

Our volume-limited sample is ideally suited for unbiased population studies of
local L and T~dwarfs, including atmospheric characteristics, properties of
binaries and companions, space densities, the luminosity function, and
assessment of the IMF and birth history underlying the formation of nearby
ultracool dwarfs, topics that will be addressed in upcoming papers.  Our sample
also provides a stepping stone to the expanded volume-limited samples that the
upcoming Rubin Observatory Legacy Survey of Space and Time (LSST) will provide.

\vspace{20 pt}

We thank Didier Saumon and Mark Marley for providing the synthetic population
data used in SM08 and for helpful comments on the manuscript.
This research has benefitted from the SpeX Prism Library and the SpeX Prism
Library Analysis Toolkit, maintained by Adam Burgasser at
\url{http://www.browndwarfs.org/spexprism}.
This work has made use of data from the European Space Agency (ESA) mission
Gaia (\url{http://www.cosmos.esa.int/gaia}), processed by the Gaia Data
Processing and Analysis Consortium (DPAC,
\url{http://www.cosmos.esa.int/web/gaia/dpac/consortium}). Funding for the DPAC
has been provided by national institutions, in particular the institutions
participating in the Gaia Multilateral Agreement.
This publication makes use of data products from the Two Micron All Sky Survey
(2MASS), which is a joint project of the University of Massachusetts and the
Infrared Processing and Analysis Center/California Institute of Technology,
funded by the National Aeronautics and Space Administration and the National
Science Foundation. This research has made use of NASA's Astrophysical Data
System and the SIMBAD and Vizier databases operated at CDS, Strasbourg, France.
W.M.J.B. received support from NSF grant AST09-09222, and grant HST-GO-15238
provided by STScI and AURA.
W.M.J.B., M.C.L., and E.A.M. received support from NSF grant AST-1313455.
T.J.D. acknowledges research support from Gemini Observatory.
Finally, the authors wish to recognize and acknowledge the very significant
cultural role and reverence that the summit of Maunakea has always held within
the indigenous Hawaiian community. We are most fortunate to have the opportunity
to conduct observations from this mountain.

\software{
TOPCAT \citep{Taylor:2005wq}.
}

\appendix

\section{Calculation of Uncertainties for {\exvmax}}
\label{appendix.uncert.vmax}

To estimate the uncertainties on \exvmax\ for a volume-limited sample or any of
its subsets, we need to account for two contributing sources: the parallax
uncertainties for the objects that comprise our sample, and the fact that our
sample and its subsets have limited sizes and are therefore subject to
statistical variations.  We calculated the uncertainty on {\exvmax} due to the
individual parallax uncertainties using Monte Carlo resampling.
  
To assess the impact of these small-number statistical variations on the
precision of {\exvmax}, we used an approach motivated by that of K19.  They
determined 68\% confidence limits for a quantized binomial distribution of $N$
measurements of {\vmax} centered on 0.5 by treating each {\vmax} measurement as
having an equal probability (i.e., $\frac{1}{2}$) of being less than 0.5 or more
than 0.5.  They then compared their computed {\exvmax} values to these
confidence intervals to assess the distances at which their sample was
consistent with completeness.  Here, we also rely on binomial statistics, but
rather than comparing each of our {\exvmax} values to the expected dispersion of
an assumed uniform distribution, we directly determine the uncertainty for each
{\exvmax} regardless of the underlying distribution of objects.  We calculate
the uncertainty on each {\exvmax}, for a sample of $N$ objects, as the standard
deviation of a binomial distribution for $N$ trials and probability $p=\exvmax$.
The variance for such a binomial distribution is $\sigma^2=Np(1-p)$, so the
standard deviation takes the form
\begin{equation}
  \sigma_N = \sqrt{N\times\exvmax\times(1-\exvmax)}.
\end{equation}
We normalized this standard deviation to the $[0,1]$ range of {\exvmax} by
dividing by $N$, giving us the uncertainty for {\exvmax}:
\begin{equation}
  \sigma(\exvmax)_N = \sqrt{ \frac{\exvmax\times(1-\exvmax)}{N} }.
  \label{sig.vmax}
\end{equation}

In practice, we incorporated our calculation of this binomial uncertainty into
the same Monte Carlo trials we used to assess the impact of the parallax
uncertainties, replacing the {\exvmax} value calculated for each trial with a
random value drawn from a binomial distribution for $N$ events and probability
$p=\exvmax$, divided by $N$.  We then took the mean and standard deviation of
the {\exvmax} values from the Monte Carlo trials as our final {\exvmax} and
uncertainty, respectively.  For a sufficient number of Monte Carlo trials, this
produces the same result as adding the uncertainty due to the parallax errors in
quadrature with $\sigma(\exvmax)_N$ from Equation~\ref{sig.vmax}.

We note that \citet{Avni:1980dx} determined that for a homogeneously distributed
sample of $N$ objects subject to Poisson statistics, {\exvmax} has dispersion
\begin{equation}
  \sigma(\exvmax)_N = \sqrt{ \frac{1}{12N} }.
  \label{sig.avni}
\end{equation}
However, as we did not know a priori if our sample and its subsets were
homogenously distributed, we adopted our approach (Equation~\ref{sig.vmax}),
which directly estimates the uncertainty for each of our subsets without
requiring them to be uniformly distributed --- as indeed they cannot be if
$\exvmax<0.5$.  The uncertainties we calculated using our approach were
typically a little less than twice as large as those estimated using the method
of \citet{Avni:1980dx}.

\section{Calculation of Poisson and Binomial Uncertainties for Space Density}
\label{appendix.uncert.density}

As in Appendix~\ref{appendix.uncert.vmax}, to estimate the uncertainties on our
space density calculations for our volume-limited sample, we needed to account
for (1) the parallax uncertainties of the individual objects that comprise our
sample and (2) the statistical fluctuations due to our finite sample.  The
latter can be assessed in more than one way depending on the statement we wish
to make based on our volume-limited sample.

As a volume-limited sample corrected for incompleteness, our sample is
representative of L0--T8 in our neighborhood of the Galaxy (in the Galactic
midplane, excluding clusters and star-forming regions).  Even in this limited
neighborhood, however, our sample contains only a fraction of the total
population of L0--T8 dwarfs.  A collection of samples with the same volume as
ours from around this neighborhood would therefore be expected to show variation
in numbers of objects, described by Poisson statistics.  So, to make a statement
about the space density of L0--T8~dwarfs in the local Galaxy based on our
sample, we needed to incorporate Poisson errors into our uncertainties, i.e.,
$\sigma_N=\sqrt{N}$ for a sample of size $N$.  For the resulting space density
uncertainty, we thus have
\begin{equation}
  \sigma_{\rm Poisson} = \frac{\sqrt{N}}{V},
  \label{sig.poisson}
\end{equation}
where $V$ is the volume of our sample.

However, if we want to make a statement about the space density of L0--T8~dwarfs
specifically within 25~pc of the Sun, we need a different assessment.  Our
sample does not cover the entire 25~pc volume, so Poisson-like statistical
variations are possible, but because our sample covers a substantial fraction of
the sky, such variations will be smaller than predicted by Poisson theory.  In
this context, binomial statistics accounting for the fraction of sky covered by
our sample (68.3\%) provide a more appropriate estimate of the space density
uncertainty.  For a set of $N_{\rm full}$ objects drawn from the full sky, each
object will be either inside or outside of our sample volume (these are the only
two options), and the probability of finding a given object inside our sample
area is the same ($p=0.683$) for every object.  Given this situation, the
binomial distribution describes the statistics of our sample.  The expectation
value for the number of objects in our sample area is $pN_{\rm full}$, with
standard deviation $\sqrt{ p(1-p) N_{\rm full} }$.  Thus, for our observed
sample of $N$ objects drawn from a fraction $p$ of the sky, we would expect that
$N = pN_{\rm full}$ with uncertainty
\begin{equation}
  \sigma_N = \sqrt{ N(1-p) }.
  \label{sig.binomial.num}
\end{equation}
This gives us an uncertainty for the space density of 
\begin{equation}
  \sigma_{\rm binomial} = \frac{\sqrt{ N(1-p) }}{V},
  \label{sig.binomial}
\end{equation}
where $V$ is again the volume of our sample.

\begin{figure}
  \centering
  \includegraphics[width=0.5\columnwidth]{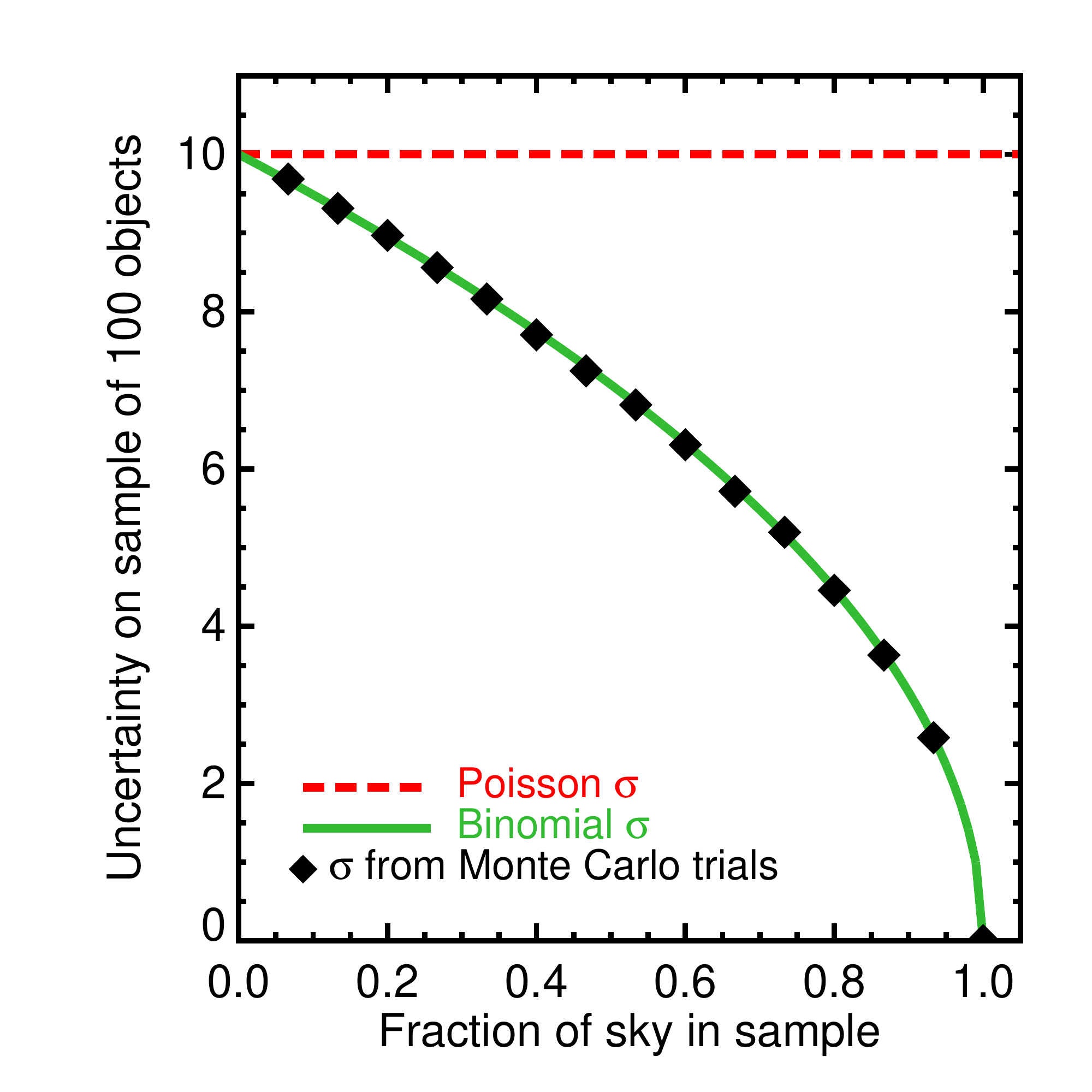}
  \caption{Uncertainties on the number of objects in a sample of 100 objects due
    to statistical fluctuations, for a range of coverage fractions of the sky,
    calculated using Monte Carlo trials (black diamonds). The green curve shows
    the uncertainties derived from the binomial distribution, which clearly
    match the Monte Carlo results. The red dashed line marks the Poisson
    uncertainty of $\sqrt{100}$, which is accurate only when the sample is a
    small fraction of the whole population. Our sample covers 68.3\% of the sky,
    so our estimate of the space density in the full 25~pc volume has an
    uncertainty 0.56 times the Poisson uncertainty.}
  \label{fig.poisson.binomial}
\end{figure}

To demonstrate this approach, we ran Monte Carlo trials to determine the scatter
among the sizes of samples covering different fractions of the sky when the mean
sample size is 100.  Figure~\ref{fig.poisson.binomial} shows that the scatter
values are predicted by the binomial distribution, i.e.,
$\sigma_N = \sqrt{ N(1-p) }$.  Figure~\ref{fig.poisson.binomial} also compares
these binomial uncertainties to the fixed Poisson uncertainty for a sample of
100 objects.  The Poisson distribution approximates the binomial distribution
when $N$ is large and $p$ is small \citep[the Poisson limit
theorem;][]{Poisson:1837wh}, i.e., when the sample in question is a small
fraction of a large full population.  As $p$ increases, the binomial uncertainty
falls steadily below the Poisson value and reaches zero when the sample covers
the whole sky (i.e., a volume-complete sample has no uncertainty for determining
the number of objects in that specific volume).  For our volume-limited sample
with $p=0.683$, the binomial uncertainty is 0.56 times the Poisson uncertainty.

As with our {\exvmax} uncertainty calculations in
Appendix~\ref{appendix.uncert.vmax}, we incorporated our calculation of the
Poisson and binomial space density uncertainties into Monte Carlo trials along
with the parallax uncertainties.  To incorporate the Poisson uncertainties into
our calculations, we replaced the number of objects $N$ in each Monte Carlo
trial with a random number drawn from a Poisson distribution having mean value
$N+0.5$ \citep[this more accurately accounts for Poissonian likelihoods when $N$
is small;][]{Metchev:2008gx}.  To incorporate the binomial errors into our
calculations, we replaced the number of objects $N$ in each Monte Carlo trial
with a random number drawn from a binomial distribution for $N/p$ events and sky
fraction $p$.  We then computed the mean and standard deviation of the Monte
Carlo trials and divided by our sample volume to obtain the final space density
uncertainties.

Binomial uncertainties are also the appropriate choice for comparing the space
densities of two samples that cover the same volume of space.  For example, to
compare our volume-limited sample, which covers 68.3\% of the full 25~pc volume,
to a sample that covers a different (possibly overlapping) portion of the full
25~pc volume, binomial uncertainties should be used for the
completeness-corrected space densities of both samples.  To compare the space
density of our sample with that of a volume-limited sample that covered the
entire 25~pc volume, one would use the binomial uncertainties for our sample and
no uncertainties for the full-volume sample.

In summary, we have described two distinct ways to estimate the uncertainty on
the space density of our volume-limited sample, each appropriate for different
uses for our sample.  To describe how well our sample represents L and T~dwarfs
in our general neighborhood of the Galaxy, of which our sample is a small
portion, Poisson uncertainties ($\sigma_{\rm Poisson}$,
Equation~\ref{sig.poisson}) are appropriate.  When we narrow our context to the
25~pc volume around the Sun, of which our sample covers a much larger portion,
binomial uncertainties ($\sigma_{\rm binomial}$, Equation~\ref{sig.binomial})
should be used to describe how precisely our volume-limited sample represents
the full 25~pc volume.  Binomial uncertainties are also the appropriate
uncertainties to use when comparing space densities for two different samples
that represent the same volume of space.

\section{Additional MKO Photometry}
\label{appendix.mko.phot}

In Table~\ref{tbl.mko.phot} we list NIR \ymko, \jmko, \hmko, and
{\kmko} photometry for {\varnnewphot} objects, compiled during the development
of our volume-limited sample.  Table~\ref{tbl.mko.phot} contains a mixture of
new and previously published photometry, but for each of these objects the
photometry is new in at least one band.  This table includes all of the new
photometry presented for \varnnewphotvollim~L0--T8 dwarfs in our volume-limited
sample (Table~\ref{tbl.sample}), along with photometry for an additional
\varnnewphotother~M, L, and T~dwarfs.  Most of the new photometry we present
here is either synthetic or converted from 2MASS using the methods described in
Section~\ref{vollim.photometry}.  In addition, we present new {\ymko} photometry
for 26 objects, {\jmko} photometry for 2 objects, and {\hmko} photometry for 3
objects from observations on multiple nights in 2010--2013 using WFCAM
\citep{Casali:2007ep} on the 3.8-meter United Kingdom InfraRed Telescope
(UKIRT).  The observations were conducted and data were reduced in standard
fashion as described in \citet{Best:2015em}.

\begin{longrotatetable}
\begin{deluxetable}{lcCCCCr}
\centering
\tablecaption{New MKO Photometry \label{tbl.mko.phot}}
\tabletypesize{\scriptsize}
\tablewidth{0pt}
\tablehead{   
  \colhead{Object} &
  \colhead{Spectral Type\tablenotemark{a}} &
  \colhead{\ymko} &
  \colhead{\jmko} &
  \colhead{\hmko} &
  \colhead{\kmko} &
  \colhead{References} \\
  \colhead{} &
  \colhead{(Optical/NIR)} &
  \colhead{(mag)} &
  \colhead{(mag)} &
  \colhead{(mag)} &
  \colhead{(mag)} &
  \colhead{(Disc; SpT; Phot)}
}
\startdata
WISEA J000131.93$-$084126.9 & .../L1pec & $[16.65\pm0.07]$ & $[15.65\pm0.05]$ & $[15.12\pm0.09]$ & $[14.67\pm0.09]$ & 154; 154; 1 \\
SDSS J000250.98+245413.8 & .../L5.5 & $[18.34\pm0.23]$ & $[17.09\pm0.22]$ & $[16.14\pm0.22]$ & $[15.64\pm0.22]$ & 43; 43; 1 \\
WISEA J000430.66$-$260402.3 & .../T2 (blue) & $[17.28\pm0.06]$ & $16.18\pm0.02$ & $[15.83\pm0.06]$ & $[15.57\pm0.05]$ & 99; 99; 1,16 \\
2MASS J00054844$-$2157196 & M9/... & $[14.07\pm0.06]$ & $[13.24\pm0.03]$ & $[12.68\pm0.02]$ & $[12.18\pm0.03]$ & 191; 189; 1 \\
2MASSI J0006205$-$172051 & L2.5/L3.0 & $[16.87\pm0.08]$ & $[15.61\pm0.06]$ & $[14.72\pm0.05]$ & $[13.99\pm0.05]$ & 108; 108,7; 1 \\
WISEA J000622.67$-$131955.2 & .../L5pec & $[17.64\pm0.14]$ & $[16.59\pm0.13]$ & $[15.63\pm0.13]$ & $[15.11\pm0.13]$ & 154; 154; 1 \\
SDSS J000649.16$-$085246.3 & M9/... & $[14.98\pm0.06]$ & $[14.10\pm0.03]$ & $[13.61\pm0.04]$ & $[13.11\pm0.04]$ & 226; 226; 1 \\
2MASS J00070787$-$2458042 & M7/... & $[13.94\pm0.05]$ & $[13.09\pm0.02]$ & $[12.50\pm0.02]$ & $[12.04\pm0.02]$ & 191; 47; 1 \\
\enddata
\end{deluxetable}
\end{longrotatetable}

\end{document}